\pdfoutput=1
\documentclass[aps,nofootinbib,twocolumn,groupedaddress,floatfix,eqsecnum]{revtex4}
\usepackage{graphicx}
\usepackage{epsfig}
\usepackage{epstopdf}
\usepackage[mathscr]{eucal}
\usepackage{amsmath,amssymb,bm,graphics}
\usepackage[pdftex,letterpaper=true]{hyperref}

\begin{document}

\def\nn{\nonumber}
\def\ts{\textstyle}

\def\Nfour{\mathcal N\,{=}\,4}
\def\Ntwo{\mathcal N\,{=}\,2}
\def\Nc{N_{\rm c}}
\def\Nf{N_{\rm f}}
\def\x{\bm x}
\def\q{\bm q}
\def\f{\bm f}
\def\v{\bm v}
\def\t{\bm t}
\def\S{\mathcal S}
\def\T{\mathcal T}
\def\O{\mathcal O}
\def\E{\mathcal E}
\def\p{\mathcal P}
\def\H{\mathcal H}
\def\K{\mathcal K}
\def\w{\omega}
\def\uh{u_h}
\def\del{\nabla}
\def\eps{\hat \epsilon}
\def\inf{\epsilon}
\def\cs{c_{\rm s}}

\def\half{{\textstyle \frac 12}}
\def\twothirds{{\textstyle \frac 23}}
\def\third{{\textstyle \frac 13}}
\def\coeff#1#2{{\textstyle \frac{#1}{#2}}}

\title
{The stress-energy tensor of a quark moving through a strongly-coupled 
\boldmath $\Nfour $ supersymmetric Yang-Mills plasma:
comparing hydrodynamics and AdS/CFT}

\author{Paul~M.~Chesler\footnotemark}
\author{Laurence~G.~Yaffe\footnotemark}

\affiliation
    {Department of Physics,
    University of Washington,
    Seattle, WA 98195-1560, USA}

\date{\today}

\begin{abstract}
The stress-energy tensor of a quark moving through a
strongly coupled $\Nfour$ supersymmetric Yang-Mills plasma is evaluated
using gauge/string duality.
The accuracy with which the resulting wake, in position space,
is reproduced by hydrodynamics is examined.
Remarkable agreement is found between hydrodynamics and
the complete result down to distances less than
$2/T$ away from the quark.  In performing the gravitational analysis,
we use a relatively simple formulation of the bulk to boundary problem
in which the linearized Einstein field equations are fully decoupled.
Our analysis easily generalizes to other sources in the bulk.
\end{abstract}


\maketitle
\iftrue
\def\thefootnote{\fnsymbol{footnote}}
\footnotetext[1]{Email: \tt pchesler@u.washington.edu}
\footnotetext[2]{Email: \tt yaffe@phys.washington.edu}
\def\thefootnote{\arabic{footnote}}
\fi

\section{Introduction}
The discovery that the quark-gluon plasma produced 
in heavy ion collisions at RHIC
behaves as a nearly ideal fluid \cite{Shuryak,Shuryak:2004cy}
has prompted much interest in understanding the dynamics of
strongly coupled non-Abelian plasmas.
Much recent theoretical work has explored the dynamics of
maximally supersymmetric Yang-Mills ($\Nfour$ SYM) plasma.
(See, for example, Refs.~\cite
    {%
    Policastro:2001yc,
    Policastro:2002tn,
    CaronHuot:2006te,
    Herzog:2006gh,
    Casalderrey-Solana:2006rq,
    Caceres:2006dj,
    Janik:2006gp,
    Liu:2006ug,
    Peeters:2006iu,
    CasalderreySolana:2007qw,
    Gubser:2006nz,
    Hatta:2007cs,
    Bak:2007fk,
    Heller:2007qt,
    Peeters:2007ti}.)
Thanks to gauge/string (or AdS/CFT) duality \cite{Aharony:1999ti,Maldacena:1997re},
the properties of a strongly coupled $\Nfour$ $SU(\Nc)$ SYM plasma,
in the limit of large $\Nc$,
are under much better theoretical control than is the case
for a strongly coupled QCD plasma.
Available evidence
from the equation of state \cite{Aharony:1999ti},
screening lengths \cite{Bak:2007fk},
and viscosity \cite{Policastro:2001yc,Meyer:2007ic},
suggests that a strongly coupled, large-$\Nc$, $\Nfour$ SYM plasma mimics the
properties of a real QCD plasma in the temperature range relevant
to heavy ion collisions (roughly 1.5 to 4 times the deconfinement
temperature) sufficiently well that one may view
$\Nfour$ SYM plasma as a useful model system for QCD.

Heavy quarks produced by hard processes during the early stages 
of a heavy ion collision may transfer much of their energy and momentum
to the medium if they travel a sufficient distance through
the fireball before potentially escaping (and hadronizing).
Analysis of the distribution and correlations in produced jets
can provide information about the dynamics of the plasma,
including the rates of energy loss and momentum broadening
of quarks traversing the plasma
\cite{Shuryak:2006ii,Adler:2005ee,Leitch:2006ex}.  

A heavy quark moving through a plasma will disturb the surrounding medium
and its motion will, in turn, be influenced by the medium.
As the quark moves, frictional forces will transfer energy
and momentum from the quark to the plasma.
A natural question to consider is
where does the energy and momentum lost by the quark go?
In other words, what is the form of the wake,
as defined by the change in the expectation
value of the stress-energy tensor,
\begin{equation}
    \Delta T^{\mu\nu}(x)
    \equiv
    \langle T^{\mu\nu}(x) \rangle_{\rm with \; quark}
    -
    \langle T^{\mu\nu}(x) \rangle_{\rm w/o \; quark} .
\end{equation}
For distances asymptotically far from the quark,
one may address this question using hydrodynamic approximations.
This approach was used in Ref.~\cite{CasalderreySolana:2006sq}.
However, in this work the form of the effective sources
for hydrodynamics were not fully specified.

Gauge/string duality
allows one to compute many observables
probing non-equilibrium dynamics of strongly coupled 
$\Nfour$ supersymmetric Yang-Mills theory,
including the rate of energy loss of a heavy
quark moving through an SYM plasma \cite{Herzog:2006gh}.
(See also Refs.~\cite {%
    Casalderrey-Solana:2006rq,%
    Caceres:2006dj,%
    Herzog:2006se,%
    %Friess:2006aw,%
    Friess:2006fk,%
    CasalderreySolana:2007qw,%
    Gubser:2006nz,%
    Gubser:2007nd,%
    Yarom:2007ni%
    %Yarom:2007ap%
    }
and references therein.)
Recent studies have calculated the energy density 
\cite{Chesler:2007an, Gubser:2007xz}
and energy flux \cite{Gubser:2007ga,Gubser:2007ni}
associated with a heavy quark moving through
a strongly coupled SYM plasma.  
These studies have shown that qualitative features of the quark wake,
such as the formation of a Mach cone for supersonic motion,
match what is expected from hydrodynamics.
However the comparison between hydrodynamics and 
the exact result for the stress-energy tensor has not yet been performed in a 
quantitative fashion.
In particular, the range of validly of hydrodynamics 
has not been addressed quantitatively.

Hydrodynamics is valid only on length (or time) scales which are large
compared to the mean free path (or time) of typical excitations in a fluid.
In a weakly coupled relativistic plasma, the mean free path of quasiparticle
excitations (quarks or gluons) is parametrically large compared to their
de Broglie wavelength,
$
    \ell_{\rm mfp} \sim 1/(T \lambda^2 \ln \lambda^{-1} )
    \gg
    \ell_{\rm de\;Broglie} \sim 1/T
$
\cite{Arnold:2002zm}.
The 't Hooft coupling $\lambda \equiv g^2 \Nc$ is the appropriate measure
of the interaction strength.
As the size of the coupling increases,
this separation of scales shrinks,
and hydrodynamics becomes valid on progressively shorter distance scales.
When $\lambda \gtrsim 1$, a description in terms of weakly interacting
quasiparticles is no longer valid.
For an ultra-relativistic non-Abelian plasma in this regime,
the minimum length scale $\ell_{\rm hydro}$ on which hydrodynamics is valid
must formally be of order $1/T$, as there is no other relevant scale.
But whether, in practice, one needs
$\ell_{\rm hydro} \gtrsim 1/T$ or
$\ell_{\rm hydro} \gtrsim 100/T$
in order for hydrodynamics to reproduce results of the full theory,
to a given accuracy,
cannot be determined without a quantitative comparison.
Performing such a comparison is one goal of this paper.

To apply hydrodynamics to our situation of interest,
in which a moving heavy quark is transferring energy and momentum to the plasma,
one must first formulate appropriate effective sources to be used with
the hydrodynamic equations.
Just as in other applications of long distance effective field theories,
this requires matching the effective theory to the underlying
microscopic theory at the required level of precision.
We show, on general grounds, that the appropriate effective sources for
hydrodynamics can be simply expressed in terms of the
drag force acting on the quark.
Using gauge/string duality, we then compute the energy density and
energy flux (or momentum density) associated with a heavy quark moving through
a strongly coupled $\Nfour$ SYM plasma,
and compare the exact results to those obtained with hydrodynamics.
Comparisons are made in both momentum space and position space.
We find that the hydrodynamic approximation to the quark wake agrees
with the exact result remarkably well even
at distances less than $2/T$ away from the quark.  
Although our comparison applies to an $\Nfour$ SYM plasma
(in the limit of large $\lambda$ and large $\Nc$)
the remarkably good performance of hydrodynamics in this context
can only bolster the hope that it is sufficient to use hydrodynamics
to model the transport of energy and momentum lost by high energy
particles traversing
a real quark-gluon plasma \cite{Betz:2007ie,Baeuchle:2007qw}
(despite the fact that gradients in the hydrodynamic variables
can be disconcertingly large).

The outline of the remainder of this paper is as follows.
In Section \ref{defs} we introduce various definitions and conventions
that we will use throughout the analysis.
In Section \ref{Hydro} we discuss the hydrodynamic description of
perturbations in the SYM stress-energy tensor.
This includes the formulation of effective sources for the hydrodynamic equations
and the separation of the energy flux into contributions from sound and diffusion modes.
In Section \ref{gravitational} we turn to the dual gravitational formulation.
Gauge/string duality maps the problem of computing the perturbation
in the stress-energy due to the quark,
$\Delta T^{\mu\nu}(x)$,
into the problem of computing the perturbation to the geometry
caused by an open string (dual to the heavy quark) moving through
a five dimensional anti-de Sitter/Schwarzschild spacetime.
We introduce a convenient set of gauge invariant variables which encode
the metric perturbation, show how their use allows one to decouple
completely the linearized Einstein equations, and then,
by analyzing the on-shell gravitational action,
show how to reconstruct the expectation value of the SYM stress-energy tensor
using our chosen gauge invariant variables.
We also describe our technique for solving, numerically,
the ordinary differential equations satisfied by the gauge invariant variables,
which is based on the construction of appropriate Green's functions
from numerically computed homogeneous solutions.
We present, in Section \ref{Results},
the results of the computation of the perturbation
in the energy density and energy flux, in position space, 
due to a moving quark, at several different velocities.
We also examine the small momentum asymptotics and compare results,
in several ways, with hydrodynamics.
Section \ref{Discussion} discusses the interpretation
of our results.
We conclude in Section \ref{Conclusions}.
Two appendices contain details of the analysis of the boundary action,
and the extraction of small momentum asymptotics.

We have endeavored to make 
the presentation relatively self-contained.
Readers interested in the results but not the 
details of the gravitational calculation
should feel free to skip over Section \ref{gravitational}.

%%%%%%%%%
\section{Definitions and Conventions}
\label{defs}

We use the Minkowski space metric $\eta_{\mu \nu} \equiv {\rm diag}(-1,1,1,1)$.
Five dimensional AdS coordinates will be denoted by $X_M$ while
four dimensional Minkowski coordinates will be denoted  by $x_\mu$.
Upper case Latin indices $M,
N, P, \dots$ run over $5d$ AdS coordinates,
while Greek indices run over the $4d$ 
Minkowski space coordinates. 
We choose coordinates such that the metric of the AdS-Schwarzschild 
(AdS-BH) geometry is
\begin{equation}
\label{metric}
    ds^2 = \frac{L^2}{u^2}
    \left [-f(u) \, dt^2 + d \x^2 + \frac{du^2}{f(u)} \right ] ,
\end{equation}
where $f(u) \equiv 1-(u/u_h)^4$ and $L$ is the AdS curvature radius.
The coordinate $u$ is an inverse radial coordinate;
the boundary of the AdS-BH spacetime is at $u = 0$
and the event horizon is located at $u=u_h$,
with $T \equiv (\pi u_h)^{-1}$ the temperature of the SYM plasma.

When introducing a Fourier transform over the $4d$ Minkowski space 
coordinates, we will often decompose vectors and tensors in terms of an
orthonormal basis of polarization vectors
$
    \{ \hat {\q}, \,
	\hat {\bm \epsilon}_{1}, \,
	\hat {\bm\epsilon}_{2} \}
$.
The quark's velocity $\v$ defines a preferred direction so a 
natural choice of polarization vectors is
\begin{equation}
\label{eps1}
    \hat {\bm\epsilon}_1 =
    \frac{q}{q_{\perp}} \,
    \hat {\q} \times (\hat {\v} \times \hat {\q}) \,,
\quad
    \hat {\bm\epsilon}_2 =
    \frac{q}{q_{\perp}} \, \hat {\v} \times \hat {\q} \,,
\end{equation}
where
$q \equiv |\q|$ and
$q_{\perp} = |\q - (\hat {\v} \cdot \q) \,  \hat {\v}|$
is the magnitude of the component of $\q$ orthogonal to $\v$.
Lower case Latin indices $a,b,\dots = 1,2$
will be used to refer to the {\it transverse} spatial components 
of vectors and tensors.
We will decompose the Fourier transform $A^M(\omega,\q)$ 
of any given vector field as follows,
\begin{subequations}
\begin{align}
A^0 &\equiv \mathcal A^0, \\
A^5 &\equiv \mathcal A^5, \\
A^i &\equiv \hat q^i  \mathcal A^{q} + \epsilon_a^{i}  \, \mathcal A^{a} \,,
\end{align}
\end{subequations}
with a sum over repeated indices implied.
The components
$
    \mathcal A^{M} \equiv \{
    \mathcal A^0,
    \mathcal A^a,
    \mathcal A^q,
    \mathcal A^5 \}
$
will be referred to as 
the components of the vector field $A$ in the polarization frame.
Higher rank tensors will be represented by polarization frame
components in the analogous fashion.

It will also prove convenient to use the notation
\begin{equation}
    \vec Z_1 \equiv Z^a_1 \> \hat{\bm \epsilon}_a \,,\qquad
    \tensor Z_2 \equiv Z^{ab}_2 \> \hat{\bm \epsilon}_a \otimes \hat{\bm \epsilon}_b \,,
\end{equation}
for quantities which transform with helicity one or two under SO(2) rotations
about the $\hat {\q}$ axis.

In four dimensions, any symmetric tensor field $T^{\mu \nu}$
which satisfies a conservation equation,
\begin{equation}
\partial_\mu \, T^{\mu \nu} = V^{\nu} \,,
\end{equation}
and the trace condition,
\begin{equation}
T^{\mu}_{\ \mu} = \beta \,,
\end{equation}
for some given $V^{\mu}$ and $\beta$,
contains only five independent degrees of freedom.
A convenient representation of these independent degrees of freedom
is provided by the following helicity zero, one, and two
components of the Fourier transformed tensor:
\begin{subequations}
\begin{align}
\T_0 &\equiv \T^{00},
\\ 
\vec \T_1 &\equiv \T^{0a} \, \eps_a,
\\
\tensor \T_2 &\equiv \left(\T^{ab}-\half \, \T^{cc} \, \delta^{ab} \right)
\eps_a\otimes\eps_b \,,
\end{align}
\end{subequations}
where $\T^{\mu \nu}$ are the components of $T^{\mu \nu}$ in the
polarization frame.
The reconstruction of the original tensor components
$T^{\mu \nu}$ is given by
\begin{subequations}
\label{t00}
\begin{align}
T^{00} &= \T_0 \,,
\\ 
T^{0i} &= \eps_a^i \, \T_1^a + \hat q^i \, \T^{0q} \,,
\\
T^{ij} &=
    \eps_a^i \eps_b^j \, \T^{ab}_2
    + \hat q^{(i} \eps^{j)}_a \, \T^{aq}
    + \coeff 32 \left(\hat q^i \hat q^j - \coeff 13 \,\delta^{ij} \right)\T^{qq}
\nn\\ &\kern 1.7cm {}
    + \coeff 12 \left( \delta^{ij} - \hat q^i \hat q^j \right)
    		\left( \beta + \T^{00} \right) ,
\end{align}
\end{subequations}
where $v^{(i} u^{j)} \equiv v^i u^j + v^j u^i$
and
\begin{subequations}
\label{taq}
\begin{align}
\T^{0q} &\equiv \left (\omega \T_0 -i \mathcal V^0 \right ) / q \,,
\\
\T^{aq}&\equiv \left ( \omega \T_1^a - i \mathcal V^a \right ) / q \,,
\\
\T^{qq} &\equiv \left (\omega \T^{0q} -i \mathcal V^q \right ) / q
\nn\\ &=
\left[ \omega^2 \T_0-i (\omega \mathcal V^0 + q \mathcal V^q)  \right ] / q^2\,,
\end{align}
\end{subequations}
with $\mathcal V^{\mu}$ the components of $V^{\mu}$ in the polarization
frame.
We will refer to the quantities
$\{ \T_s \}$ as {\it helicity variables} and the representation of
$T^{\mu \nu}$ in terms of $\T_s$ as the
{\it helicity decomposition} of $T^{\mu \nu}$.
We emphasize that the helicity decomposition is complete only when the helicity 
variables $\T_s$ {\it and} both $\beta$ and $V^{\mu}$ are known.

\section{Hydrodynamic Description}
\label{Hydro}

We consider a fundamental representation quark of mass $M$ 
moving with constant velocity $\v$ through
an $\Nfour$ SYM plasma at temperature $T$.
We assume that both the quark mass $M$, and its kinetic energy,
are large compared to $T$.%
\footnote
    {%
    Having kinetic energy large compared to $T$ implies that
    Brownian motion of the quark,
    induced by thermal fluctuations in the plasma,
    may be neglected \cite{Herzog:2006gh}.
    Our analysis in the gravitational dual will require
    a stronger condition on the quark mass,
    $M \gg \sqrt\lambda \, T$,
    where $\lambda$ is the (large) 't Hooft coupling.
    Strongly-coupled SYM with massive fundamental hypermultiplets
    has deeply bound mesons whose masses scale as the quark mass $M$
    divided by $\sqrt\lambda$, so the condition
    $M \gg \sqrt\lambda \, T$ implies
    that both mesons and quarks are heavy compared to $T$.
    }

We assume that the quark has been moving at the velocity $\v$
for an arbitrarily long time
$\Delta t \rightarrow \infty$.
Due to its interaction with
the SYM gauge (and scalar) fields,
the quark will perturb the surrounding plasma
and the plasma will exert a friction force, or drag, on the quark.
The drag force on the quark is minus the rate at which
momentum is transferred from the quark to the plasma.
In the absence of any external force, the drag exerted by the plasma
would cause the quark to lose momentum and slow down.
To maintain a constant velocity, 
energy and momentum must be supplied to the quark via an external force
which exactly counterbalances the plasma drag (at a given terminal velocity).
This is naturally accomplished by turning on a constant background
$U(1)$ electric field
which couples to the fundamental representation quark but not to the
adjoint representation SYM degrees of freedom.

The microscopic energy-momentum conservation equation takes the form
\begin{equation}
\label{conservation}
\partial_\mu T^{\mu \nu}(x) = F^{\nu}(x) \,,
\end{equation}
where
$T^{\mu \nu}$ is the stress-energy tensor for the system
(not including the background $U(1)$ electric field),
and
$F^{\nu}$ is the external force density,
or minus the drag force density, acting on the quark.
In the limit of large quark mass,
the quark can be arbitrarily well-localized and the force density
will have point support.
In this regime we may write
\begin{equation}
\label{Fmu}
F^{\mu}(t,\x) = f^{\mu} \, \delta^3(\x - \v t) \,,
\end{equation}
where $f^{\mu} \equiv dp^\mu_{\rm quark}/dt$
is the external force (or minus the drag force) acting on the quark,
and coordinates are chosen so that the quark is at the origin at time $t = 0$.%
\footnote
    {%
    Note that $f^\mu$ is the four-momentum transfer per unit coordinate time,
    not proper time;
    the covariant four-force $dp^\mu_{\rm quark}/d\tau$
    equals $f^\mu/\sqrt{1-v^2}$.
    }
For strongly coupled SYM,
the magnitude of the drag force has been evaluated
(using gauge/string duality) and one finds
\cite{Herzog:2006gh, Casalderrey-Solana:2006rq}
\begin{subequations}
\label{fmu}
\begin{align}
    %\frac{d \bm p_{\rm quark}}{dt}
    \f
    &= \frac{\pi}{2} \, \sqrt{\lambda} T^2 \frac{\v}{\sqrt{1-v^2}} \,,
\\
    %\frac{d p^0_{\rm quark}}{dt}
    f^0
    &=
    %\frac{d \bm p_{\rm quark}}{dt}
    \f
    \cdot \v \,.
\end{align}
\end{subequations}

For distances $d \equiv |\x - \v t|$ of order $1/T$
or less,
gradients in the stress-energy tensor are large,
non-hydrodynamic degrees of freedom are important,
and the complete microscopic theory is needed to compute
the transport of energy and momentum.
But as the disturbance in the plasma, induced by the quark,
propagates out to larger distances,
dissipative effects will decrease the size of gradients.
At sufficiently large 
distances from the quark, the transport of energy and 
momentum can be described by neutral fluid hydrodynamics.
This is the appropriate effective theory for the long wavelength,
slowly relaxing degrees of freedom in a non-Abelian plasma.%
\footnote
    {%
    In Abelian plasmas,
    the appropriate effective theory is
    magneto-hydrodynamics as the magnetic field can have
    an arbitrarily long correlation length.
    But in non-Abelian plasmas, even static magnetic fluctuations
    of the gauge field develop develop a finite correlation length.
    }
The relevant hydrodynamic variables are the locally conserved
energy-momentum densities $T^{0\mu}(x)$.
In particular, as we next discuss,
long wavelength
perturbations in the spatial stress tensor can be expressed entirely
in terms of these conserved densities.

Instead of working directly with the conserved densities $T^{0\mu}(x)$,
it is conventional, and convenient, to introduce
the proper energy density $\epsilon(x)$
and the fluid four-velocity $u^{\mu}(x)$.  
The fluid four-velocity is defined as the velocity of a local reference frame
in which the spatial momentum density vanishes,
and the proper energy density is the energy density in this
local fluid rest frame.  

We will denote with a bar the components of the stress-energy tensor
in the local fluid rest frame (at the location $x$),
If the fluid were in perfect equilibrium, with a globally-defined rest frame,
then in that rest frame the stress-energy tensor would have the form,
\begin{subequations}
\begin{align}
\overline T^{00}_{\rm eq} &\equiv \epsilon \,,
\\
\overline T^{0i}_{\rm eq} &\equiv 0 \,,
\\
\overline T^{ij}_{\rm eq} &\equiv p \> \delta_{ij}  \,,
\end{align}
\end{subequations}
with the pressure $p$ and energy density $\epsilon$ related by
the equilibrium equation of state of the fluid,
$p = p(\epsilon)$.
When the fluid is not in perfect equilibrium,
the definition of the local fluid rest frame allows one to write
\begin{subequations}
\begin{align}
\overline T^{00}(x) &\equiv \epsilon(x) \,,
\\
\overline T^{0i}(x) &\equiv 0 \,,
\\
\overline T^{ij}(x) &\equiv p(x) \, \delta_{ij} + \tau_{ij}(x) \,,
\end{align}
\end{subequations}
where $p(x) \equiv p(\epsilon(x))$ is the equilibrium value of the pressure
which corresponds (via the equation of state) to the
local energy density $\epsilon(x)$,
and all non-equilibrium effects are contained in
the dissipative contribution to the stress tensor, $\tau_{ij}$.

For sufficiently long wavelength variations in the energy and momentum
density (or proper energy density and fluid velocity),
the dissipative contribution to the stress may be expanded in
terms of spatial gradients of the hydrodynamic variables.
This, by definition, is the hydrodynamic regime.

It is straightforward to construct the gradient expansion of $\tau_{ij}$.  
However this exercise is simplified when
(i) the underlying theory is conformal, and
(ii) the variations in the energy and momentum density are parametrically
small compared to their equilibrium value.
In conformal theories, such as $\Nfour$ SYM,
the stress-energy tensor is traceless.
Consequently, $\epsilon = 3p$ and $\tau_{ii} = 0$.
If the variations in energy and momentum density are small,
then one may also expand in powers of departures from equilibrium.
This is the case in our application involving a single
fundamental representation quark interacting with a large $\Nc$ SYM plasma.
In the large $\Nc$ limit,
the equilibrium stress-energy tensor of the plasma is $\O(\Nc^2)$ while
the perturbations in $T^{\mu\nu}$
due to the addition of a single quark are $\O(\Nc^0)$.
Consequently, both the fractional change in the energy density
and the fluid velocity induced by the moving quark are of order $1/\Nc^2$.
It follows that the equations determining perturbations in
the stress-energy tensor must be linear in the large $\Nc$ limit.
More specifically, nonlinear terms in the hydrodynamic equations of motion 
are suppressed by powers of $\Nc$.
For later convenience, let
\begin{subequations}
\begin{equation}
    \E(x) \equiv \epsilon(x) - \epsilon_{\rm eq} \,,
\end{equation}
and
\begin{equation}
    \mathcal P(x) \equiv \frac{\partial p}{\partial \epsilon} \> \E(x)
    = \coeff 13\, \E(x) \,,
\end{equation}
\end{subequations}
denote, respectively,
the deviation of the energy density from its equilibrium value
and the associated deviation in the pressure.
Linearity of the hydrodynamic equations of motion implies
that only terms linear in $\bm u$ and $\E$ (and their derivatives) will be needed in
the gradient expansion of the stress tensor.
With this and the vanishing trace condition in mind, the gradient expansion
of the dissipative stress $\tau_{ij}$ has the form
\begin{align}
\label{tau}
\tau_{ij} = -& \ \eta \left( \del_i u_j + \del_j u_i
 - \twothirds \delta_{ij} \del \cdot \bm u \right)
 \nn\\
  +& \ \Theta \left(\del_i \del_j - \third \delta_{ij} \del^2 \right) \E+\cdots.
\end{align}
The coefficient $\eta$ is the shear viscosity,
while the coefficient $\Theta$ characterizes the 
first correction to viscous hydrodynamics.

The stress-energy tensor in the local fluid rest frame is related to the 
stress-energy tensor in the lab frame by the local boost (in block form)
\begin{equation}
\Lambda(x) = \left(
\begin{array}{ccc}
1 & \bm u(x)^{\rm T}
\\
\bm u(x) & \bm 1
\end{array}
\right ) + \O(\bm u(x)^2).
\end{equation}
Hence, the stress-energy tensor in the lab frame is
\begin{equation}
T^{\mu \nu}_{\rm hydro}(x) =
T^{\mu \nu}_{\rm eq} + \Delta T^{\mu \nu}_{\rm hydro}(x) \,,
\end{equation}
where
\begin{equation}
T^{\mu \nu}_{\rm eq} =
{\rm diag}(\epsilon_{\rm eq},p_{\rm eq},p_{\rm eq},p_{\rm eq}) \,,
\end{equation}
and
\begin{subequations}
\label{Tmunu}
\begin{align}
\label{T00}
\Delta T^{00}_{\rm hydro} &= \E \,,
\\ 
\label{T0i}
\Delta T^{0i}_{\rm hydro} &= (\epsilon_{\rm eq} + p_{\rm eq}) \, u_{i} \,,
\\\nn
\Delta T^{ij}_{\rm hydro} &=
\mathcal P \, \delta_{ij}
- \eta \left( \del_i u_j + \del_j u_i - \twothirds \delta_{ij}
\del \cdot \bm u \right)
 \\
 &\;\qquad\quad {}
 +  \Theta \left(\del_i \del_j - \third \delta_{ij} \del^2 \right) \E
 +\cdots \,,
\label{Tij}
\end{align}
\end{subequations}
(neglecting nonlinear corrections subleading in $1/\Nc^2$).
Further derivative corrections to the above formulas
only enter in Eq.~(\ref{Tij}), which is the constitutive relation
expressing the spatial stress tensor in terms of the hydrodynamic
variables.

\subsection{Equations of Motion and Effective Source}

The equations of motion for the hydrodynamic fluctuations follow
from applying the energy-momentum conservation equation (\ref{conservation}) 
to $T^{\mu\nu}_{\rm hydro}$.
However, the hydrodynamic decomposition of the stress tensor
is not valid in the near zone close to where the source $F^{\nu}$
is non-vanishing.
In this region,
gradients become large and truncating the gradient expansion 
in Eq.~(\ref{tau}) is unjustified, as is the neglect
of all non-hydrodynamic degrees of freedom.
However, in the spirit of effective field theory,
one may formulate effective sources for the hydrodynamic equations,
which encapsulate the dynamics in the near zone.  
Doing so, we write the
energy conservation relation for the hydrodynamic fluctuations as
\begin{equation}
\label{hydrocons}
\partial_\mu T^{\mu \nu}_{\rm hydro}(x) = J^{\nu}(x) \,,
\end{equation}
where $J^{\nu}$ is the effective source.  
Because hydrodynamics is 
a long distance effective theory, the effective source
can be regarded as having point support at the location of the quark.
That is, $J^{\mu}$
may be expanded in terms of delta functions and their derivatives.%
\footnote
  {%
  In other words, the spatial Fourier transform of $J^\nu$
  may be expanded in a Taylor series in the wavevector $\bm q$.
  This is completely analogous to the situation in
  electromagnetism where,
  given localized charge and current densities,
  one may similarly represent the charge and current densities
  as a sum of delta functions and their derivatives,
  and this directly leads to the multipole expansion
  for the electric and magnetic fields in the far zone.
  }
To leading order in gradients we have
\begin{equation}
\label{jmu}
J^\mu(t,\x) = j^\mu_{(0)} \, \delta^{3}(\x - \v t) + \cdots \,.
\end{equation}
The vector $j^\mu_{(0)}$ can in principle depend on time.  However,
during the (assumed large) interval of time in which the quark is moving at
constant velocity,
$j^\mu_{(0)}$ will be time independent.

The effective source $J^{\mu}(x)$
can be determined by calculating the long wavelength limit 
of the stress-energy tensor via the full quantum theory,
and then matching its form
to hydrodynamics.  We do this in Section \ref{Results}
and find, in particular,
that the coefficient $j^\mu_{(0)}$ of the leading term
in the expansion (\ref{jmu})
is simply equal the rate at which the quark transfers four-momentum
to the plasma,
so that $j^\mu_{(0)}$ coincides with the external force (\ref {fmu}).
This value for the leading term in the effective source can also be
deduced with the following simple argument, which nicely highlights
the fact that this relation between the effective source $j^\mu_{(0)}$
and the microscopic drag force emerges even though hydrodynamics
is not applicable in a near zone surrounding the quark.

Assume that the electric field pulling the quark is turned on
at time $t_i$, and turned off at some later time $t_f$.
The total four-momentum transfered from the quark to the plasma is
\begin{equation}
\label{deltap}
\Delta p^{\mu} = \int_{t_i}^{t_f} dt \> f^{\mu}(\v(t)) \,,
\end{equation}
where $f^{\mu}(\v)$ is the external force (\ref{fmu})
(or minus the drag force)
for velocity $\v$.
For times $t$ much later than $t_f$,
all of the four-momentum transfered to the
plasma will have been transported to distances far from the quark.
(As discussed below, since the relevant dynamics is diffusive,
the characteristic distance
$d$ is proportional to $\sqrt{t {-} t_f}$ and diverges for large $t$.)
The resulting four-momentum stored in the far zone is computable
using hydrodynamics and,
for late times $t \gg t_f$, this far-zone four-momentum 
must match the four-momentum (\ref{deltap}) lost by the quark.

In the limit in which the quark has been moving at
constant velocity $\v$ for an arbitrarily long period of time, 
the total four-momentum
transfered to the plasma
(neglecting irrelevant endpoint corrections sensitive to how
the electric field is ramped up and down) is just
\begin{equation}
\label{deltap2}
\Delta p^{\mu} \approx (t_f - t_i) \,  f^{\mu}(\v) \,.
\end{equation}
In the hydrodynamic description,
the total momentum transfered to the far zone at time $t \gg t_f$ 
is given by
\begin{equation}
\Delta p_{\rm hydro}^{\mu}(t) = \int d^3 x \> \Delta T^{0 \mu}_{\rm hydro}(t,\x) \,.
\end{equation}
Rewriting this as
\begin{align}
\Delta p_{\rm hydro}^{\mu}(t)
= \int_{- \infty}^t dt' \>
\frac \partial{\partial {t'}}  \int d^3 x \> \Delta T^{0 \mu}_{\rm hydro}(t',\x) \,,
\end{align}
and then using the effective energy-momentum conservation relation
(\ref{hydrocons}), yields
\begin{align}
\label{deltaphydro2}
\Delta p^{\mu}_{\rm hydro}(t) = \int_{- \infty}^t dt' \int d^3 x
\left [- \del_i \, \Delta T^{i \mu}_{\rm hydro} + J^\mu \right ].
\end{align}
The first term in the integral can be converted to a surface
integral over the sphere at spatial infinity.
This surface term vanishes by causality --- if the quark has only been
moving for a finite period of time,
then perturbations in the stress-energy tensor must vanish at spatial infinity.
In the second term,
the spatial integral of $J^\mu$
just gives the leading term in the expansion (\ref{jmu}) of
the effective source,
$\int d^3x \> J^\mu = j^\mu_{(0)}$ and, as noted above,
$j^{\mu}_{(0)}$ will be time independent during the long interval
during which the quark moves with constant velocity.
Consequently,
for $t \gg t_f$,
we have
\begin{equation}
\label{deltaphydro3}
\Delta p^{\mu}_{\rm hydro}(t) \approx (t_f - t_i) \, j^{\mu}_{(0)} \,.
\end{equation}
Demanding that this reproduce the microscopic result
(\ref{deltap2}) shows that
\begin{equation}
j^{\mu}_{(0)} = f^\mu \,,
\label{eq:j=f}
\end{equation}
as asserted above.

Higher order gradient corrections to $J^\mu$,
which we evaluate in Section \ref{Results},
lead to corrections to the stress-energy tensor
which are suppressed by additional inverse powers
of $| \x|$ in the far zone.  
We note however, that
in the large $\Nc$ limit
\begin{equation}
J^{0}(t,\x) \equiv F^0(t,\x)
=
\f \cdot \v \; \delta^3(\x - \v t) \,,
\end{equation}
with no additional derivative corrections.
This follows from the fact that the linearized hydrodynamic expressions
for the conserved densities
$\Delta T^{0\nu}_{\rm hydro}$,
as given in Eqs.~(\ref{T00}) and (\ref{T0i}),
do not receive any derivative corrections
(and all non-linear terms are suppressed by additional factors of $1/\Nc$).
The form of the gradient corrections to $\bm J$ are also constrained.
These constraints are conveniently formulated in Fourier space.
If the duration of time $\Delta t$ in which the quark has been moving at
constant velocity $\v$ is sent to infinity,
then $\bm J(\omega,\q)$ will be proportional to
$2 \pi \delta(\omega - \v \cdot \q)$.
Furthermore, $\bm J(\omega,\q)$ can only
depend on the vectors $\v$, $\q$, which implies that
$\bm J(\omega,\q)$ must
lie in the plane
spanned by these vectors.%
\footnote
  {%
  In principle one can consider an additional component of $\bm
  J$ proportional to $\v \times \q$.
  Since this is a pseudovector, its coefficient function must
  be a pseudoscalar.
  But there is no pseudoscalar that can be constructed out of $\v$ and
  $\q$ alone, so this term is not allowed.
  }
We therefore can write 
\begin{align}
\bm J(\omega,\q) =
\left [ \v \, \phi_v(\omega,q^2) + i \q  \, \phi_q(\omega,q^2) \right ]
2 \pi \delta(\omega {-} \v \cdot \q) \,.
\label{eq:Jdef}
\end{align}
The functions $\phi_v(\omega,q^2)$ and $\phi_q(\omega,q^2)$ must
be analytic in both $\omega$ and $q^2$,
in a neighborhood of $\omega = \q = 0$,
so that $\bm J(t,\x)$ has an
expansion in terms of derivatives of delta functions 
with point support at the location of the quark.
The condition (\ref{eq:j=f}) implies that
$\phi_v(0,0) = \f \cdot \v / v^2$.

\subsection{Sound and Diffusion Modes}

To compare with the gravitational results presented in Section
\ref{gravitational} it will be advantageous to perform a spacetime 
Fourier transform and express the stress-energy tensor
in terms of the helicity decomposition discussed in Section~\ref{defs}.  In particular,
$\Delta T^{\mu \nu}_{\rm hydro}$ can be reconstructed from the
helicity variables $\Delta \T_0^{\rm hydro}$, $\Delta \vec \T_1^{\rm hydro}$ and $\Delta \tensor \T_2^{\rm hydro}$,
as summarized in
Eqs.~(\ref{t00})--(\ref{taq}) with $V^{\mu} = J^\mu$ and $\beta = 0$.
The ansatz given in Eq.~(\ref{Tmunu})
and the effective conservation relation (\ref{hydrocons}) determine the
functional form of the helicity variables.  
Let $\bm J_{\rm L}$ and $\bm J_{\rm T}$ be the transverse and longitudinal components
of $\bm J$.
Ignoring the second order $\Theta$ term and higher order derivative corrections
in Eq.~(\ref{Tij}),
one finds that the helicity variables are given by
\begin{subequations}
\begin{align}
\label{soundmode}
 \Delta \T_0^{\rm hydro} &= 
 \frac{\rho}{\omega^2 - \cs^2 \, q^2 + i \gamma q^2 \omega } \,,
\\
\label{diffusionmode}
\Delta \vec \T_1^{\rm hydro} &= \frac{\bm J_{\rm T}}{-i \omega + D q^2  }\,,
\\
\label{hydrotensor}
\Delta \tensor \T_2^{\rm hydro} &= 0 \,,
\end{align}
\label{eq:hydroT}
\end{subequations}
where
\begin{equation}
\label{rhodef}
\rho \equiv i \q \cdot \bm J +i \omega \, J^{0} - \gamma \,q^2 J^{0} \,.
\end{equation}
Here $\cs^2 = \partial \epsilon/\partial p$ is the speed of sound,
$\gamma = 4 \eta/ 3(\epsilon+p)$ is the sound attenuation constant, and
$D = \eta/(\epsilon+p)$ is the transverse momentum diffusion constant.
For strongly coupled SYM \cite{Policastro:2002tn, Kovtun:2004de}
\begin{subequations}
\label{eq:cs,gam,D}
\begin{align}
\cs^2 &= 1/3 \,,
\\
\gamma &= 1/(3\pi T) \,,
\\ 
D &= 1/(4 \pi T) \,.
\end{align}
\end{subequations}

The lack of any helicity two component,
Eq.~(\ref{hydrotensor}),
remains true (in the large $\Nc$ limit)
even when higher order gradient corrections are included.
This is a consequence of the fact that $\Delta T^{ij}_{\rm hydro}$
is linear in the perturbations $\E$ and $u^{i}$.
With this constraint,
regardless of the number of spatial derivatives,
one cannot construct a nonzero transverse traceless component
of the spatial stress.

It is straightforward to reconstruct the perturbation in the energy density
$\E$ and energy flux $S_i \equiv \Delta T^{0i}_{\rm hydro}$
from the helicity variables and the effective energy-momentum conservation relation (\ref{hydrocons}).
The energy density and energy flux, in spacetime, are given by
\begin{align}
\label{hydroenergy}
\E(t,\x) =& \int \frac{d \omega}{2 \pi}\frac{d^3 q}{(2 \pi)^3} \>
\frac{\rho}{\omega^2 - \cs^2 \, q^2 + i \gamma \, q^2 \omega } \>
e^{i Q \cdot x },
\\
\bm S_{\rm L}(t, \x) =&
    \int \frac{d \omega}{2 \pi}\frac{d^3 q}{(2 \pi)^3}\>
    \frac{\cs^2 \, i \q J^0+ i\omega \bm J_{L}}
    {\omega^2 - \cs^2 \, q^2 + i \gamma \, q^2 \omega } \,
    e^{i Q \cdot x },
\label{hydrofluxL}
\\ 
\bm S_{\rm T}(t, \x) =&
    \int \frac{d \omega}{2 \pi}\frac{d^3 q}{(2 \pi)^3}\>
    \frac{- \bm J_{\rm T}}{i \omega -D \, q^2   }\,
    e^{i Q \cdot x } \,,
\label{hydrofluxT}
\end{align}
with
$
\bm S(t, \x) \equiv
\bm S_{\rm L}(t, \x) +
\bm S_{\rm T}(t, \x)
$
and the four-vector $Q \equiv (\omega, \q)$.
These expressions show
that the energy density
and the longitudinal component $\bm S_{\rm L}$ of the energy flux satisfy  
a diffusive wave equation,
while the
transverse component $\bm S_{\rm T}$ of the energy flux
satisfies a diffusion equation.
The energy density obeys the inhomogeneous wave equation
\begin{equation}
\label{energysound}
\left(-\partial_0^2 +\cs^2 \, \del^2 +\gamma \del^2 \partial_0 \right ) \E
= \rho \,,
\end{equation}
which describes damped sound waves,
traveling at speed $\cs = 1/\sqrt 3$,
with a source $\rho$ given by Eq.~(\ref{rhodef}).

It is instructive to separate the energy flux into sound and diffusion modes.
Since the longitudinal part of the flux $\bm S_{\rm L}$ satisfies a wave
equation while the transverse part $\bm S_{\rm T}$ satisfies a diffusion
equation, one would naturally expect that $\bm S_{\rm L}$ should be
identified with the energy flux carried by sound waves, while
$\bm S_{\rm T}$ characterizes diffusive energy flux.
But, in a steady-state situation where the quark has been moving
for an arbitrarily long time, things are not so simple.
The effective source $\bm J(\omega,\q)$, Eq.~(\ref{eq:Jdef}),
equals $\delta(\omega - \v \cdot \q)$,
times a function which is regular as $\omega$ and $\q \to 0$.
But its longitudinal projection has a directional singularity at $q = 0$,
\begin{equation}
    \bm J_{\rm L}
    =
    \frac{\q \, (\q \cdot \bm J)}{q^2}
    =
    \q
    \left[
	\frac {\v \cdot \q}{q^2}  \, \phi_v(\omega,q^2)
	+ \cdots
    \right]
    2\pi \delta(\omega{-}\v \cdot \q) \,,
\label{eq:JLsing}
\end{equation}
with $\cdots$ denoting terms regular as $\q \to 0$.
Therefore, the Fourier transform of $\bm J_{\rm L}$
is not localized at the quark, but rather falls off in space like the
inverse cube of the distance away from the quark.
As a result, the longitudinal energy flux $\bm S_{\rm L}$,
defined by Eq.~(\ref{hydrofluxL}),
contains the term%
\footnote
    {%
    \label{fn:Snonloc}%
    To see this,
    use the frequency delta function to
    rewrite the factor of $\omega$ multiplying
    $\bm J_{\rm L}$ in Eq.~(\ref{hydrofluxL}) times the
    $(\v \cdot \q)$ factor in the singular term
    (\ref{eq:JLsing}) as
    $
	\omega^2
	=
	(\omega^2 - \cs^2 q^2 + i\gamma q^2\omega)
	+
	q^2 (\cs^2 - i\gamma\omega)
    $.
    The first term cancels the denominator of Eq.~(\ref{hydrofluxL}),
    producing the above result for $\bm S_{\rm nonlocal}$.
    The second term is no longer singular when divided by $q^2$
    and is properly viewed as a contribution to the energy flux
    carried by sound.
    In the resulting expression (\ref{eq:Snonlocal}),
    if one writes
    $
	\phi_v(\omega,q^2)
	=
	\phi_v(\omega,0)
	{-}
	\left[ \phi_v(\omega,q^2) - \phi_v(\omega,0) \right]
    $
    then, strictly speaking, it is only the first $\phi_v(\omega,0)$ term
    which gives a non-local contribution.
    But including the additional
    $
	\left[ \phi_v(\omega,q^2) - \phi_v(\omega,0) \right]
    $
    term in the definition of $S_{\rm nonlocal}$ leads to simpler
    expressions for the effective sound and diffusion sources
    discussed below.
    }
\begin{align}
    \bm S_{\rm nonlocal}
    =
    \int \frac{d \omega}{2 \pi} \frac{d^3 q}{(2 \pi)^3}  \,
    \frac{ i \q}{q^2} \, \phi_v(\omega,q^2) \>
    2 \pi \delta(\omega {-} \v \cdot \q) \, e^{i Q \cdot x} \,.
\label{eq:Snonlocal}
\end{align}
This gives a Coulomb-field-like flux at large distance,
\begin{equation}
    \bm S_{\rm nonlocal}(t,\x)
    \sim
    \del \left(\frac 1{4 \pi v | \x - \v t|}\right)
    \frac{\f \cdot \v}{v^2}
    \,.
\end{equation}
This is co-moving with the quark and does not propagate like a wave.
It makes little sense to regard this as a contribution to the
energy flux carried by sound waves.
(Note, in particular, that $\bm S_{\rm nonlocal}$
is non-vanishing in front of the quark
even when the quark is moving supersonically.)

Since the transverse part of the effective source, $\bm J_{\rm T}$
equals $\bm J$ (which is regular as $\q \to 0$)
minus $\bm J_{\rm L}$,
a completely parallel argument shows that the transverse flux
$S_{\rm T}$, as defined in Eq.~(\ref{hydrofluxT}),
contains a non-diffusing contribution which is precisely
$-\bm S_{\rm nonlocal}$.

The presence of these equal and opposite Coulomb-like contributions
naturally suggests that the energy flux which is properly
associated with sound and diffusion can be obtained by
suitably adding and subtracting this term.
To this end, we write
\begin{equation}
\bm S = \bm S_{\rm sound} + \bm S_{\rm diffusion},
\end{equation}
with
\begin{subequations}
\begin{align}
\label{Ssound}
\bm S_{\rm sound} &\equiv \bm S_{\rm L}-\bm S_{\rm nonlocal}\,,
\\ \label{Sdiffusion}
\bm S_{\rm diffusion} &\equiv \bm S_{\rm T}+\bm S_{\rm nonlocal}\,.
\end{align}
\end{subequations}
Using Eq.~(\ref{hydrofluxL}),
it is straightforward to see that $\bm S_{\rm sound}$ satisfies the damped 
wave equation
\begin{equation}
\label{soundflux}
    \left(-\partial_0^2 +\cs^2 \, \del^2 +\gamma \del^2 \partial_0 \right )
    \bm S_{\rm sound}
= \bm J_{\rm sound} 
\end{equation}
with a source
\begin{align}
\bm J_{\rm sound}(t,\x)
&\equiv 
\int \frac{d \omega}{2 \pi}\frac{d^3 q}{(2 \pi)^3} \;
i \q \;
(2 \pi ) \delta(\omega - \v \cdot \q) \,e^{i Q \cdot x}
\nn \\ &\times
    \left[
    \cs^2 f^0 + ( \cs^2  {-} i \gamma \omega ) \phi_v(\omega,q^2)
    + i  \omega \, \phi_q(\omega,q^2)
    \right]_{\strut} .
\end{align}
Unlike $\bm J_{\rm L}(\omega,\q)$, the integrand defining
the modified source $\bm J_{\rm sound}$ is now regular as $\q \to 0$.
Hence $\bm J_{\rm sound}$ is localized at the quark
(it may be expanded in terms of delta functions and their derivatives).
Similarly, using Eq.~(\ref{hydrofluxT}), one finds that
$\bm S_{\rm diffusion}$ satisfies the inhomogeneous
diffusion equation
\begin{equation}
\label{diffusionflux}
\left(\partial_0 - D \del^2\right ) \bm S_{\rm diffusion}
= \bm J_{\rm diffusion}  \,,
\end{equation}
with the source
\begin{align}
    \label{Jdiffusion}
    \bm J_{\rm diffusion}
    \equiv  &
    \int \frac{d \omega}{2 \pi}\frac{d^3 q}{(2 \pi)^3} \>
    (2 \pi) \delta(\omega {-} \v \cdot \q) \, e^{i Q \cdot x}
\nn\\ & {} \times 
    ( \v + i \q D ) \, \phi_v(\omega,q^2) \,,
\end{align}
which is also localized at the quark.
Because the redefined sources $\bm J_{\rm sound}$ and
$\bm J_{\rm diffusion}$ are localized,
the corresponding energy fluxes,
$\bm S_{\rm sound}$ and $\bm S_{\rm diffusion}$,
may be regarded as a sensible decomposition of the flux
into contributions from
propagating and diffusing degrees of freedom.%
\footnote
    {%
    Since $\del \cdot \bm S_{\rm nonlocal}$ is localized
    at the quark,
    the fluxes
    $\bm S_{\rm sound}$ and
    $\bm S_{\rm diffusion}$ do provide a decomposition of the
    energy flux into pieces which are, respectively,
    longitudinal and transverse
    everywhere in space excluding the position of the quark,
    up to terms which fall faster than any inverse power of distance
    from the quark.
    Because hydrodynamics is only valid sufficiently far from the quark,
    one may regard $\bm S_{\rm sound}$ and
    $\bm S_{\rm diffusion}$ as only being defined in
    $\mathbb R^3\backslash \mathcal B$,
    where $\mathcal B$ is a ball surrounding the quark.
    In such a non-contractible region, the decomposition of a vector
    field into transverse and longitudinal components is not unique.
    The above decomposition is a choice for which the
    resulting effective sources $\bm J_{\rm sound}$
    and $\bm J_{\rm diffusion}$ are localized at the quark
    ({\em i.e.}, have Fourier transforms analytic in $\q$).
    With that additional condition,
    this decomposition is effectively unique ---
    alternative choices
    [such as the replacement of
    $\phi_v(\omega,q^2)$ by $\phi_v(\omega,0)$ in the definition
    (\ref {eq:Snonlocal}) of $\bm S_{\rm nonlocal}$]
    merely correspond to adding to $\S_{\rm sound}$
    (and subtracting from $\S_{\rm diffusion}$)
    a localized contribution which falls exponentially with distance
    from the quark.
    }

At leading order in the gradient expansion of $\bm J$,
the sound and diffusion sources are given by
\begin{subequations}
\begin{align}
\label{Jsound0}
    \bm J_{\rm sound} &=
    \cs^2 \,\Bigl (1+ \frac{1}{v^2} \Bigr ) \, \del F^0 \,,
\\ \label{Jdiff}
    \bm J_{\rm diffusion} &= \bm F \,,
\end{align}
\end{subequations}
while the leading behavior of the source (\ref{rhodef}) for
the energy density is
\begin{equation}
    \rho = \del \cdot \bm F - \partial_0 F^0 \,.
\end{equation}
Using these sources and the corresponding wave and diffusion equations
(\ref{energysound}),
(\ref{soundflux}), and (\ref{diffusionflux}),
the long range behavior of the sound and diffusion modes
is completely specified. 
We note that in the large $\Nc$ limit the magnitude of the drag force 
appears as an overall normalization of the perturbation in the stress-energy tensor.

\section{Gravitational Description}
\label{gravitational}

According to gauge/string duality, $\Nfour$ SYM (on $\mathbb R^4$)
is equivalent to type IIB string theory on $AdS_5 \times S^5$
\cite{Aharony:1999ti,Maldacena:1997re}.
Turning on a non-zero temperature corresponds,
in the dual description, to introducing a black brane
(a black hole with a flat horizon) into the $AdS_5$ space,
leading to the AdS-Schwarzschild geometry described by the metric
(\ref{metric}).%
\footnote
    {%
    Turning on a temperature does not deform the internal $S^5$.
    Only the AdS-Schwarzschild spacetime will be relevant for the
    our purposes;
    the five-sphere will play no role and may be ignored.
    }

The addition of massive fundamental representation fields
(an $\Ntwo$ hypermultiplet)
to the $\Nfour$ SYM theory corresponds,
in the dual gravitational description, to the addition
of a D7 brane wrapping an $S^3$ of the internal $S^5$ and covering
the five dimensional asymptotically AdS space from the boundary down to a
minimal radial position $u_m$, which is inversely related to
the hypermultiplet mass $M$ for large mass \cite{Herzog:2006gh}.

A single quark moving through the $\Nfour$ SYM plasma
corresponds, under gauge/string duality, to an open string
which runs from the D7 brane down to the black hole horizon
\cite{Karch:2002sh,Herzog:2006gh}.
The presence of the string perturbs the geometry via Einstein's equations
and the behavior of the metric perturbation near the AdS boundary encodes the
change in the SYM stress-energy tensor.%
\footnote
    {%
    The D7 brane also perturbs the geometry.
    This is a small correction of order $1/\Nc$ provided
    the number of flavors $\Nf$
    (which is the same as the number of D7 branes)
    is held fixed as $\Nc \to \infty$.
    This correction encodes the $\O(\Nc)$ contribution
    of fundamental representation fields to
    the equilibrium pressure and energy density of the SYM plasma.
    We ignore this correction
    (which is subleading relative to the $\O(\Nc^2)$
    contribution of adjoint representation fields)
    as our goal is the non-equilibrium
    contribution due to a moving quark.
    }

In the $\Nc \to \infty$ limit,
the $5d$ gravitational constant becomes parametrically small
and consequently the presence of the string
acts as a small perturbation on the AdS-BH geometry.
To obtain leading order results in $\Nf/\Nc \ll 1$,
we write the full metric as $G_{MN} = G^{(0)}_{MN} + h_{MN}$,
where $G^{(0)}_{MN}$ is the metric of the AdS-BH geometry given in Eq.~(\ref{metric}),
and then linearize the resulting Einstein equations in the
perturbation $h_{MN}$.

According to the AdS/CFT correspondence, the on-shell gravitational action $S_{\rm G}$ 
is the generating functional for the boundary stress-energy tensor 
\cite{Gubser:1998bc, Witten:1998qj, Skenderis:2000in}. The metric
$G_{M N}$ induces a metric $g_{\mu \nu}$ on the boundary
of the AdS-BH geometry \cite{Skenderis:2000in}.
The boundary metric is related to the boundary value of $G_{\mu \nu}$
by a scaling function of the AdS radial coordinate.  
More specifically, because the metric $G_{MN}$ has a second order pole
at the boundary, the boundary metric, which must be regular at the boundary,
may be defined as
\begin{equation}
g_{\mu \nu} \equiv \frac{u^2}{L^2} \> G_{\mu \nu} \Bigr |_{u = \inf}\,,
\end{equation}
where the infinitesimal $\inf$ will be sent to zero after
the required boundary terms needed to properly define
the gravitational action are added.
The expectation value of the boundary stress
tensor is then given by \cite{Witten:1998qj, Skenderis:2000in}
\begin{equation}
    T^{\mu \nu}(x)
    = \lim_{\inf \rightarrow 0} \frac{2}{\sqrt{-g(x,\inf)}}\,
    \frac{\delta S_{\rm G}}{\delta g_{\mu \nu}(x,\inf)} \,,
\label{eq:TfromS}
\end{equation}
with $g$ denoting the determinant of $g_{\mu\nu}$.
Defining for later convenience
\begin{equation}
H_{\mu \nu} \equiv \frac{u^2}{L^2} \> h_{\mu \nu} \,,
\end{equation}
the boundary stress-energy tensor may also be expressed as
\begin{equation}
\label{genfcn}
T^{\mu \nu}(x) =
2 \lim_{\inf \rightarrow 0}
\frac{\delta S_{\rm G}}{\delta H_{\mu \nu}(x,\inf)} \,.
\end{equation}
(Since the boundary is flat Minkowski space, the factor of
$\sqrt{-g}$ in Eq.~(\ref{eq:TfromS})
approaches one as $\inf \to 0$ and may be ignored.)

The action for the entire gravitational system is
given by
\begin{equation}
    S_{\rm G} =
    S_{\rm EH} + S_{\rm GH} + S_{\rm DBI} + S_{\rm NG} +S_{\rm CT} \,.
    \label{SG}
\end{equation}
The first term is the Einstein-Hilbert action,
\begin{equation}
    S_{\rm EH} \equiv \frac{1}{2 \kappa_5^2}
    \int d^5 x \> \sqrt{-G} \> (R {+} 2  \Lambda ) \,,
\end{equation}
where $\kappa_5^2 \equiv 4 \pi^2 L^3/\Nc^2$ is the 
$5d$ gravitational constant,
$G$ is the determinant of $G_{\mu\nu}$
and
$\Lambda \equiv {6}/{L^2}$
is the cosmological constant.
The next term is the Gibbons-Hawking action,
\begin{equation}
    S_{\rm GH}
    \equiv \frac{1}{2 \kappa_5^2}
    \int d^4 x \> \sqrt{-\gamma} \; 2 K \,,
\end{equation}
with $\gamma_{\mu \nu}$ the induced metric
on  the slice $u = \inf$ (and $\gamma$ its determinant),
and $K$ the trace of the extrinsic curvature of the slice.
This term is needed to obtain an action which
makes the Dirichlet problem well defined
({\em i.e.}, an action depending only on first derivatives of the metric)
\cite{Gibbons:1976ue}.  
The remaining terms are
the Dirac-Born-Infeld action $S_{\rm DBI}$ for the D7 brane,
the Nambu-Goto action $S_{\rm NG}$ for the string,
and a counter-term action $S_{\rm CT}$ 
defined on the boundary which is required for holographic
renormalization \cite{Skenderis:2000in}.
This term cancels poles in $1/\inf$ which are generated by the
bare gravitational action;
the form of the counter-terms is constrained by the requirement of
diffeomorphism invariance of the gravitational theory.
In Eq.~(\ref{SG}) all boundary integrals are evaluated on the
$u = \inf$ slice, with $\inf \rightarrow0$ after all terms are combined.

\subsection{Heavy Quark Effective Theory}

In the limit that the quark mass becomes arbitrarily large,
its presence in the plasma merely serves as a external source for
the SYM fields.
This can be made explicit by constructing a heavy quark effective
theory (HQET) from the field theory Lagrangian
(as given explicitly in Ref.~\cite{Chesler:2006gr}).
As we now outline, this procedure has a natural counterpart
in the gravitational dual.

\begin{figure}
\includegraphics[scale=0.45]{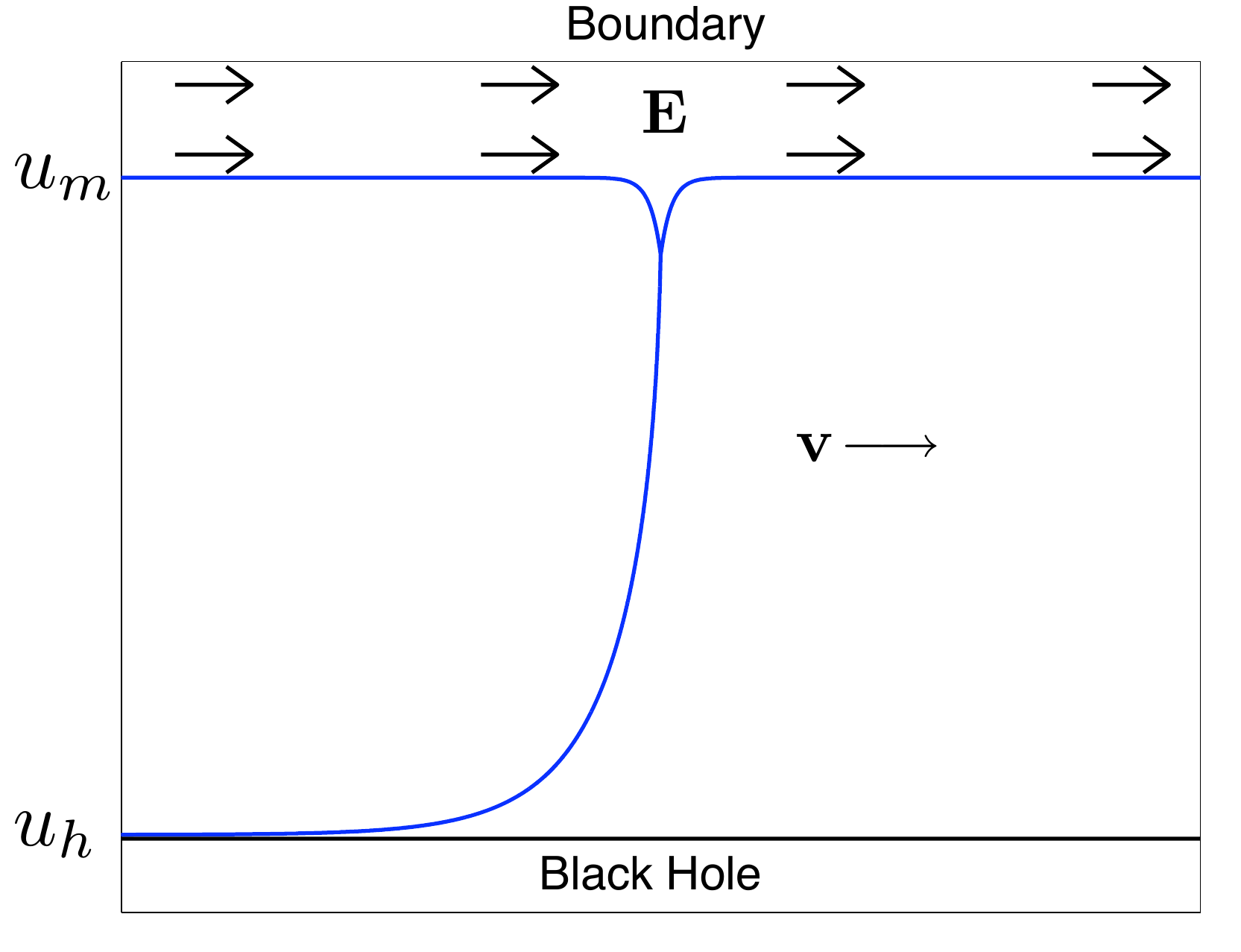}
\caption{\label{Dbrane}
    A cartoon of the string plus D7 brane system in the large mass limit. 
    The D7 brane covers the asymptotically AdS space down to a minimal
    radial position, away from the string, denoted $u_m$.
    The trailing string is moving to the right
    at constant velocity $\v$.  An energy flux flows down the 
    string toward the black hole horizon,
    which is located at radial coordinate $\uh$.
    This energy is supplied by a constant $U(1)$ electric field
    living on the D7 brane.
    The D7 brane is deformed in the neighborhood of the endpoint of the
    string over a length scale of order $1/M$.
    }
\end{figure}

The interaction between the string and the D7 brane is governed by
the DBI action and the Nambu-Goto action.  Fig.~\ref{Dbrane}
shows a cartoon of the string plus D-brane system.  As the string moves, a 
flux of energy and momentum flows down the string to the black hole horizon 
\cite{Herzog:2006gh, Casalderrey-Solana:2006rq}.
This flux of energy and momentum is responsible for the drag force
on the quark in the dual boundary theory and is supplied by a $U(1)$
gauge field living on the D7 brane --- the strength of which
is tuned to match the drag force acting on the quark and thereby
maintain a constant velocity.  
The presence of the string pulls on the D7 brane, deforming it.  
However, in the large mass limit the trailing  string will only
deform the D7 brane over length scales of order $1/M$.
As $M \rightarrow \infty$,
the string endpoint approaches the boundary and the 
size of the region in which the D7 brane is significantly deformed
shrinks to zero.%
\footnote
    {%
    This is a bit oversimplified.
    For a finite mass $M$, the back-reaction of the string on the
    D7 brane is expected to cause the D-brane embedding to develop
    a thin narrow tube which reaches all the way down to the horizon
    \cite{Callan:1997kz,Gibbons:1997xz}.
    In other words, the fundamental string will be ``puffed up''
    into a tiny tube.
    However, the resulting dynamics of a sufficiently thin tube 
    is indistinguishable from that of a fundamental string.
    In the $M \to \infty$ limit,
    the width of the tube (at any non-zero value of $u$) will
    shrink to zero and this issue may be ignored.
    }

Instead of solving the string plus D7 brane system in its entirety,
just as in HQET one may consider the large mass limit
and construct an effective gravitational theory
for the perturbations in the SYM stress-energy tensor.
In the effective theory we send the quark
mass $M \rightarrow \infty$ before performing holographic renormalization
of the gravitational action.
In doing so, we take $u_{m} \rightarrow 0$ so  the D-brane
no longer extends inward from the boundary.
Furthermore, the deformation of the D-brane due to the 
string shrinks to a vanishingly small region
about the string endpoint (which approaches the boundary).

This suggests that the dynamics of the D7 brane should be irrelevant
in the heavy quark limit.
This is not quite true, however.
Because general relativity is a gauge theory,
the construction of a heavy quark effective theory is constrained by the
requirement that the effective theory be invariant under infinitesimal 
diffeomorphisms
\begin{equation}
\label{diffeomorphisms}
X_{M} \rightarrow X_{M} + \xi_M(X) \,,
\end{equation}
where $\xi_M$ is an arbitrary infinitesimal vector field.  
Because an energy-momentum flux flows from the boundary down the
trailing string, 
gauge invariance at the boundary requires 
that the $U(1)$ electric field on the D7 brane not be neglected.
We therefore construct the effective theory for the perturbations
in the SYM stress-energy tensor with the brane effective action 
\begin{equation}
S_{\rm DBI}^{\rm eff} = S_{\rm EM} \,,
\end{equation}
where $S_{\rm EM}$ is the Maxwell action for the $U(1)$ electromagnetic field
which resides on the boundary.

To complete the definition of the effective theory, we must specify the
counter-term action.
The counter-terms must cancel $1/\inf$ poles in the gravitational 
action and must be diffeomorphism invariant.  
The presence of the string ending on the boundary induces $1/\inf$ 
poles in the gravitational action which have point
support at the location of the quark.
The remaining $1/\inf$ poles in the gravitational 
action are simply those obtained in pure gravity.
The appropriate counter-term needed to offset 
the pure gravity divergences is well known
\cite{Liu:1998bu, Skenderis:2000in, Policastro:2002tn}
and given by
\begin{equation}
\label{counterterms}
S_{\rm CT}
= -\frac{1}{2 \kappa_5^2}\, \frac{6}{L} \int d^4 x \> \sqrt{ \gamma}\,.
\end{equation}
This term is simply proportional to the area of the boundary.  

Point-like $1/\inf$ divergences with support at the location of the quark 
are fundamentally different from the other UV divergences ---
they are physical as the
boundary stress-energy tensor {\it should} contain a $1/\inf$ divergence at the
location of the quark. 
This simply reflects the contribution to the stress-energy tensor of an
infinitely massive quark,
\begin{equation}
\label{quarkstress}
T^{\mu \nu}_{\rm quark} \equiv M \, U^{\mu} U^{\nu}\sqrt{1- v^2}  \;
 \delta^3 (\x {-} \v t) \,,
 \end{equation}
where $U^{\mu}$ is the quark's four velocity.
The (divergent) coefficient $M$ can
either be determined by a matching to the divergent point-like
terms in the boundary action or, alternatively, by integrating the
energy density of the trailing string and
isolating the divergent contribution to find its bare mass.
Taking this approach, we find 
\cite{Herzog:2006gh}
\begin{equation}
    M = \frac{\sqrt{\lambda}}{2 \pi \inf} \,.
\label{eq:Mdef}
\end{equation}
Instead of adding local counter-terms to the gravitational action to 
offset the point-like divergence, we simply
define the boundary stress-energy tensor via 
\begin{equation}
\label{genfcn2}
T^{\mu \nu}(x) = \lim_{\inf \rightarrow 0} \left [
2 \frac{\delta S_{\rm G}}{\delta H_{\mu \nu}(x,\inf)} 
-T^{\mu \nu}_{\rm quark}(x) 
\right ],
\end{equation}
with $M$ and $\inf$ related by (\ref{eq:Mdef}).
The resulting stress-energy tensor is finite as $\inf \rightarrow 0$ and
everywhere traceless, including at the position of the quark.%
\footnote
  {%
  So the Fourier transform
  $T^{\mu \nu}(\omega,\q)$
  of the stress-energy tensor is traceless for all momenta.
  }

\subsection{Gauge Invariants}
\label{gaugeinvariants}

The continuity equation (\ref{conservation}) and vanishing trace
conditions satisfied by
the $\Nfour$ SYM stress-energy tensor imply,
as discussed in section \ref{defs},
that the stress-energy tensor contains five independent degrees 
of freedom which are conveniently isolated by performing
a spacetime Fourier transform and decomposing the stress-energy tensor in terms
of helicity variables $\T_0$, $\vec \T_1$ and $\tensor \T_2$.

The five independent degrees of freedom contained in $T_{\mu \nu}$
may be contrasted with the 15 degrees of
freedom contained in the metric perturbation $h_{M N}$.
Not all of these degrees of freedom are physical, however.
The linearized gravitational field equations are invariant under the 
infinitesimal diffeomorphisms (\ref{diffeomorphisms}).
Under such transformations, the metric perturbation transforms as
\begin{equation}
\label{gaugetrans}
    h_{M N} \rightarrow
    h_{MN}-D_{M} \, \xi_{N} - D_{N} \, \xi_{M} \,,
    \end{equation}
where $D_{M}$ is the covariant derivative with respect to
the background metric $G^{(0)}_{M N}$.
Physical degrees of freedom carried by $h_{MN}$ must be invariant 
under the above gauge transformations.
This limits the number of \textit{independent} physical degrees of freedom
carried to the boundary by the metric perturbation $h_{MN}$ to five,
matching that of the SYM stress tensor \cite{Chesler:2007an}.

The correspondences between the number of independent
gauge invariant degrees of freedom contained in the metric perturbation
$h_{MN}$ and the number of
independent degrees of freedom in the SYM stress tensor suggests
that the bulk to boundary problem can be formulated directly in terms of gauge 
invariant degrees of freedom.  Using this approach has several advantages
\cite{Kovtun:2005ev}.
As we show below,
the gauge invariants can be chosen to satisfy decoupled equations 
of motion.  Furthermore, as we show in Section \ref{SYMstress}, 
the on-shell gravitational action can be expressed in
terms of gauge invariant degrees of freedom plus computable boundary terms
(which are due to the non-conservation of the SYM stress tensor at the location of
the quark).

Gauge invariants can be constructed out of the Fourier mode amplitudes 
$h_{MN}(u;\omega,\q)$ and classified according to their helicity 
under rotations about the $\hat q$ axis
\cite{Kovtun:2005ev}.
As we shall discuss in detail below,
the determination of the SYM stress-energy tensor requires the 
construction of a scalar gauge invariant $Z_{0}$, a vector gauge invariant 
$\vec Z_1$, and a tensor gauge invariant $\tensor Z_2$.  
The choice of gauge invariants $Z_s$ is not unique.
However, all gauge invariants of a common helicity carry the same
information to the boundary.%
\footnote
  {%
  This may be verified by analyzing the asymptotic behavior of both
  the linearized Einstein field equations and the 
  various gauge invariants of the same helicity near the boundary.  Doing so,
  one finds that the boundary values of gauge invariants
  (with the same helicity) are linearly related.
  }
We are therefore free to chose any convenient set of
gauge invariants.

Let $\H_{MN}$ denote the components of $H_{MN}$ in the polarization frame. 
Using Eq.~(\ref{gaugetrans})
it is easy to see that the following five degrees of freedom are gauge invariant
\begin{subequations}
\begin{align}
\nn
    Z_0 &\equiv
	q^2 \,\H_{00} + 2 \omega q \, \H_{0q}
	+ \omega^2 \, \H_{qq} \,,
\\
         &\qquad\qquad {} +
	\half \left [(2 {-} f) \, q^2{-} \omega^2 \right]
	\H_{aa} \, ,
    \label{sound}
\\ 
    \label{vector}
    \vec Z_1 &\equiv
    \left ( \H'_{0a} - i \omega \,\H_{a5} \right)  \eps_{a} \, ,
\\ 
     \label{tensor}
     \tensor Z_2 & \equiv \left (\H_{ab}-\half \H_{cc} \,\delta_{ab}\right )
     \eps_{a} \otimes \eps_{b} \,,
\end{align}
\end{subequations}
where sums over repeated indices are implied and primes
denotes differentiation with respect to $u$.%
\footnote
    {%
    Using the differential helicity one invariant (\ref{vector})
    instead of an equivalent non-differential invariant
    turns out to be more convenient for technical reasons involving
    the numerical stability of the resulting ordinary differential equation.
    }

%%%%%%%%%%

\subsection{Equations of Motion}
The equations of motion for the invariants $Z_s$
follow from the linearized Einstein field equations.  
The full Einstein field equations are
\begin{equation}
\label{fieldeq}
R_{MN} - \coeff{1}{2}  G_{MN}  \left(R+ 2 \Lambda \right )
= \kappa_5^2 \; t_{MN} \,,
\end{equation}
where $t_{MN}$ is the $5d$ stress-energy tensor of the trailing string.
Writing 
\begin{equation}
G_{MN}= G^{(0)}_{MN} +h_{MN} \,,
\end{equation}
where $ G^{(0)}_{MN}$ is the AdS-BH metric, and expanding the left hand side
of the field equations (\ref{fieldeq}) in the perturbation $h_{MN}$,
produces the linearized equations,
\begin{align}
    &- D^2 \, h_{MN} +2 D^{P} D_{(M} h_{N) P} 
      -D_{M}D_{N} \, h +\coeff{8}{L^2} \, h_{MN} 
\nonumber
\\ \label{lin1} &{}
    + \left (D^2 h -D^{P} D^{Q} \, h_{P Q} -\coeff{4 }{L^2} \, h \right)
	G_{MN}^{(0)}  =  2 \kappa_5^2 \; t_{MN} \, , 
\end{align}
where $h \equiv h^{M}_{\ M}$.  

The trailing string profile is determined by minimizing the 
Nambu-Goto action for a stationary, constant velocity profile.
The result is \cite{Herzog:2006gh, Casalderrey-Solana:2006rq}
\begin{equation}
\x(t,u) = \v \, t +  \x_{\rm string} (u) \,,
\end{equation}
with
\begin{equation}
  \x_{\rm string} (u)   \equiv
    \frac{\v \, u_h}{2} \!
    \left[
	\tan^{-1}\Bigl(\frac{u}{u_h}\Bigr)
	+ \half \ln \Big( \frac{u_h{-} u}{u_h{+}u} \Big)
    \right].
\end{equation}
The $5d$ stress-energy tensor for the 
trailing string is \cite{Friess:2006fk}
\begin{subequations}
\label{stringstress}
\begin{align}
t_{00} &= s \left (f + v^2 u^4 u_h^{-4} \right ) , &
t_{55} &=s \, (f + v^2) \, f^{-2} \,,
\\
t_{0i} &= -s \, v_i \,,&
t_{ij} &=  s \, v_i v_j \,, 
\\
t_{05} &= -s \, v^2 \, f^{-1} u^2 u_h^{-2} \,,&
t_{i5} &= s \, v_i \, f^{-1} u^2 u_h^{-2} \,,
\end{align}
\end{subequations}
where 
\begin{equation}
s(u) \equiv \frac{u \, \sqrt{\lambda }}{2 \pi  L^3  \sqrt{1{-}v^2}} \;
\delta^{3} (\x - \v t -\x_{\rm string}(u) ) \,.
\end{equation}
We denote the components of $t_{\mu \nu}$
in the polarization frame 
by $\t_{\mu \nu}$.

Because the background geometry is invariant under spatial rotations, 
the differential operators on the LHS of Eq.~(\ref{lin1}) cannot couple 
linear combinations of $\H_{MN}$ which transform with different helicities 
under rotations about the $\hat \q$ axis.
Consequently, the field equations (\ref{lin1}) reduce to three coupled
sets of ordinary differential equations labeled
by their helicity.
Although it is not necessary,
working in the gauge in which $h_{5 N} = 0$ for all $N$
simplifies the extraction of decoupled equations for the
gauge invariants $Z_s$.
The scalar set of equations then reduces to four second order and
three first order equations,
while the vector set of equations reduces to two second order 
and one first order equation.
However, not all of these
equations are independent ---
some of the second order equations 
can be derived from the first order equations.
One may use this observation to reduce the 
number of required second order equations of motion.
 
\subsubsection{Tensor Mode}

The components of $\H_{\mu \nu}$ which transform with helicity two
under rotations about the $\hat q$ axis are the traceless part of $\H_{ab}$.
The $ab$ components of the linearized equations (\ref{lin1}) may be
written explicitly as
\begin{align}
\label{H_ab}
    - f \, \mathcal K_{ab}''
    -  \frac{u f' {-} 3 f}{u} \, \mathcal K_{ab}'
    + \frac{q^2 f  {-} \omega^2}{f} \, \mathcal K_{ab}
&
\\
\nonumber
 -  \left [
	 f' \, \big (\H_{00}/2 f \big)'
	 - \widetilde Q^{\mu} \widetilde Q^{\nu} \, \H_{\mu \nu} 
     \right ] \delta_{ab} 
&= 2 \kappa_5^2 \; \t_{ab} \,,
\end{align}
where $\widetilde Q^{\mu}\equiv (\omega/f, \q)$ and
\begin{equation}
\mathcal K_{ab} \equiv \H_{ab}+(\H_{00}/f -\H_{ii}) \, \delta_{ab} \,.
\end{equation}
The equation of motion for the helicity 2 gauge invariant $\tensor Z_2$
immediately follows from the traceless part of Eq.~(\ref{H_ab}),
\begin{equation}
\label{eqm2}
\tensor Z''_2+A_2 \, \tensor Z'_2 + B_2 \, \tensor Z_2 = \tensor S_2 \,,
\end{equation}
where
\begin{subequations}
\begin{align}
A_2 &\equiv \frac{u f' - 3f}{u f} \,,
\\ 
B_2 &\equiv -\frac{q^2 f - \omega^2}{f^2} \,,
\\
\label{S2general}
\tensor S_2 &\equiv
- \frac{2 \kappa_5^2}{f} \,
(\t_{ab} - \half \, \t_{cc} \, \delta_{ab} ) \> \eps_a \otimes \eps_b \,.
\end{align}
\end{subequations}
Inserting the 
string stress-energy tensor (\ref{stringstress}) into
expression (\ref{S2general}) for the source $\tensor S_2$ gives
\begin{align}
    \tensor S_2
    &=
    -\frac{\kappa_5^2 \sqrt{\lambda}}{2 \pi L^3} \,
    \frac{u v^2 q_{\perp}^2 }{q^2 f \sqrt{1 - v^2} }  \>
    (2 \pi) \delta(\omega {-} \v \cdot \q) 
\nonumber
\\
    &{} \times
    e^{-i \q \cdot \x_{\rm string}} \,
    \left(\eps_1 \otimes \eps_1 - \eps_2 \otimes \eps_2\right) \,.
\end{align}

\subsubsection{Vector Mode}

The components of $\H_{\mu \nu}$ which transform with helicity one
under rotations
about the $\hat q$ axis are $\H_{0a}$, $\H_{aq}$ and $\H_{a5}$
(which has been set to zero).
The equations of motion for these quantities contain two second order
equations and one first order equation.
For the derivation of the equation of motion
for $\vec Z_1$, it is sufficient to use one second order equation (which is independent
of the first order equation) and the first order equation.
We use the following components of the linearized field equations (\ref{lin1})
\begin{align}
    -f \,\H''_{0a}+\frac{3 f}{u} \,\H'_{0a}+q^2 \,\H_{0a} + q \omega \,\H_{aq}
    &= 2 \kappa_5^2 \, \t_{0a} ,
\\ 
\label{vec1}
    \frac{1}{f} \left (i \omega \, \H'_{0a}+i q f \, \H'_{aq} \right )
    &= 2 \kappa_5^2 \, \t_{a5} \,.
\end{align}
Using these equations, we find the equation of motion for 
the helicity one invariant $\vec Z_1$
\begin{equation}
\label{eqm1}
\vec Z''_1+A_1 \, \vec Z'_1 + B_1 \, \vec Z_1 = \vec S_1 \,,
\end{equation}
where
\begin{subequations}
\begin{align}
A_1 &\equiv \frac{u f'-3 f}{ uf} \,,
\\ 
B_1 &\equiv \frac{3 f^2-u \left(u q^2+3 f'\right) f+u^2 \omega ^2}{u^2 f^2} \,,
\\
\label{S1general}
    \vec S_1 &\equiv
    \frac{2 \kappa_5^2}{f} \, \Big [ \t'_{0a}+i \omega \, \t_{a5} \Big ] \,
    \eps_a \,.
\end{align}
\end{subequations}
Substituting the 
string stress-energy tensor (\ref{stringstress}) into Eq.~(\ref{S1general}), gives
\begin{equation}
    \vec S_1 =
    \frac{ \kappa_5^2 \sqrt{\lambda }}{ \pi L^3 } \,
    \frac{ v q_{\perp}}{ q f \sqrt{1{-}v^2}}  \,
    (2 \pi) \delta(\omega {-} \v \cdot \q ) \,
    e^{- i \q \cdot \x_{\rm string}} \;
    \eps_1 \,.
\end{equation}

\subsubsection{Scalar Mode}

The (non-vanishing) components of $\H_{\mu \nu}$ which transform
like scalars under rotations
about the $\hat q$ axis are $\H_{00}, \ \H_{0q}, \ \H_{qq}$ and $\H_{aa}$. 
The equations of motion for these quantities contain four second order
equations and three first order equations.
For the derivation of the equation of motion
for $Z_0$, it is sufficient to use one second order equation (which is independent
of the three first order equations) and the three first order equations.
By taking suitable linear combinations of the components of the linearized 
field equations (\ref{lin1}), we find the following equations of motion:
\begin{subequations}
\begin{align}
\label{sound1}
&
    -\H''_{00}
    + \frac{6 f {+} u  f'}{ 2 u f} \, \H'_{00} 
    +\frac{f'}{3} \, \H_{ii}'
    + \frac{4 q^2 f {-} 3 f'^2}{6 f^2} \, \H_{00}  
\nonumber\\&\qquad{}
    +\frac{q^2 f + 2 \omega^2}{3 f} \, \H_{ii}
     - \frac{q^2 }{3} \, \H_{qq}
    +\frac{4 q \omega}{3f} \, \H_{0q}
\nonumber\\&\qquad{}
  = \frac{2 \kappa_5^2}{3f} \,
  \big (  2\, \t_{00} {+}  f \, \t_{ii} \big) \,,
\\&
\label{sound3}
    3 \H'_{00}
    +\half ({u f'{-}6f}) \,\H'_{ii}
    +\frac{u \omega^2}{f} \, \H_{ii}
    +u q^2 \, \H_{aa}
\nonumber \\&\qquad{}
    +\frac{q^2 u {-} 3 f'}{f} \, \H_{00}
    + \frac{2 u \omega q}{f} \, \H_{0q}
    =  2 u f \kappa_5^2 \, \t_{55} \,,
\\&
\label{sound4}
    i \omega \, \H'_{0q}
    +i q \, \H'_{00}
    -i q f \, \H'_{aa}
    -\frac{i q f'}{2 f} \, \H_{00} 
    =  2 f \kappa_5^2 \, \t_{q5} \,,
\\&
\label{sound5}
     i \omega \, \H_{ii}'
     + i q \, \H_{0q}'
     - \frac{i \omega f'}{2f} \, \H_{ii}
    - \frac{i q f'}{f} \, H_{0q}
    = 2 \kappa_5^2 \, \t_{05} \,.
\end{align}
\end{subequations}
It is straightforward (but tedious) to derive the equation of motion of
$Z_0$ from the above equations.
The result has the form
\begin{equation}
\label{eqm0}
Z''_0+A_0 \, Z'_0 + B_0 \, Z_1 = S_0 \,.
\end{equation}
The coefficients $A_0$ and $B_0$ and the source $S_0$ can be determined by
substituting the definition of $Z_0$
into the above ansatz and then using the above equations of motion
(\ref{sound1})--(\ref{sound5}) to eliminate all derivatives except $\H'_{00}$.
Demanding that the coefficient $\H'_{00}$
vanish then determines $A_0$.
Once $A_0$ is determined, it is straightforward to read off the
coefficient $B_0$ and the source $S_0$.
One finds
\begin{subequations}
\begin{align}
   A_0
    &=
    \frac{1}{u}
    \Bigl[
	1
	+
	\frac{uf'}{f}+
	\frac{24 \left(q^2 f{-}\omega ^2\right)}
	    { q^2\left(u f'{-}6 f\right)+6 \omega^2}
    \Bigr] \,,
\\
    B_0 &=
    \frac 1f
    \Bigl[
	-q^2 + \frac{\omega^2}{f}
	- \frac {32 \, q^2 u^6 u_h^{-8}}{q^2(u f' {-} 6f) + 6 \omega^2}
    \Bigr] \,.
\end{align}
\end{subequations}
The expression for the source $S_0$ in terms of a general
string stress-energy tensor $\t_{MN}$ is sufficiently lengthy
that we will not give it here.
But substituting the explicit form of the string stress-energy
tensor (\ref{stringstress}) produces a relatively compact result,
\begin{align}
S_0 \equiv{}&
     \frac {\kappa_5^2 \sqrt\lambda} {6\pi L^3 } \;
     \frac{q^2 \left(v^2{+}2\right){-}3 \omega ^2} {q^2 \sqrt{1{-}v^2}} \,
\nonumber\\ &{}\times {}
     \frac{ u [q^4 u^8+48 i q^2 \omega  u_h^2 u^5-9 (q^2{-}\omega ^2)^2 u_h^8]}
	  {f \left(f q^2+2 q^2-3 \omega ^2\right) u_h^8} \>
\nonumber\\ &{}\times {}
	 (2 \pi) \delta(\omega {-} \v \cdot \q) \,
	 \, e^{- i \q  \cdot \x_{\rm string}} \, .
\end{align}

\subsection{SYM Stress Tensor}
\label{SYMstress}

The variation of the on-shell gravitational action may be expressed as
\begin{equation}
\delta S_{\rm G} = \delta S_{\rm horizon} + \delta S_{\rm B} \,,
\end{equation}
where $S_{\rm B}$ is a surface term at the boundary, and
$S_{\rm horizon}$ is a surface term at the horizon.
Following Ref.~\cite{Son:2002sd}, we neglect $S_{\rm horizon}$.%
\footnote
  {%
  This can be understood as follows.
  As we discuss below, causality implies that near the horizon the gauge invariants
  behave as
  $Z_s \sim (u{-}\uh)^{-i \omega \uh/4}$.
  To make the $u\to\uh$ limit of this quantity (and $S_{\rm horizon}$)
  meaningful, the frequency can be infinitesimally analytically continued
  into the complex plane.  Causality dictates the required direction of
  continuation.  For retarded boundary conditions, one must send
  $\omega$ into the upper half plane,
  $\omega \rightarrow \omega + i \delta$.
  With such an infinitesimal continuation understood,
  the gauge invariants $Z_s$ vanish at the horizon.
  }
We show in Appendix \ref{BoundaryAction} that
\begin{align}
\nonumber
    \delta S_{\rm B}
    &=
    \int_{u = \inf} \frac{d^4q}{(2 \pi)^4} 
    \Bigl [\,
    \mathcal A_0  \, \delta Z_0^\dagger \, \partial_u \, Z_0
    +\mathcal A_1  \, \delta \vec {\mathcal Z}_1^\dagger  \, \vec Z_1 
\\ & \qquad\qquad\qquad {}
    + \mathcal A_2 \, {\rm tr} \, \delta \tensor Z_2^\dagger \,
	    \partial_u \, \tensor Z_2
    + \half \, \delta \H_{\mu \nu}^{\dagger} \, \T_{\rm eq}^{\mu \nu}
\nonumber
\\[4pt] & \qquad\qquad\qquad {}
    + \half \, \delta \H_{\mu \nu}^{\dagger} \, \mathscr J^{\mu \nu}
    \Big ]
    + \delta S_{\rm EM} +\O(\inf)\,,
\label{bdaction}
\end{align}
with
\begin{subequations}
\begin{align}
    \mathcal A_0 &= \frac{L^3}{6 \kappa_5^2 (q^2 {-} \omega^2)^2} \,
    \frac 1{u^3} \,,
\\
    \mathcal A_1 &=  -\frac{L^3}{2 \kappa_5^2 \, q}  \, \frac 1{u^3} \,,
\\
    \mathcal A_2 &=  \frac{L^3}{4 \kappa_5^2} \, \frac 1{u^3} \,,
\end{align}
\end{subequations}
and 
\begin{equation}
\vec {\mathcal Z_1} \equiv (q \H_{0a} + \omega \H_{aq} ) \, \eps^{a} \,.
\end{equation}
In the fourth term of (\ref{bdaction}), $T^{\mu\nu}_{\rm eq}$ denotes the
stress-energy tensor of the equilibrium $\Nfour$ SYM plasma,
\begin{equation}
    T^{\mu \nu}_{\rm eq}
    = \coeff{3}{8} \Nc^2 \pi^2 T^4 \;
    {{\rm diag}(1,\coeff{1}{3},\coeff{1}{3},\coeff{1}{3} ) } \,,
\end{equation}
with $\T^{\mu \nu}_{\rm eq}$ its components in the polarization
frame.
In the fifth term of (\ref{bdaction}),
the symmetric tensor 
$\mathscr J^{\mu \nu}$ is a linear combination of components of the string
stress-energy tensor evaluated at the boundary.%
\footnote
  {%
  See Eqs.~(\ref{jaq}) and (\ref{jmunu})
  for the definitions expressing
  $\mathscr J^{\mu \nu}$ in terms of $t_{MN}$.
  }
Explicitly,
\begin{subequations}
\begin{align}
\mathscr J^{00} &= \mathcal C 
    \left\{
    \frac{q^2}{u}-\frac{i \omega
	\left[ q^2 \left(5 v^2{+}1\right)
	    - 3 \omega^2 \left(v^2{+}1\right) \right]}
       {\uh^2 \left(1{-}v^2\right) \left(q^2{-}\omega ^2\right)}
    \right\},
\\
\mathscr J^{qq} &= \mathcal C 
\left \{
\frac{\omega ^2}{u}-\frac{i \omega  \left[3 q^2
   \left(v^2{+}1\right)-\left(v^2{+}5\right) \omega ^2\right]}{\uh^2
   \left(1{-}v^2\right) \left(q^2{-}\omega ^2\right)}
\right \},
\\
\mathscr J^{0q} &= \mathcal C
\left \{
\frac{q \omega }{u}-\frac{i \left[3 v^2 q^4+\left(1{-}v^2\right) \omega
   ^2 q^2-3 \omega ^4\right]}{q \uh^2 \left(1{-}v^2\right)
   \left(q^2{-}\omega ^2\right)}
 \right \},
\\
\mathscr J^{aa} &= \mathcal C \left [
-\frac{q^2{-}\omega ^2}{u}+\frac{i \omega }{\uh^2} \right ],
\label{eq:Jaa}
\\
\mathscr J^{aq} &= \mathcal C  \left [
    -\frac{3 i \, v \, q_{\perp} \left(q^2{-}\omega ^2\right)}
	    {q^2 \uh^2 \left(1{-}v^2\right)}
\right ],
\end{align}
\end{subequations}
with
\begin{equation}
    \mathcal C \equiv \frac{\sqrt{ \lambda} }{6 \pi} \,
    \frac{ \sqrt{1{-}v^2}}{q^2 {-} \omega^2} \;
    (2 \pi) \delta(\omega {-} \v \cdot \q).
\end{equation}
All other components of $\mathscr J^{\mu \nu}$ vanish.
In Eq.~(\ref{eq:Jaa}), no sum on $a$ is implied.

On the boundary, the action of diffeomorphisms
on the metric perturbation takes the simple form
\begin{equation}
H_{\mu \nu} \rightarrow H_{\mu \nu} - \partial_\mu \xi_\nu -\partial_\nu \xi_\mu
\,.
\end{equation}
The first four terms in the boundary action (\ref{bdaction}) are manifestly 
invariant.
The fifth term however violates diffeomorphism invariance.
One may easily verify that
$\mathscr J^{\mu \nu}$ satisfies
\begin{equation}
i q_{\mu} \, \mathscr J^{\mu \nu} = \mathcal F^{\nu} \,,
\end{equation}
where 
\begin{subequations}
\begin{align}
\mathcal F^{0} &= \frac{v^2 \sqrt{\lambda }}{2 \pi  \uh^2 \sqrt{1{-}v^2}} \,
(2 \pi) \delta(\omega {-} \v \cdot \q) \,,
\\
\mathcal F^{q} &= \frac{(\hat q \cdot \v) \sqrt{\lambda } }{2 \pi \uh^2 \sqrt{1{-}v^2}} \,
(2 \pi) \delta(\omega {-} \v \cdot \q) \,,
\\
\mathcal F^{a} &= \frac{ (\eps_a \cdot \v) \sqrt{\lambda }}{2 \pi  \uh^2 \sqrt{1{-}v^2}} \,
(2 \pi) \delta(\omega {-} \v \cdot \q) \,.
\end{align}
\end{subequations}
These are precisely the components of the external force density
$F^{\nu}$ in the polarization basis.
The violation of diffeomorphism (or gauge) invariance only occurs
at the location of the string endpoint.
The presence of $\mathscr J^{\mu \nu}$ in the boundary action
reflects the flux of energy-momentum from the boundary into the string.
This flux comes from the $U(1)$ electric field living on the D7 brane.
By definition,
\begin{equation}
    \delta S_{\rm EM} = \int \frac{d^4 q}{(2 \pi)^4} \>
    \half \, \delta \H_{\mu \nu} \, \T_{\rm EM}^{\mu \nu} \,,
\end{equation}
where $\T_{\rm EM}^{\mu \nu}$ are the components of the electromagnetic
stress-energy tensor in the polarization frame.
By construction, the strength of the $U(1)$ electric field is tuned to offset the 
drag force and maintain a constant velocity of the quark.
It therefore follows that we must have
\begin{equation}
i q_\mu \T^{\mu \nu}_{\rm EM} = - \mathcal F^{\nu}.
\end{equation}
Consequently, the total boundary action is gauge (diffeomorphism) invariant.

Given the boundary action (\ref{bdaction}), one can evaluate the
boundary stress-energy tensor via the definition (\ref{genfcn2}).
As discussed earlier, it is convenient to express the SYM stress-energy
tensor in terms of its helicity variables $\T_s$.
Let $\T_s^{\rm quark}$ and $\T_s^{\rm eq}$
denote the helicity variables of 
$T^{\mu \nu}_{\rm quark}$ and $T^{\mu \nu}_{\rm eq}$, respectively.
Neglecting the contribution from the $U(1)$ electromagnetic field,
a short exercise yields
\begin{subequations}
\begin{align}
\T_0 &= \lim_{u \rightarrow 0} \left [
2 q^2 \mathcal A_0 \, \partial_u Z_0 
 + \mathscr J^{00} 
 + \T_0^{\rm eq}-\T_0^{\rm quark} \right ],
\label{T0}
\\
\label{T1}
\vec \T_1 &= \lim_{u \rightarrow 0}
\left [ q \mathcal A_1 \vec Z_1 +\vec \T_1^{\rm eq}-\vec \T_1^{\rm quark} \right ],
\\ 
\label{T2}
\tensor \T_2 &= \lim_{u \rightarrow 0}
\left [ 2 \mathcal A_2 \, \partial_u \tensor  Z_2
    +\tensor \T_2^{\rm eq}-\tensor \T_2^{\rm quark} \right ].
\end{align}
\end{subequations}

We now simplify the above expressions for the helicity variables $\T_s$.  
The gauge invariants are solutions of their
respective differential equations (\ref{eqm2}), (\ref{eqm1}) and (\ref{eqm0}),
and vanish at the boundary.
This implies that they
have the power series expansions,
\begin{subequations}
\begin{align}
\label{expansions}
Z_0 &=  u^3 Z^{(3)}_0 + u^4 Z^{(4)}_0+\cdots,
\\ 
\vec Z_1 &=  u^2 \vec Z^{(2)}_1 + u^3 \vec Z^{(3)}_1+\cdots,
\\
\tensor Z_2 &=  u^3 \tensor Z^{(3)}_2 + u^4 \tensor Z^{(4)}_2+\cdots.
\end{align}
\end{subequations}
The leading terms in these expansions are temperature independent.  
Plugging the expansions into the respective equations
one finds
\begin{subequations}
\begin{align}
\nonumber
Z^{(3)}_0 &= \frac{\kappa_5^2 \sqrt{\lambda} }{6 \pi L^3} \,
\frac{\left[q^2 \left(v^2{+}2\right){-}3 \omega ^2\right]
    \left(q^2{-}\omega ^2\right)}{q^2 \sqrt{1{-}v^2}}
\\ 
& \kern 2.7cm {}
\times (2 \pi) \delta(\omega- \v \cdot \q) \,,
\\
\vec Z^{(2)}_1 &= -\frac{\kappa_5^2 \sqrt{\lambda} }{\pi L^3} \,
 \frac{v q_{\perp}}{q  \sqrt{1{-}v^2}} \,
(2 \pi) \delta(\omega{-} \v \cdot \q) \, \eps_1 \,,
\\ \nn
\tensor Z^{(3)}_2 &= \frac{\kappa_5^2 \sqrt{\lambda} }{6 \pi L^3} \,
\frac{v^2 q_{\perp}^2 }{q^2 \sqrt{1-v^2}}  \,
(2 \pi)  \delta(\omega{-} \v \cdot \q)
\\
& \kern 2.3cm {}
    \times \left(\eps_1 \otimes \eps_1 - \eps_2 \otimes \eps_2\right).
\end{align}
\end{subequations}
{}From the definition (\ref{quarkstress}) of $T^{\mu \nu}_{\rm quark}$
and our choice (\ref{eps1}) of polarization vectors,
it is easy to see that
\begin{subequations}
\begin{align}
\T_0^{\rm quark} &= \frac{M}{\sqrt{1{-}v^2}} \,
(2 \pi) \delta(\omega {-} \v \cdot \q),
\\ 
\vec \T_1^{\rm quark} &= \frac{M}{ \sqrt{1{-}v^2}} \,
\frac{v q_{\perp}}{q} \,
(2 \pi) \delta(\omega {-} \v \cdot \q) \,
\eps_1,
\\ 
\nonumber
\tensor \T_2^{\rm quark} &= \frac{M}{\sqrt{1{-}v^2}} \,
\frac{(v q_{\perp})^2}{2 q^2} \,
(2 \pi) \delta(\omega {-} \v \cdot \q)
\\
&\kern 2cm {} \times\left(\eps_1 \otimes \eps_1 - \eps_2 \otimes \eps_2\right) ,
\end{align}
\end{subequations}
where $M\equiv\sqrt{\lambda}/(2 \pi \inf)$ is the quark mass which is
being sent to infinity.
The helicity components of the equilibrium stress-energy tensor are just
\begin{equation}
    \T_0^{\rm eq} = \frac{3}{8} \, \Nc^2 \pi^2 T^4 \,, \quad
    \vec \T_1^{\rm eq} =  0 \,,\quad
    \tensor \T_2^{\rm eq} = 0 \,.
\end{equation}
Putting it all together we have%
\footnote
    {%
    These expressions for the $\T_s$ can also be
    obtained by solving the complete set of field equations
    for the metric perturbation $H_{\mu \nu}$.
    Near the boundary,
    $
	H_{\mu \nu} = u^3 \, H_{\mu \nu}^{(3)} + u^4 \, H_{\mu \nu}^{(4 )}+\cdots
    $.
    In a gauge in which $h_{5 M} = 0$, the SYM stress-energy tensor is
    given by
    $
	T_{\mu \nu} =
	T_{\mu \nu} ^{\rm eq}
	+ \frac{2 L^3}{\kappa_5^2} \, H_{\mu \nu}^{(4)}
    $
    \cite{Skenderis:2000in}.
    This approach was taken in
    Refs.~\cite{Chesler:2007an, Friess:2006fk, Lin:2007pv}.
    Our treatment above has the virtue of being self-contained
    and very explicit about the treatment of contributions from the quark
    and the external electric field.
    }
\begin{subequations}
\begin{align}
\label{t0}
    \T_0 &=
    \frac{4 q^2 L^3}{3 \kappa_5^2 (q^2{-}\omega^2)^2} \; Z_0^{(4)} 
    +\mathcal D
    +\T_0^{\rm eq}
    \,,
\\ \label{t1}
\vec \T_1 &=
    %\vec \T_1^{\rm eq}
    - \frac{L^3}{2 \kappa_5^2} \; \vec Z_1^{(3)}  \,,
\\ \label{t2}
    \tensor \T_2 &=
    %\tensor \T_2^{\rm eq} +
    \frac{2 L^3}{ \kappa_5^2} \; \tensor Z^{(4)}_{2} \,.
\end{align}
\end{subequations}
where
\begin{align}
    \mathcal D
    &\equiv
    -\mathcal C \, \frac{i \omega  \left[q^2 \left(5 v^2{+}1\right)-3
   \left(v^2{-}1\right) \omega ^2\right]}{\uh^2 \left(1{-}v^2\right)
   \left(q^2{-}\omega ^2\right)}
\nonumber\\
    &=
    -\frac{\sqrt{ \lambda} }{6 \pi} \,
    { \sqrt{1{-}v^2}} \,
    (2 \pi) \delta(\omega {-} \v \cdot \q)
\nonumber\\ & \kern 1cm {} \times
    \frac{i \omega  \left[q^2 \left(5 v^2{+}1\right)-3
   \left(v^2{-}1\right) \omega ^2\right]}{\uh^2 \left(1{-}v^2\right)
   \left(q^2{-}\omega ^2\right)^2} \,.
\end{align}
Note that, after subtracting $\T^{\rm quark}_s$,
all divergent quantities have canceled.

Below, we will focus on the
temperature dependent perturbation in the stress-energy tensor
due to the moving quark,%
\footnote
    {%
    The zero-temperature stress energy tensor is completely
    determined, up to the overall normalization,
    by conformal invariance.  In the rest frame of the quark
    it is given by
    $
	T^{\mu\nu}(x)\Big|_{T=0}
	= \sqrt{\lambda} \, {\rm diag}(1,\third,\third,\third) \big/
	(12 \pi^2 \x^4 )
	%= frac{\sqrt{\lambda}}{12 \pi^2 \x^4 } \,\rm{diag}(1,\third,\third,\third)
    $
    \cite{Gubser:2007nd}.
    }
\begin{equation}
\label{DT}
\Delta T^{\mu \nu} \equiv
T^{\mu \nu} -T^{\mu \nu}_{\rm eq}- T^{\mu \nu}\Big|_{T = 0} \,.
\end{equation}
That is, we subtract the zero temperature stress-energy of the quark
as well as the stress-energy of the equilibrium plasma.
We also define
\begin{subequations}
\begin{align}
\Delta \E &\equiv \Delta T^{00},
\\
\Delta S_i &\equiv \Delta T^{0i},
\end{align}
\end{subequations}
which are the temperature dependent perturbations in the
energy density and energy flux, respectively.

As discussed in section~\ref{defs},
the symmetric tensor $\Delta T^{\mu\nu}$ is determined by
the energy-momentum conservation equation (\ref{conservation}) it obeys
together with the perturbations in the associated helicity variables,
\begin{subequations}
\begin{align}
\Delta \T_0 &=
\frac{4 q^2 L^3}{3 \kappa_5^2 (q^2{-}\omega^2)^2} \; \Delta Z_0^{(4)} 
+
\mathcal D \,,
\\
\Delta\vec \T_1 &= - \frac{L^3}{2 \kappa_5^2} \; \Delta \vec Z_1^{(3)} \,,
\\
\Delta\tensor \T_2 &=\frac{2 L^3}{ \kappa_5^2} \; \Delta \tensor Z^{(4)}_{2} \,,
\end{align}
\label{eq:delta Ts}
\end{subequations}
where
$\Delta Z_0^{(4)}$, $\Delta \vec Z_1^{(3)}$, and $ \Delta \tensor Z^{(4)}_{2}$
are the values of $Z_0^{(4)}$, $\vec Z_1^{(3)}$ and $\tensor Z^{(4)}_{2}$,
respectively,
at temperature $T$ minus zero temperature.

\subsection{Numerical Solutions}
\label{Numerics}

One must solve the inhomogeneous differential equations for the $Z_s$,
Eqs.~(\ref{eqm2}), (\ref{eqm1}) and (\ref{eqm0}),
with appropriate boundary conditions at the horizon and at the boundary.
To do so, we construct Green's functions $G_s(u,u')$ out
of homogeneous solutions,
\begin{equation}
    G_s(u,u') = g^{<}_s(u_{<}) \> g^{>}_s(u_>) \bigm/ W_s(u') \,,
\end{equation}
with $W_s(u)$ the Wronskian of $g^<_s$ and $g^>_s$,
and convolve with the source,
\begin{equation}
    Z_s(u) = \int_0^{u_h} du' \> G_s(u,u') \, S_s(u') \,.
\end{equation}
The appropriate homogeneous solutions to use
are dictated by the boundary conditions.
The differential operators in
Eqs.~(\ref{eqm2}), (\ref{eqm1}) and (\ref{eqm0}) have singular points at $u=0$ and $u=u_h$.  
The indicial exponents at 
$u = 0$ are non-negative integers and those at $u = u_h$ are $\pm i \omega u_h/4$.
Vanishing of the metric perturbation at the boundary requires that
$g^{<}_s(0) = 0$,
while the requirement that the black hole not radiate \cite{Son:2002sd} implies that
$g^{>}_s(u) \sim (u{-}u_h)^{-i \omega u_h/4}$ as $u \to u_h$.

\subsubsection{Tensor Mode}

The indicial exponents of the helicity two differential equation (\ref{eqm2})
at $u = 0$ are 0 and 4.
Hence the homogeneous solution $g_2^<$ must have the asymptotic behavior
$g^{<}_2(u) \sim u^4$ as $u\to0$.
We fix the (arbitrary) overall normalization by requiring that
$\lim_{u\to0} g^{<}_2(u)/u^4 = 1$.

At zero temperature, the homogeneous solutions to the helicity two equation
are
\begin{subequations}
\begin{align}
g^{>}_{2,T=0}(u) &= u^2 \, K_2( u \sqrt{q^2 {-} \omega^2} ) \,,
\\
g^{<}_{2,T=0}(u) &= \frac{8 u^2}{q^2 {-}\omega^2} \,
	I_2(u \sqrt{q^2 {-} \omega^2} ) \,,
\end{align}
\end{subequations}
where $K_2$ and $I_2$ are modified Bessel functions.
The corresponding zero 
temperature Wronskian is
\begin{equation}
    W_{2,T=0}(u) = -\frac{ 8 u^3}{q^2 {-} \omega^2} \,.
\end{equation}

The radial coordinate of zero temperature AdS space
(using the metric (\ref{metric}) with $u_h = \infty$)
runs over the interval $(0, \infty)$.
To subtract zero temperature contributions to the stress-energy
tensor it will be convenient to map this interval
onto $(0,\uh)$ with a change of variables
\begin{equation}
\xi(u) = \frac{u}{f(u)} \,.
\label{eq:changeofvar}
\end{equation}

Putting everything together, we have
\begin{align}
\label{grnsfcnrep2}
    \Delta \tensor Z^{(4)}_2 & =
    \int_{0}^{u_h} du \> \Bigr \{\frac{g^{>}_2(u)}{W_2(u)} \> \tensor S_2(u)
\\ \nonumber
    &+
    J(u) \, \frac{q^2{-}\omega^2}{8 \, \xi(u)}
    K_2 \!\left (\xi(u) \sqrt{q^2 {-} \omega^2} \right )
   \tensor  S_{2,T=0}(\xi(u)) \Bigr \} \, ,
\end{align}
where $J \equiv d\xi/d u = (f - u f')/f^2$ is the Jacobian for
the change of variables (\ref{eq:changeofvar}).

\subsubsection{Vector Mode}

The indicial exponents of the helicity one differential equation (\ref{eqm1})
at $u = 0$ are 0 and 3.
Vanishing of the metric perturbation at the boundary thus requires that
$g^{<}_1(u) \sim u^3$ as $u\to0$.
The overall
normalization of $g^<_1$ is fixed by requiring
$\lim_{u\to0} g^{<}_1(u)/u^3 \equiv 1$.

At zero temperature, the homogeneous solutions to the helicity one equation
are
\begin{subequations}
\begin{align}
g^{>}_{1,T=0}(u) &= u^2 K_1( u \sqrt{q^2 {-} \omega^2} ) \,,
\\
g^{<}_{1,T=0}(u) &= \frac{2 u^2}{\sqrt{q^2 {-}\omega^2}} \,
I_1(u \sqrt{q^2 - \omega^2} ) \,,
\end{align}
\end{subequations}
and their Wronskian is
\begin{equation}
W_{1,T=0}(u) = -\frac{ 2 u^3}{\sqrt{q^2 {-} \omega^2}} \,.
\end{equation}
We therefore find
\begin{align}
\label{grnsfcnrep1}
    \Delta \vec Z^{(3)}_1 &=
    \int_{0}^{u_h} du \> \Bigr \{\frac{g^{>}_0(u)}{W_0(u)} \> {\vec S_1}(u)
\\ \nonumber
    &+
    J(u) \frac{\sqrt{q^2{-}\omega^2}}{2 \, \xi(u)}
    K_1 \!\left (\xi(u) \sqrt{q^2 {-} \omega^2} \right)
    \vec S_{1,T=0}(\xi(u)) \! \Bigr \} .
\end{align}

\subsubsection{Scalar Mode}

Finally,
the exponents of the helicity zero differential equation (\ref{eqm0})
at $u = 0$ are 0 and 4.
Hence $g_0^<$ must satisfy
$g^{<}_0(u) \sim u^4$ as $u\to0$,
and the normalization is fixed by requiring
$\lim_{u\to0} g^{<}_0(u)/u^4 \equiv 1$.

At zero temperature, the homogeneous solutions to the helicity zero equation
are
\begin{subequations}
\begin{align}
g^{>}_{0,T=0}(u) &= u^2 K_2( u \sqrt{q^2 {-} \omega^2} ) \,,
\\
g^{<}_{0,T=0}(u) &= \frac{8 u^2}{q^2 {-}\omega^2} \,
I_2(u \sqrt{q^2 {-} \omega^2} ) \,,
\end{align}
\end{subequations}
and their Wronskian is
\begin{equation}
W_{0,T=0}(u) = -\frac{ 8 u^3}{q^2 {-} \omega^2} \,.
\end{equation}
Consequently,
\begin{align}
\label{grnsfcnrep0}
    \Delta Z^{(4)}_0& =
    \int_{0}^{u_h} du \> \Bigr \{\frac{g^{>}_0(u)}{W_0(u)} \> S_0(u)
\\ \nonumber
    &+
    J(u) \, \frac{q^2{-}\omega^2}{8 \, \xi(u)} \,
    K_2 \!\left (\xi(u) \sqrt{q^2 {-} \omega^2} \right )
    S_{0,T=0}(\xi(u)) \Bigr \} \,.
\end{align}

\subsubsection{Numerics}

For a given momentum $\q$, the homogeneous solutions
$g^{>}_s$ are evaluated by numerically integrating 
the appropriate homogeneous differential equations
(\ref{eqm2}), (\ref{eqm1}) and (\ref{eqm0})
(without sources)
outward from the horizon.
Then the solutions $g^{<}_s$ are evaluated by numerically integrating 
the same equations (without sources)
inward from the boundary.
Given these numerically determined homogeneous solutions,
the perturbations in the helicity variables $\Delta \T_s$
are evaluated by numerically performing the
radial integrals in Eqs.~(\ref{grnsfcnrep2}),  (\ref{grnsfcnrep1})
and (\ref{grnsfcnrep0}),
and inserting the results in Eq.~(\ref{eq:delta Ts}).

The reconstruction of the stress-energy tensor components from
the helicity variables is given in Eqs.~(\ref{t00}) and (\ref{taq}).
In particular,
the perturbation in the energy density is
\begin{equation}
\label{de}
    \Delta\mathcal E(t,\x)
    =
    \int \frac{d^4q}{(2 \pi)^4} \; \Delta \T_0(\omega,\q) \,
    e^{i Q \cdot x} \,,
\end{equation}
while the perturbation in the energy flux is
\begin{align}
\label{ds}
\Delta\bm S(t,\x) &= \int \frac{d^4q}{(2 \pi)^4}
    \bigg [
	 \frac{\q}{q^2} \big( \omega \Delta\T_0 - i \mathcal F^{0} \big)
	+\Delta\vec \T_1(\omega,\q) 
    \bigg ]
\nonumber\\ & \qquad {}  \times
    e^{i Q \cdot x} \,.
\end{align}
The frequency integrals in these expressions are trivial as the integrands are
proportional to $2 \pi \delta(\omega {-} \v \cdot \q)$.
We write the remaining $3d$ integrals in spherical coordinates
$(q,\theta,\phi)$ with $\theta$
the polar angle
measured relative to the direction of the velocity $\v$.
In this coordinate system we have
\begin{equation}
    q_{\|} = q \cos\theta \,,\qquad q_{\perp} = q \sin\theta \,,
\end{equation}
where $q_{\|}$ is the component of $\q$ parallel to the quark's velocity.
The integration over $\phi$ is easily done analytically,
as the physical problem is cylindrically symmetric about the direction $\v$.
The remaining integrals over $q$ and $\theta$ are evaluated numerically.

\begin{figure}[t]
\includegraphics[scale=0.3]{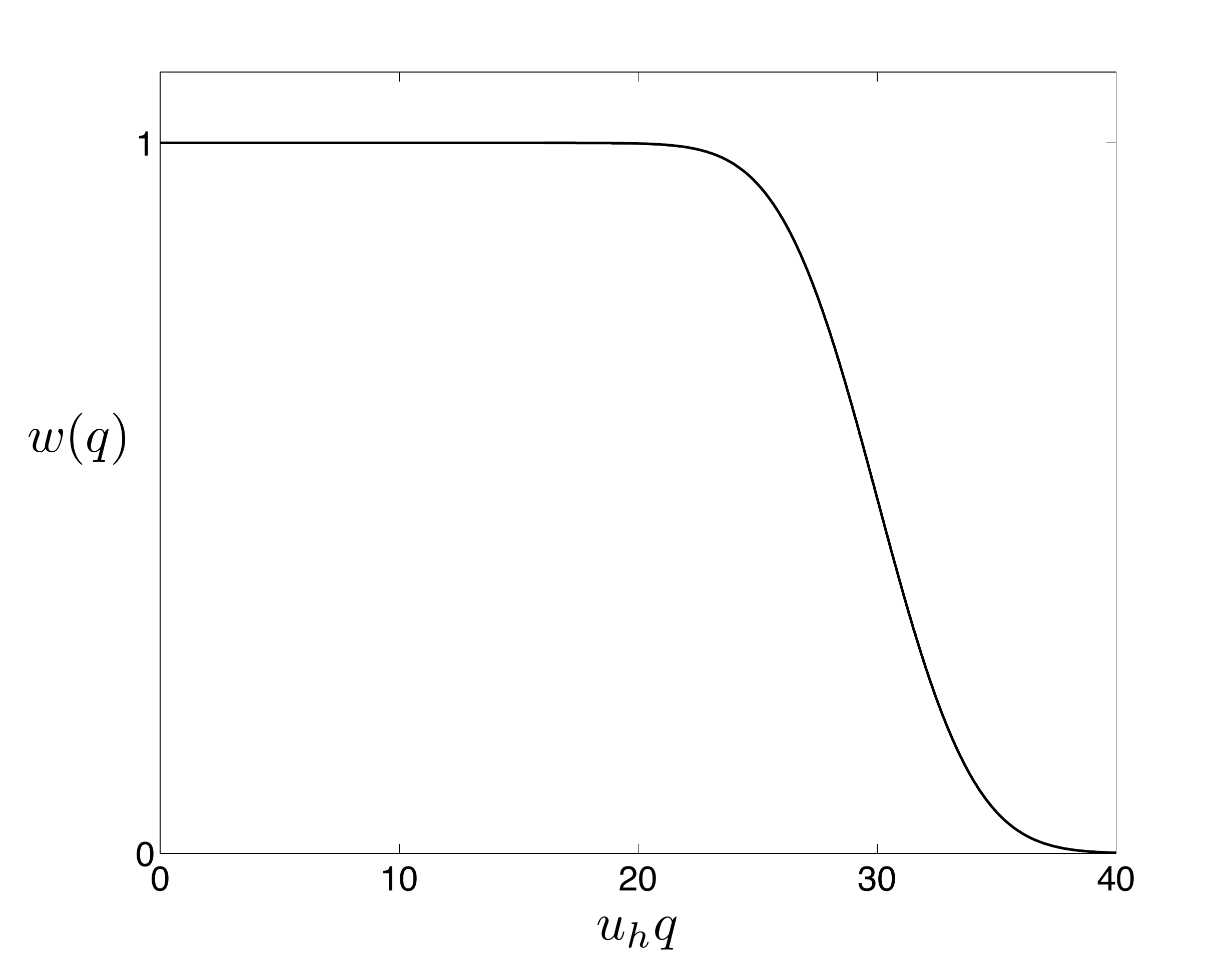}
\caption{\label{window}
A plot of the window function $w(q)$.  The window function does not modify the 
Fourier space data for $\uh q < 20$.
}
\end{figure}

The integrands (with the factor of $q^2$ from the measure)
scale like $q$ times an oscillatory function in the large $q$ limit. 
This reflects on the fact that $\Delta T^{\mu \nu}(t,\x)$ diverges like $T^2/|\x - \v t|^2$ in the vicinity of the quark.
To deal with this divergence
we multiply the Fourier space data by the window function
\begin{equation}
    w(q) = \frac{1}{2}
    \left [1-\text{erf} \, \Bigl ( \frac{\uh q-30}{4.5} \Bigr ) \right ].
\end{equation}
This window function is shown in Fig.~\ref{window}.  From the figure we
see that the window function does not significantly modify the Fourier space
data for $\uh q <20$.  Correspondingly the window function alters the real space
energy density and energy flux over length scales $\sim \uh/20$, which
is much smaller than the position space grids used to make the plots
shown below.
With this window function,
the integral over $q$ is evaluated with an upper limit of $\uh q = 40$.

\begin{figure*}[p]
\includegraphics[scale=0.47]{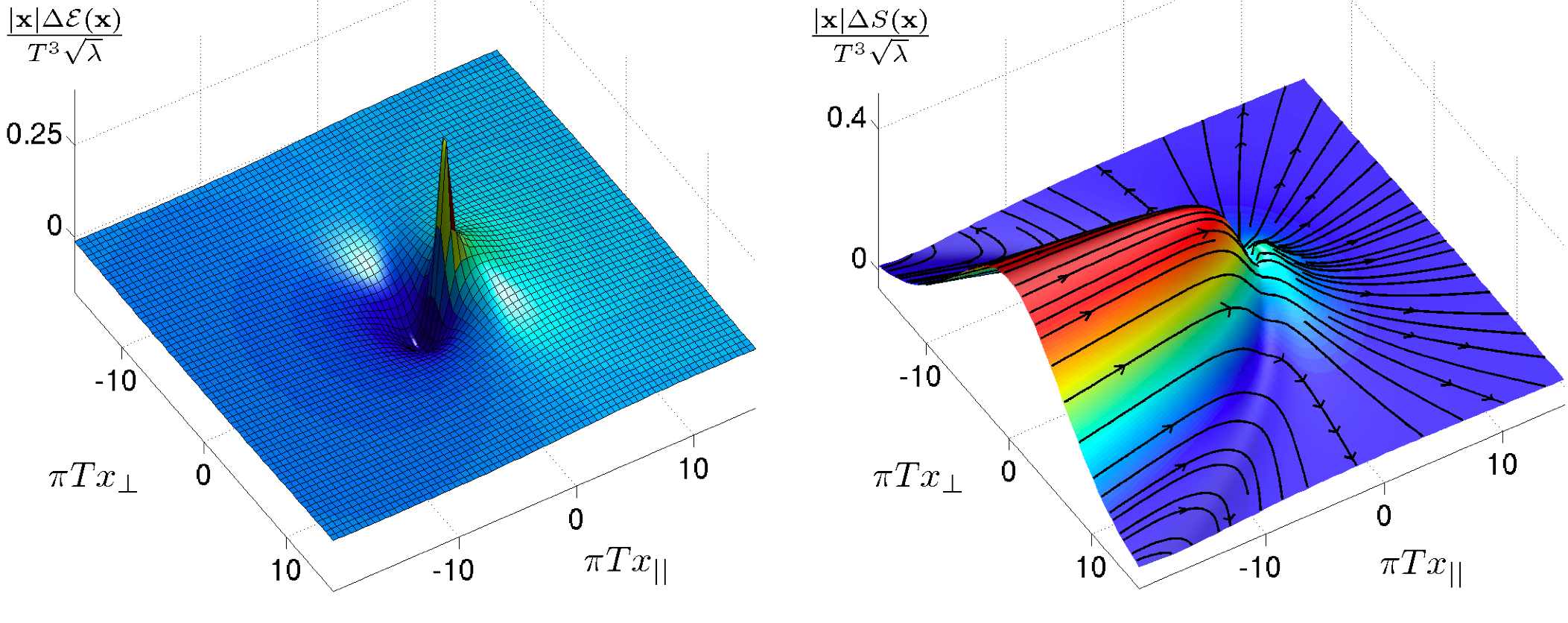}
\caption{\label{v25}
Left---Position space plot of $| \x| \Delta\E(\x)/(T^3 \sqrt{\lambda})$ for $v = 1/4$.
Right---Position space plot of $| \x| \Delta S(\x)/(T^3 \sqrt{\lambda})$ for $v = 1/4$.
The flow lines on the surface are the flow lines of the energy flux
$\Delta \bm S(\x)$.
There is a surplus of energy in front of the quark and a deficit behind it.
Correspondingly, trailing the quark there is a stream of energy flux
which moves in the same direction as the quark.
Note the absence of structure in $\Delta\E(\x)$ for distances $| \x| \gg 1/ (\pi T)$.
}
\end{figure*}

\begin{figure*}[p]
\includegraphics[scale=0.51]{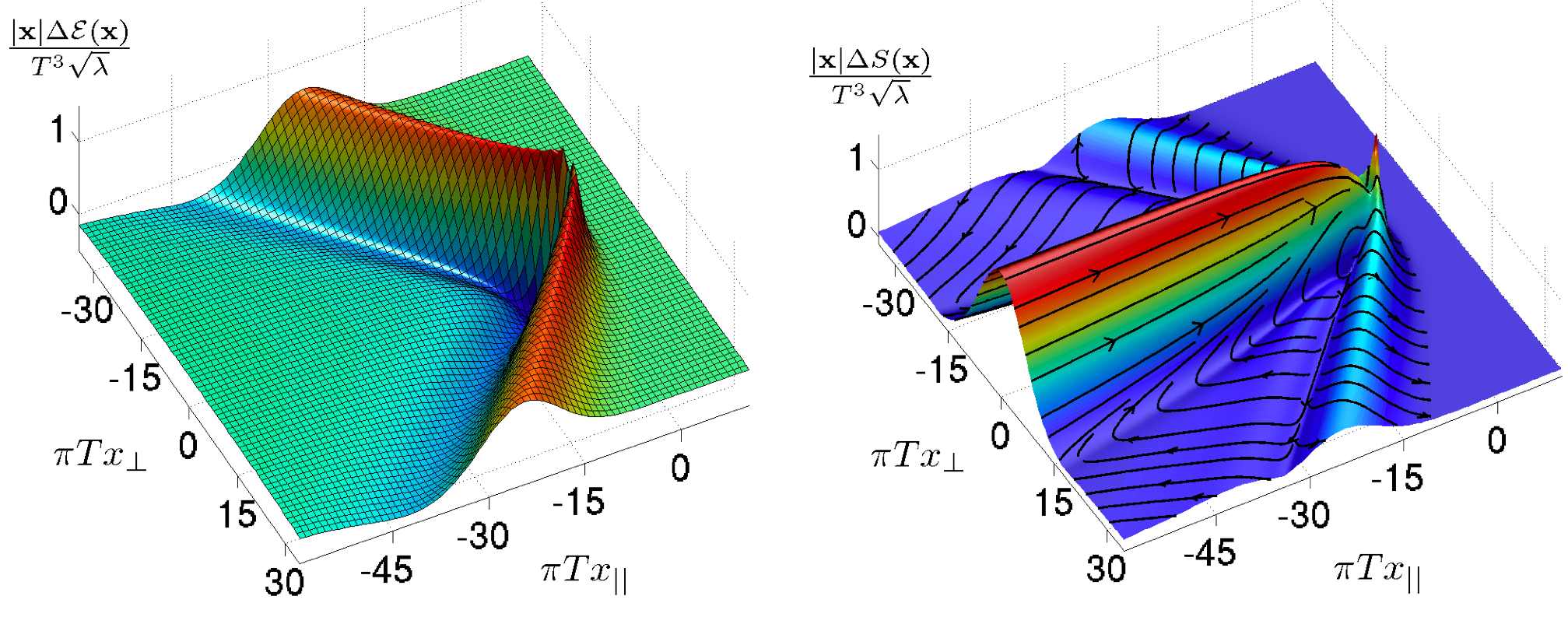}
\caption{\label{v75}
Left---Plot of $| \x| \Delta \E(\x)/(T^3 \sqrt{\lambda})$ for $v =3/4$.
Right---Plot of $| \x| \Delta S(\x)/(T^3 \sqrt{\lambda})$ for $v = 3/4$.
The flow lines on the surface are the flow lines of $\Delta\bm S(\x)$.
There is a surplus of energy in front of the quark and a deficit behind it.
Correspondingly, trailing the quark there is a narrow stream of energy flux
which moves in the same direction as the quark.
A Mach cone, with opening half angle $\theta_M \approx 50^{\circ}$
is clearly visible in both the energy density and the energy flux.
Near the Mach cone, 
the bulk of the energy flux flow is orthogonal to the
wavefront.
}
\end{figure*}

\section{Results}
\label{Results}

For small distances
$d \equiv |\x{-}\v t| \ll 1/T$
away from the moving quark,
the dominant contributions to the stress-energy tensor
come from the $T = 0$ stress-energy tensor,
which scales like $1/d^4$.
To highlight the medium dependent perturbations
in the stress-energy tensor we defined
$\Delta T^{\mu \nu}$
[in Eq.~(\ref{DT})]
to be the temperature dependent
perturbation to $T^{\mu \nu}$.

%%%%%%%%

\begin{figure*}[t]
\includegraphics[scale=0.47]{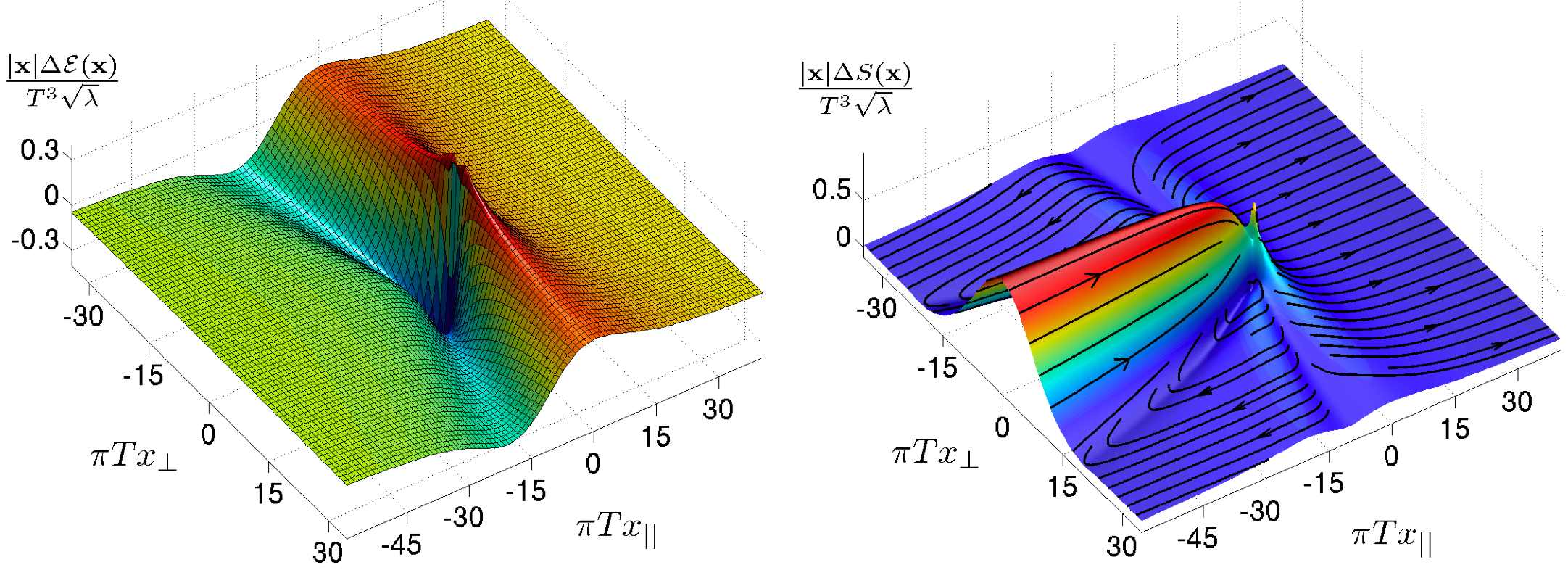}
\caption{\label{vvs}
Left---Plot of $| \x| \Delta\E(\x)/(T^3 \sqrt{\lambda})$ for $v =\cs$.
Right---Plot of $| \x| \Delta S(\x)/(T^3 \sqrt{\lambda})$ for $v = \cs$.
The flow lines on the surface are the flow lines of the energy flux
$\Delta\bm S(\x)$.
A planar Mach cone is visible in 
both the energy density and the energy flux.  Near the Mach cone, 
the bulk of the energy flux flow is orthogonal to the
wavefront.}
\end{figure*}

Figures~\ref{v25}--\ref{vvs} show plots of $\Delta \mathcal E(t,\x)$ 
and $\Delta S(t,\x) \equiv |\Delta \bm S(t,\x)|$ at 
quark velocities $v = 1/4$, $v = 3/4$ and $v = 1/\sqrt{3}$,
respectively.%
\footnote
  {%
  In making these plots, we use spatial grids with resolution
  $\Delta x$ equal to one to two times  $1/(2\pi T)$.
  This limits the fidelity of these plots for distances 
  $|\x|  \sim \Delta x$ from the quark.
  As noted earlier,
  the temperature dependent energy density and energy flux behave like 
  $T^2/|\Delta\x|^2$ for distances $|\Delta\x| \ll 1/T$ from the quark.  
  See Refs.~\cite{Gubser:2007nd, Yarom:2007ap}
  for a discussion of the stress-energy tensor in the near zone. 
  }
The flow lines superimposed on the plots of 
$S(t,\x)$ are the flow lines of $\bm S(t,\x)$.  
In these plots the quark, at the time shown,
is at $\x = 0$.
We use units in which $\pi T = 1$.
Since $\Nfour$ SYM is a conformal theory,
the speed of sound is $1/\sqrt 3$.
Hence Fig.~\ref{v25} shows subsonic motion,
Fig.~\ref{v75} shows supersonic motion,
and Fig.~\ref{vvs} is precisely at the speed of sound.

As discussed earlier,
$\Delta T^{\mu \nu}$ may be reconstructed from the helicity
variables $\Delta \T_s$, combined with
the energy-momentum conservation
equation (\ref{conservation}) and the vanishing trace condition
of the SYM stress-energy tensor.
In Appendix 
\ref{smallq} we compute the small momentum limit of $\Delta \T_s$
to $\O(q^0)$ with the ratio $r \equiv \omega/q$ held fixed.
Defining 
\begin{equation}
\K \equiv \frac{\sqrt{\lambda}}{2 \pi \uh^2 \sqrt{1-v^2}} \,
(2 \pi) \delta(\omega - \v \cdot \q),
\end{equation}
one finds%
\footnote
  {%
  These results for the small momentum limit of the stress-energy tensor 
  may also be found in Ref.~\cite{Gubser:2007nd}.
  }
\begin{subequations}
\begin{align}
\Delta \T_0 &= 3 \K
\left [
\frac{-i r (1{+}v^2)}{(1{-}3 r^2) q}
+\frac{\uh r^2 (2 {-} 3r^2 {+}v^2)}{(1{-}3r^2)^2}
\right ], \ \ \
\\ 
\Delta \vec \T_1 &=  \K \,
\frac{v q_{\perp}}{q}
\left  [-\frac{1}{i r q } + \frac{\uh(1{-}4 r^2)}{4 r^2}  \right ]  \,\eps_1,
\\
\Delta \tensor \T_2 &= -\K \, \frac{\uh}{2} \,
\frac{(v q_{\perp})^2}{q^2} \,
(\eps_1 \otimes \eps_1 - \eps_2 \otimes \eps_2),
\end{align}
\end{subequations}
up to $\O(q)$ corrections. 

\begin{figure*}[t!]
\includegraphics[scale=0.35]{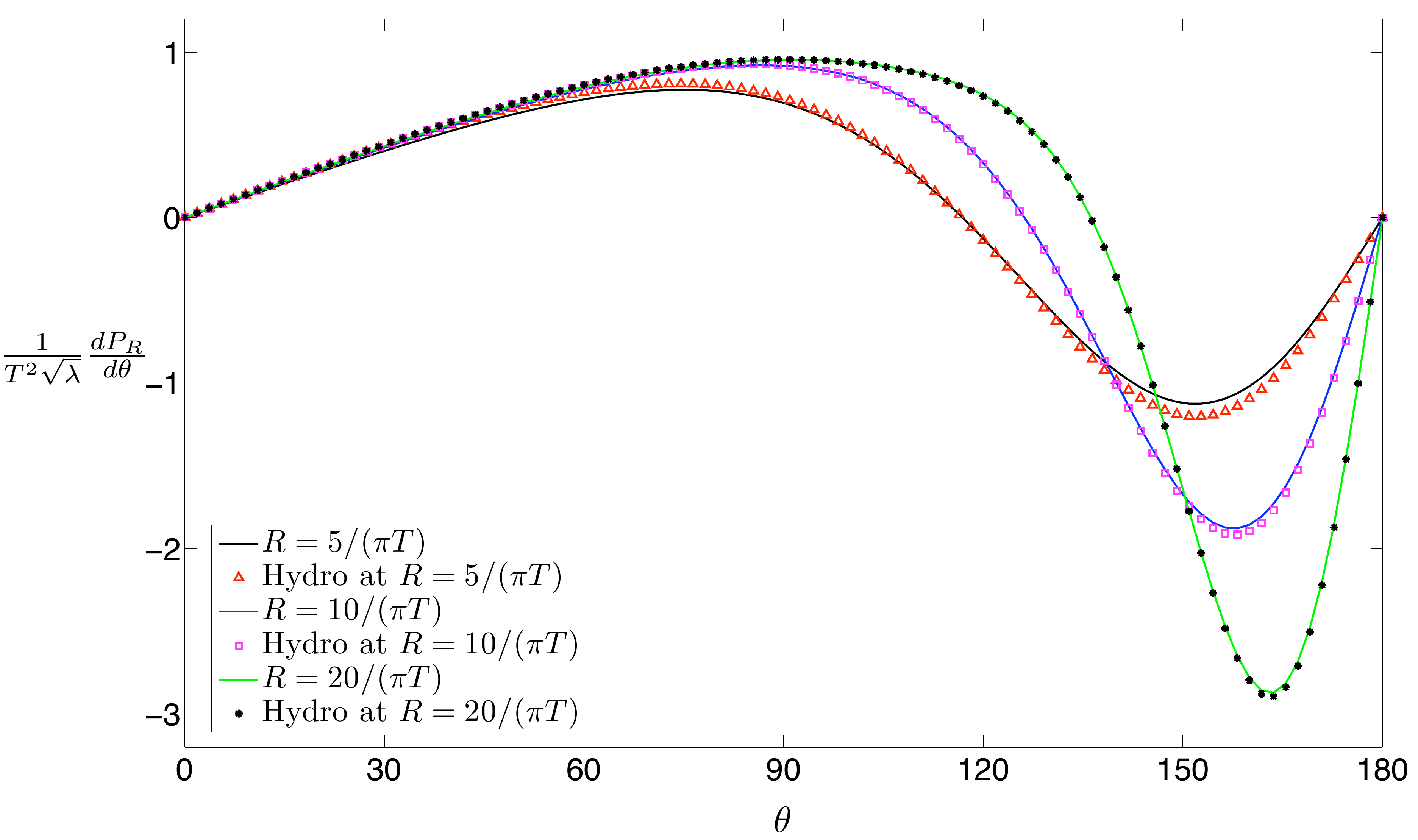}
\caption{\label{p25}
Plot of
$(T^2 \sqrt{\lambda})^{-1} {d P_R}/{d \theta}$
for $v = 1/4$ at distances $R = 5/(\pi T)$ (black curve and red triangles),
$10/(\pi T)$ (blue curve and purple squares), and
$20/(\pi T)$ (green curve and black dots).
Solid lines are the full AdS/CFT results,
while the symbols show the linearized hydrodynamic approximation.
Note the presence of energy being radiated into the forward hemisphere in
front of the quark.
Negative values in the backward hemisphere are due to the diffusion wake.
At all three distances, the hydrodynamic approximation to
$ {d P_R}/{d \theta} $ agrees very well with the AdS/CFT results.
\label{fig:angular}
}
\end{figure*}

These results for the long wavelength limit of $\Delta \T_s$ may be compared
to the analogous long wavelength limit of the hydrodynamic quantities
$\Delta \T_s^{\rm hydro}$ given in 
Eqs.~(\ref{eq:hydroT}).
To give a meaningful comparison, 
we expand the hydrodynamic helicity variables in powers of $q$ with the ratio $r = \omega/q$ fixed.  
Writing
\begin{subequations}
\begin{align}
\rho &= q \, \rho^{(1)} + q^2 \, \rho^{(2)} \cdots \,,
\\
\bm J_{\rm T}&= \bm J_{\rm T}^{(0)}+q \, \bm J_{\rm T}^{(1)} + \cdots \,,
\end{align}
\end{subequations}
we find
\begin{subequations}
\begin{align}
\label{hyd0}
\Delta \T_0^{\rm hydro} &= 
-\frac{3 \rho^{(1)}} {(1{-}3 r^2)\, q}
    - \frac{9 i  r \, \gamma\, \rho^{(1)} + 3 (1{-}3 r^2) \rho^{(2)}}{( 1{-}3 r^2)^2}\,,
\\ \label{hyd1}
\Delta \vec \T_1^{\rm hydro} &=  
-\frac{\bm J_{\rm T}^{(0)} }{i r q } + \frac{D \bm J_{\rm T}^{(0)} + i r \bm J_{\rm T}^{(1)}}{ r^2} \,,
\\ \label{hyd2}
\Delta \tensor \T_2^{\rm hydro} &= 0 \,,
\end{align}
\end{subequations}
neglecting $\O(q)$ corrections in Eqs.~(\ref{hyd0}) and (\ref{hyd1}).  
Comparing the expansions of $\Delta \T_0^{\rm hydro}$ and
$\Delta \vec \T_1^{\rm hydro}$ 
to that of $\Delta \T_0$ and $\Delta \vec \T_1$, we see they agree
provided
the transverse momentum diffusion constant and the sound attenuation constant
have the expected values given earlier,%
\footnote
  {%
  Higher order transport coefficients such as $\Theta$,
  the coefficient multiplying  second derivatives of the energy density,
  can be determined by carrying out the expansion of the
  helicity variables to higher order in $q$.
  }
namely $D = 1/(4 \pi T)$ and 
$\gamma = 1/(3 \pi T)$,
and the hydrodynamic sources have the expansions  
\begin{subequations}
\begin{align}
\label{rho}
\rho &=  \K
\left [ \, i \omega (1+v^2) -\uh \omega^2 + \O(q^3) \right ]  ,
\\ \label{JT}
\bm J_{\rm T} &= \K \,
\frac{v q_{\perp}}{q} 
\left [1 +i \uh \omega + \O(q^2) \right ] \eps_1 \,.
\end{align}
\end{subequations}
Using the fact that $J^{0} \equiv F^{0}$, Eqs.~(\ref{rhodef}) and (\ref{rho})
imply that the longitudinal component of the effective source has the expansion
\begin{equation}
\label{JL}
\bm J_{\rm L} =\K \, \frac{\q}{q^2}
\left [ \, \omega +i \uh \omega^2 -i  \gamma v^2 q^2 +\O(q^3)\right] .
\end{equation}
Combining Eqs.~(\ref{JT}) and (\ref{JL}), we find
\begin{align}
\bm J =  \K
 \left [\v
+i \uh \omega \v  -i v^2 \gamma \q  +\O(q^2) \right ].
\end{align}
Comparing with Eqs.~(\ref{Fmu}) and (\ref{fmu}),
one sees that, as expected, the leading term in the gradient 
expansion of the effective source $\bm J$ is simply
the microscopic force density $\bm F$.
When Fourier transformed to position space, subleading
corrections to the effective source generate derivatives of
delta functions at the location of the quark.

\begin{figure*}[p]
\includegraphics[scale=0.36]{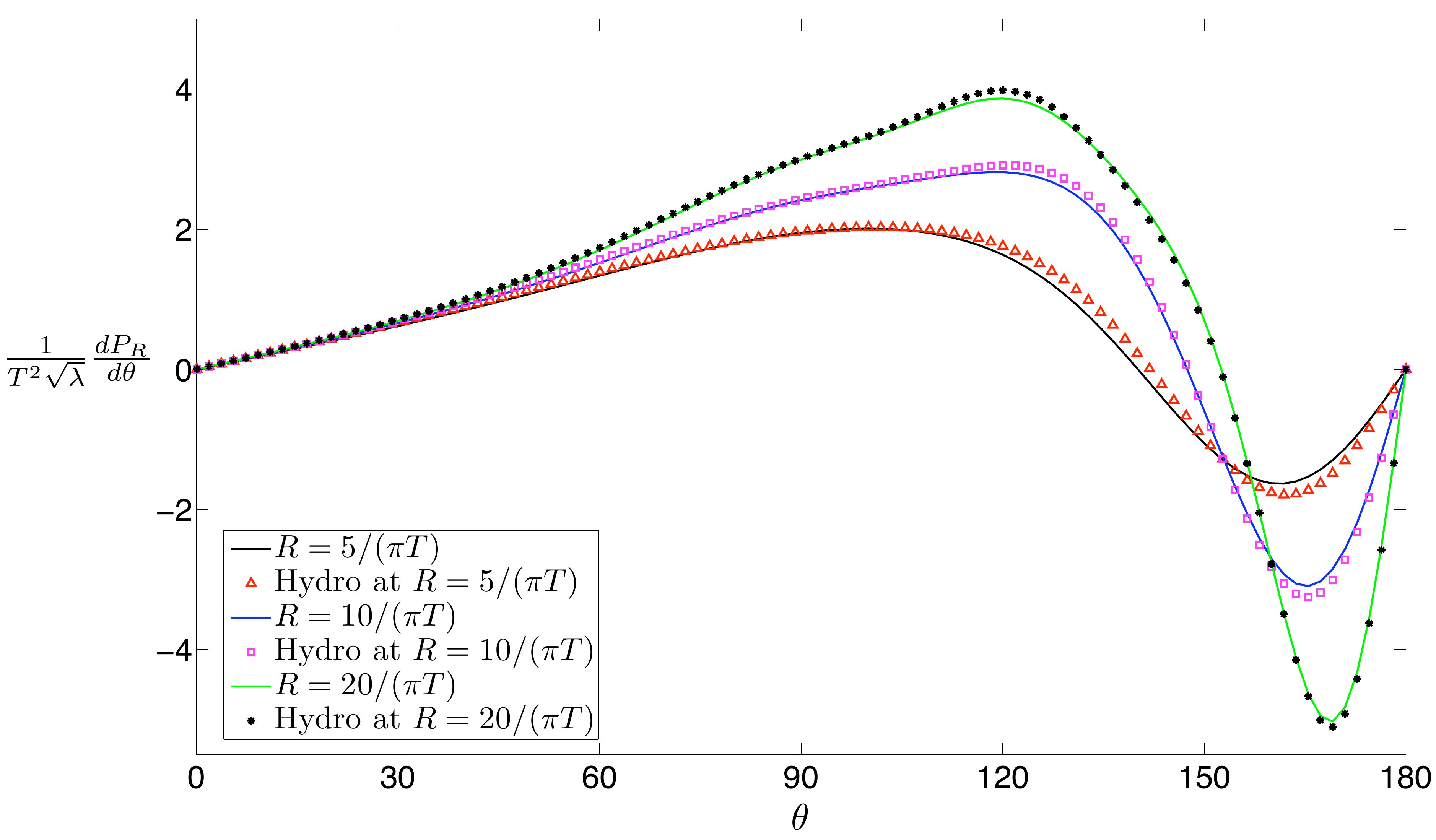}
\caption{\label{pvs}
Plot of
$(T^2 \sqrt{\lambda})^{-1} {d P_R}/{d \theta}$
at $v = \cs$ and distances $R = 5/(\pi T)$,
$10/(\pi T)$, and $20/(\pi T)$.
The labeling of curves is the same as in Fig.~\ref{fig:angular}.
Note the presence of energy being radiated into the forward hemisphere
in front of the quark, together with the inward flux associated with
the diffusion wake in the backward hemisphere.
The Mach cone is centered at $\theta = 90^{\circ}$.  
At all three distances, the hydrodynamic approximation to
${d P_R}/{d \theta}$
agrees very well with the AdS/CFT results.
}
\end{figure*}

\begin{figure*}[p]
\includegraphics[scale=0.35]{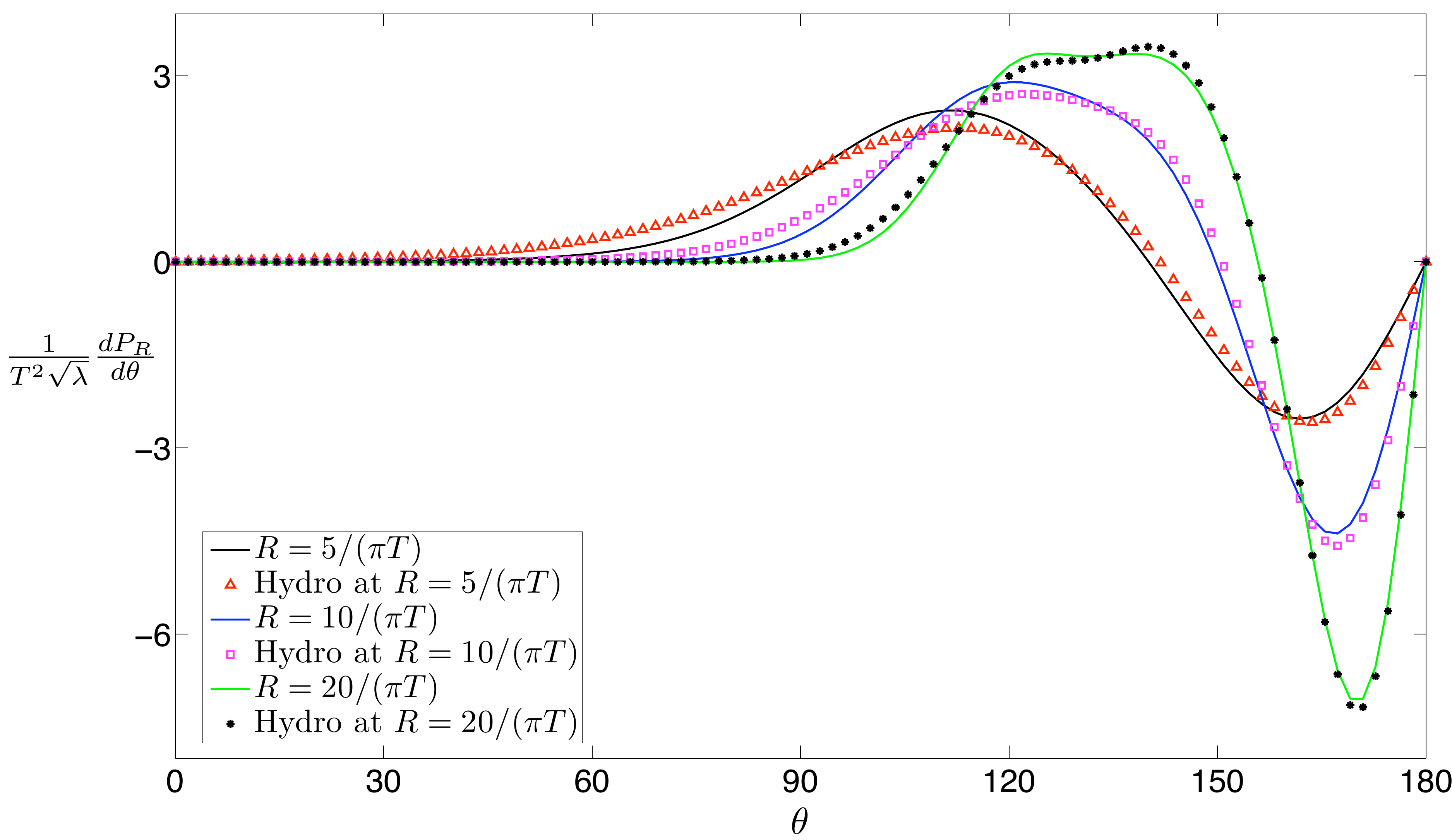}
\caption{\label{p75}
Plot of $(T^2 \sqrt{\lambda})^{-1} {d P_R}/{d \theta}$
at $v = 3/4$ and distances $R = 5/(\pi T)$,
$10/(\pi T)$, and $20/(\pi T)$.
The labeling of curves is the same as in Fig.~\ref{fig:angular}.
Note the absence of energy radiated in front of the quark,
as the quark is moving supersonically.
The Mach cone near 130$^\circ$is sharper than in the transonic case
shown in Fig.~\ref{pvs}.
The hydrodynamic approximation to ${d P_R}/{d \theta}$ improves
as $R$ grows, but is already fairly good at $R = 5/(\pi T)$.
}
\end{figure*}

As the above makes clear, the small $q$ expansions
of $\Delta \T_0$ and $\Delta \vec \T_1$
are completely consistent with
their hydrodynamic counterparts.
However the small $q$ limit of $\Delta T^{\mu \nu}$,
as reconstructed with the helicity variables $\Delta \T_s$,
does not agree with the small $q$ expansion of $\Delta T^{\mu\nu}_{\rm hydro}$.
There are two sources of discrepancy.
First, the helicity two variable
$\Delta \tensor \T_2$ is non-vanishing at $\O(q^0)$ while
$\Delta \tensor \T_2^{\rm hydro}$ vanishes identically in the large $\Nc$ limit.
Second, even with identical helicity variables, the reconstruction
of $\Delta T^{\mu \nu}$ from its helicity variables $\Delta \T_s$
differs from the reconstruction of $\Delta T^{\mu \nu}_{\rm hydro}$
from $\Delta \T_s^{\rm hydro}$,
because $\Delta T^{\mu \nu}$ and $\Delta T^{\mu \nu}_{\rm hydro}$
satisfy differing energy-momentum conservation relations.
In particular,  $\Delta T^{\mu \nu}$ satisfies the exact microscopic 
conservation relation (\ref{conservation}) 
with the force density $F^{\nu}$ on the right hand side,
while $\Delta T^{\mu \nu}_{\rm hydro}$  satisfies the
effective energy-momentum conservation relation (\ref{hydrocons})
involving the effective source $J^{\nu}$.
However, examining
the small $q$ limits of $\Delta T^{\mu \nu}$ and $\Delta T^{\mu \nu}_{\rm hydro}$, 
we find
\begin{equation}
\label{hydrocompare}
\Delta T^{\mu \nu} = \Delta T^{\mu \mu}_{\rm hydro} + \mathcal A^{\mu \nu} \,,
\end{equation}
where
\begin{subequations}
\begin{align}
\mathcal A^{0 \mu} &= 0 \,,
\\
\mathcal A^{ij}& = \K \left(v^2 \gamma \, \delta_{ij} - \uh \, v_i v_j \right).
\end{align}
\end{subequations}
When Fourier transformed back to position space, $\mathcal A^{\mu \nu}$ has 
point support at the location of the quark and satisfies
\begin{equation}
\label{A}
\partial_{\mu} \mathcal A^{\mu \nu} = F^{\mu} - J^{\mu} \,.
\end{equation}
In other words,
$\mathcal A^{\mu \nu}$ is a local contribution to the stress-energy tensor
which
is precisely tailored to compensate for the difference between the
effective source $J^{\mu}$ and the microscopic force density.
This is a necessary result.  The gradient expansions of
$\Delta T^{\mu \nu}$ and $\Delta T^{\mu \nu}_{\rm hydro}$ must agree
far from the quark.
However, in the long wavelength limit, any discrepancy between 
$\Delta T^{\mu \nu}$ and $\Delta T^{\mu \nu}_{\rm hydro}$ in the near zone
may be represented as a term with point support
at the location of the quark.
Because $\Delta T^{\mu \nu}$ and $\Delta T^{\mu \nu}_{\rm hydro}$ satisfy 
different conservation equations, $\mathcal A^{\mu \nu}$ must obey 
Eq.~(\ref{A}) to all orders in the gradient expansion of $J^{\mu}$.
It is, of course, reassuring
to see $\mathcal A^{\mu \nu}$ naturally emerge from the gravitational calculation.  
Note that information from all three gauge invariants $Z_s$ is used to 
calculate $\mathcal A^{\mu \nu}$, thus showing the non-trivial interplay between the 
decoupled gravitational equations of motion (\ref{eqm2}), (\ref{eqm1}),
and (\ref{eqm0}).

It is illuminating to compare 
hydrodynamics to the AdS/CFT results in position space.
A simple quantity to compare, which involves both the sound and diffusion modes, 
is the instantaneous  angular distribution of the energy flux at a distance $R$
from the quark.
Choosing to evaluate this at $t= 0$, when the quark is at the origin,
the angular distribution of the energy flux is
\begin{equation}
\frac{d P_R}{d \Omega} \equiv R^2 \, \hat x \cdot  \Delta \bm S(0,R \, \hat x) \,.
\end{equation}
Because of the rotational symmetry about the velocity axis, 
its sufficient to consider the energy flux
per unit polar angle $\theta$, 
\begin{equation}
\frac{dP_R}{d \theta} = 2 \pi R^2 \, \sin \theta  
\left ( \cos \theta \, \Delta S_{\|} + \sin \theta \, \Delta S_{\perp} \right ),
\end{equation}
with $\theta = 0$ corresponding to the $\hat v$ axis
and $\Delta S_{\|}$ and $\Delta S_{\perp}$ the components of the energy flux
parallel and perpendicular to the velocity vector $\v$, respectively.
We plot 
$dP_R/d \theta$ for $v = 1/4$ in Fig.~\ref{p25}, $v = \cs$ in Fig.~\ref{pvs},
and $v = 3/4$ in Fig.~\ref{p75}.
In all three figures we plot
the angular distribution of the flux
at distances $R = 5/(\pi T)$, $10/(\pi T)$, and $20/(\pi T)$.
The hydrodynamics curves were made 
by numerically integrating Eqs.~(\ref{hydrofluxL}) and (\ref{hydrofluxT})
with the leading order effective source $J^\mu = F^\mu$.
Hydrodynamics becomes increasingly accurate at longer distances, but
even at $R = 5/(\pi T)$
one sees rather good agreement between linearized hydrodynamics
and the full AdS/CFT results.

We emphasize that ${d P_R}/{d \Omega}$ is not the same as the power
radiated in the rest frame of the quark.
In the quark's rest frame, the total power radiated
through spheres of radius $R$ is independent of $R$.
However, in the rest frame of the plasma,
the total flux,
$\int d \Omega \> ({d P_R}/{d \Omega})$,
through a sphere centered on the instantaneous position of the quark
grows with increasing $R$ as
successively larger spheres capture
energy radiated ever farther back in time.

%%%%%%%%%%

\section{Discussion}
\label{Discussion}

\begin{figure}[t]
\includegraphics[scale=0.7]{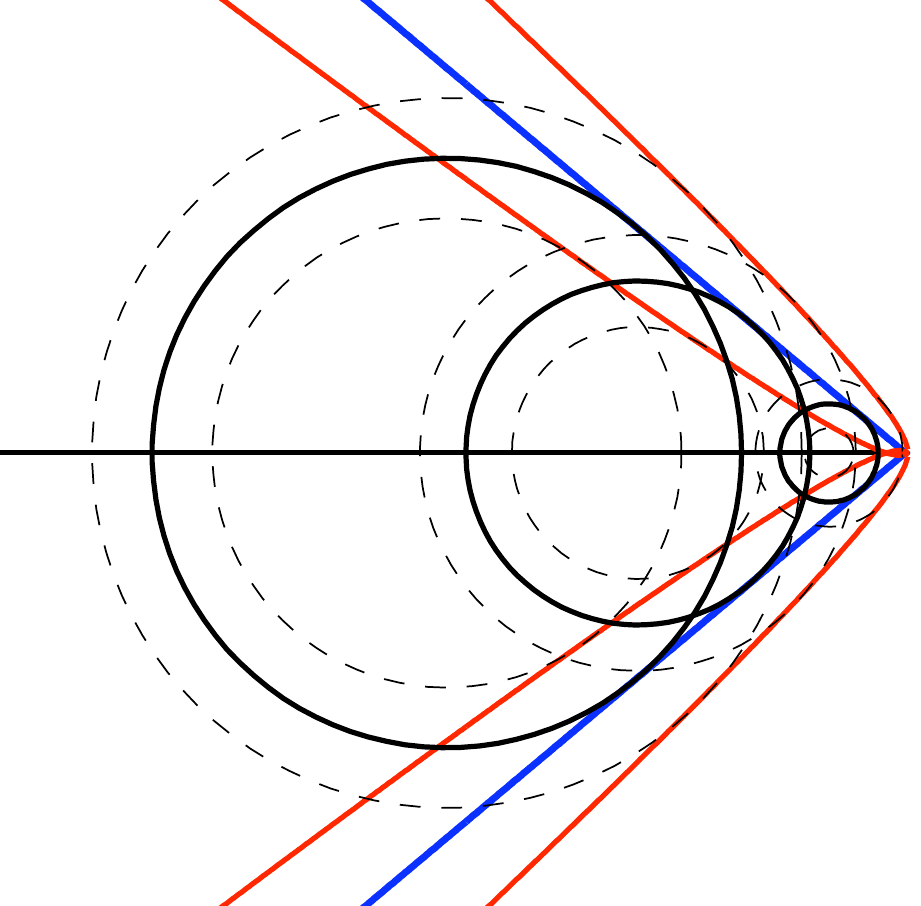}
\caption{\label{mach} A schematic representation of the Green's function
solution to the sound equations (\ref{energysound}) and (\ref {soundflux})
for supersonic motion.
As the quark moves, it creates sound disturbances which propagate
out in spherical shells.  The resulting sound waves add coherently along
the Mach cone, shown in blue.
Because the sound waves are damped, each spherical wave will broaden as it propagates.
The broadening is schematically indicated by the dashed lines.
This implies that the Mach cone broadens with increasing
distance from the quark, as indicated by the red lines.
}
\end{figure}

Much of the qualitative and quantitative structure in the plots of
$\Delta\E (t,\x)$ and
$\Delta S(t,\x)$ can be understood from hydrodynamic considerations alone.
This statement is reinforced by Eq.~(\ref{hydrocompare}), which shows that the long  
wavelength limit of the stress tensor as computed with gauge/string duality
coincides with the linearized hydrodynamics result,
provided the latter is computed with the correct effective source.
As discussed in Section~\ref{Hydro}, long wavelength perturbations
in the energy density satisfy the diffusive wave  
equation (\ref{energysound}) describing
sound waves with the dispersion relation
\begin{equation}
\omega \approx \pm \cs \, q -\coeff i2 \, \gamma \, q^2 \,,
\end{equation}
[with $\cs = 1/\sqrt{3}$ and $\gamma= 1/(3 \pi T)$].
A textbook constructive interference argument, illustrated in Fig.~\ref 
{mach}, shows that
a projectile moving supersonically will produce a Mach cone with an  
opening half-angle given by
$\sin \theta_{\rm M} = \cs/v$
(where $\tan \theta_{\rm M} \equiv -x_\perp/x_\| $).
For $v = 3/4$ this is $\theta_{\rm M}  = 50.3^{\circ}$.
As is evident in Fig.~\ref{v75}, at $v = 3/4$ the energy wake is  
concentrated, as expected, along a $50^\circ$ cone with the the bulk
of the associated energy flux flowing perpendicular
to the wave front.  Similarly, in Fig.~\ref{vvs}, at $v =\cs$ the  
energy wake is concentrated
along  the plane $x_\|=0$ with the bulk of the associated  
energy flux again flowing perpendicular to the wave front.

The damping of sound waves implies that their waveforms must  
broaden as they propagate.
This in turn implies that the Mach cone broadens with increasing distance  
from the quark.
This behavior is clearly seen in Fig.~\ref{v75}.
As one can reason from Fig.~\ref{mach},
the width $\Gamma_{\rm Mach}$ of the Mach cone,
defined as the length scale
over which the energy density attenuates (exponentially) in the forward direction,
increases with distance $d$ from the quark like
\begin{equation}
\label{Gamma}
    \Gamma_{\rm Mach}
    \sim
    %\sqrt{\frac{\gamma \, d}{\sqrt{1-\cs^2/v^2} } } \,.
    \frac {(\gamma \, d)^{1/2}}{(v^2-\cs^2)^{1/4} } \,.
\end{equation}
This attenuation length diverges as $v \rightarrow \cs$;
in this limit there is only power law falloff of the
Mach cone in front of the quark.
This reflects the fact that at $v = \cs$, sound waves emitted by
the quark arbitrarily far  
in the past can add coherently to the Mach cone
at finite distances from the quark.
If a given sound wave has to  
travel a distance $\ell$ to reach the
Mach cone, the width of the wave will have broadened to
$\sim \sqrt {\gamma \ell/\cs}$.
%% Including $\cs$ makes dimensions consistent w/o setting $c=1$.
At precisely $v = \cs$, sound waves with $\ell \rightarrow \infty$
contribute to the Mach cone at finite distances
from the quark, leading to power-law falloff of the Mach cone.

\begin{figure}[t]
\includegraphics[scale=0.25]{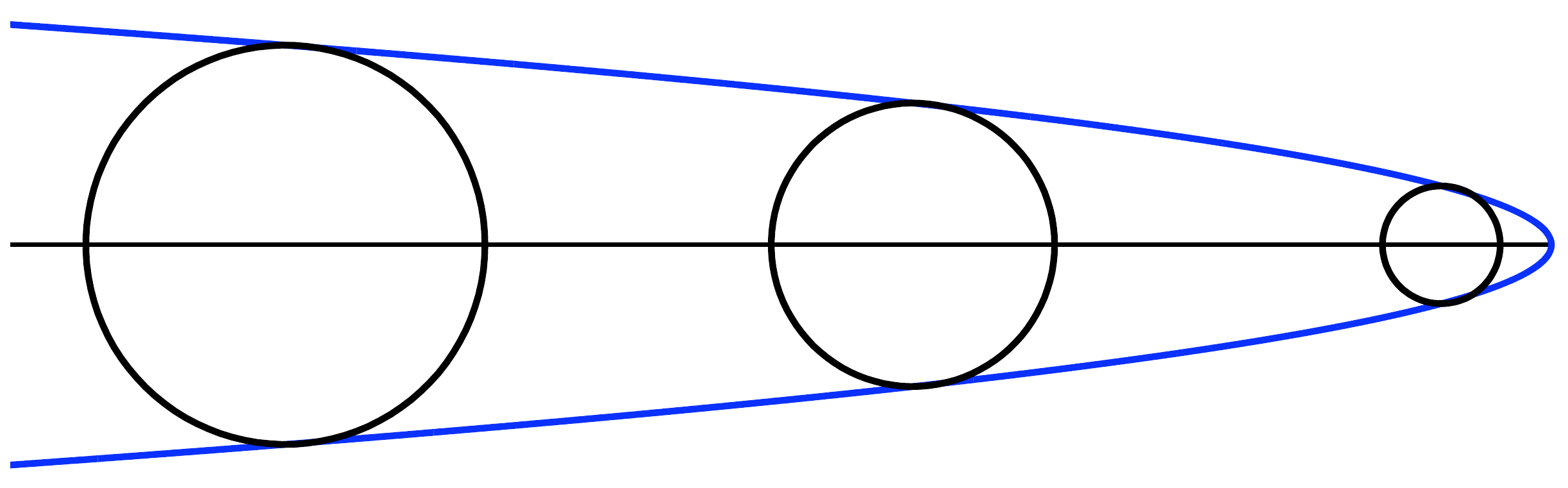}
\caption{\label{diffusionwake}
A schematic representation of the Greens function
solution to the diffusion equation (\ref{diffusionflux}).  As the  
quark moves, momentum deposited
in the diffusion channel at time $t_0$ diffuses over the length scale  
$\sqrt{ D (t - t_0)}$.
This results
in the broadening of the diffusion wake with increasing distance from the quark.
}
\end{figure}

{}From Eq.~(\ref{diffusionflux}), we see that part of the long  
wavelength perturbation in
the energy flux satisfies a diffusion equation with
diffusion constant
$D = 1/(4 \pi T)$.
Fig.~\ref{diffusionwake} shows a pictorial  
representation of the the
Greens function solution to the diffusion equation (\ref {diffusionflux}).
As the quark moves, momentum transferred to the diffusion mode at time $t$
mode will gradually diffuse outward,
resulting in a stream of momentum trailing the quark.   
The width of the corresponding
{\it diffusion wake} grows with distance $d$ from the quark like
\begin{equation}
\Gamma_{\rm diffusion} \sim \sqrt{D d/v}.
\end{equation}
This behavior is evident in the energy flux plots in
Figs.~\ref{v25}--\ref {vvs}.
The directional dependence of the diffusion wake
is dictated (in the long wavelength limit) by the directional dependence
of the effective source $\bm J_{\rm diffusion}$,
(defined in Eq.~\ref{Jdiffusion}).
To leading order in gradients, $\bm J_{\rm diffusion} \propto \v$,
so the diffusion wake should flow in the same direction as the quark's motion.
Again, this behavior is clearly seen in the energy flux plots of
Figs.~\ref{v25}--\ref{vvs}.

It is noteworthy that the diffusion wake is not seen at all in the
plots of the energy density.
This can be understood from hydrodynamic and large $\Nc$ scaling considerations.
First of all, note that the \textit{kinetic energy} associated
with bulk motion of the fluid is negligible in the large $\Nc$ limit.
This follows from the fact that the fluid velocity
which results from the motion of the quark through the plasma
is $\O(1/\Nc^{2})$.
(The fluid velocity $\bm u$ equals the $\O(\Nc^0)$ momentum density
divided by the $\O(\Nc^2)$ enthalpy, $\epsilon {+} p$.)
And hence the associated bulk kinetic energy density,
given by $(\epsilon{+}p) \, \bm u^2$, is also $\O(1/\Nc^2)$.%
\footnote
  {%
  Consequently, in the large $\Nc$ limit, all of the energy lost
  by a quark traversing the plasma is turned into heat.
  %Only an $\O(1/\Nc^2)$ fraction of the energy
  %lost by a quark is converted into bulk kinetic energy of the fluid.
  }
This means that an $\O(\Nc^0)$ momentum density need not produce any
imprint on the $\O(\Nc^0)$ energy density perturbation, exactly as
seen in Figs.~\ref{v25}--\ref{vvs}.
To see this more formally note, 
from Eq.~(\ref{Sdiffusion}), that the divergence of the  
diffusive energy flux
has point support at the quark.  That is,
$\del \cdot \bm S_{\rm diffusion} = 0$ everywhere
away from the quark.
This implies that the
momentum transported via the diffusion flux does not involve any corresponding
perturbation in the energy density.
This is analogous, in electromagnetism, to the possibility of having a
transverse electric current density without any corresponding charge density.

To put our discussion on a somewhat more quantitative footing,
we now turn to the plots of $d P_R/ d \theta$ shown in Figs.~\ref{p25}--\ref{p75}.
For subsonic and transonic motion (Figs.~\ref{p25} and \ref {pvs}),
$d P_R/d \theta$ is reproduced quite well by hydrodynamics
all the way down to distances $R = 5/(\pi T)$ from the quark.
For supersonic motion (Fig.~\ref{p75}) the agreement between
hydrodynamics and the complete result is not quite as good, quantitatively,
at $R = 5/ (\pi T)$,
but the level of agreement is still rather remarkable.
The region of largest
discrepancy occurs on or near the Mach cone.
It is natural that the hydrodynamic
approximation to the stress tensor deviates more for supersonic  
motion than for subsonic motion.
This is because gradients in the energy density and energy flux
are largest on the Mach cone.
This behavior is clearly seen in the near zone behavior of the energy  
density in Fig.~\ref{v75}.  However, in constructing the
hydrodynamic approximation to the stress tensor in
Eq.~(\ref{Tmunu}),
we have only kept the leading viscous terms in the gradient expansion and  
neglected $\Theta$ and all other higher order terms in the gradient
expansion of the stress.

We emphasize the importance of including the non-zero viscosity when
evaluating the hydrodynamic curves shown in Figs.~\ref{p25}--\ref{p75}.  Had we
neglected viscosity, the stress tensor would be discontinuous on the Mach cone
and, for $v=\cs$, would actually diverge on the Mach cone.  Furthermore, 
in the limit of zero viscosity the width of the diffusion wake vanishes,
leaving a diffusion wake proportional to $\delta^2(\bm x_{\perp}) \Theta(-x_\|+v t)$.
Therefore, neglecting viscosity yields a very poor approximation to the
stress tensor in the vicinity of both the Mach cone and the diffusion wake.

The agreement between
hydrodynamics and AdS/CFT at distances $d \gtrsim 1/T$ from the quark  
should be contrasted with the corresponding situation at weak coupling.
If the 't Hooft coupling $\lambda \ll 1$, then the
SYM plasma has a quasi-particle description.
The mean free path of quasi-particles (gluons, fermions, or scalars)
scales like \cite{Arnold:2002zm}
\begin{equation}
\ell_{\rm mfp} \sim \frac{1}{T \lambda^2 \ln \lambda^{-1}} \,.
\end{equation}
Therefore, the mean free path, at weak coupling, is parametrically longer
than $1/T$.
Hydrodynamics is only valid on spatial scales
large compared to the mean free path.
So in contrast to the strong coupling results discussed above,
in a weakly coupled plasma hydrodynamics is never valid on
distance scales comparable to $1/T$.

It is interesting to ask how the energy and momentum
lost by the quark is distributed into the sound and diffusion modes.
This question has been addressed in the steady
state limit in Refs \cite{Gubser:2007ni,Gubser:2007ga}.
However, for reasons which are outlined below, we question
the physical value of addressing this topic in the
limit where the quark has been moving forever.

Let $V$ denote a large sphere of radius $R$ whose location is fixed
and centered on the position of the quark at time $t = 0$.
The rate of change of the energy inside volume $V$ is
\begin{equation}
\label{dEdt}
\frac{d E_V}{dt} = \int_V d^3 x \; \partial_0 T^{00}(t,\x) \,.
\end{equation}
After using the energy-momentum conservation
equation (\ref{conservation}),
and separating the energy flux into
sound and diffusive pieces, Eq.~(\ref{dEdt}) becomes
\begin{equation}
\label{dEdt2}
    \frac{d E_V}{dt} =
    f^0 -\int_{\partial V} d \Sigma \cdot \bm S_{\rm sound}
    -\int_{\partial V} d \Sigma \cdot \bm S_{\rm diffusion}\,.
\end{equation}
The first term on the right is simply the rate at which the quark deposits  
energy into the plasma
while the second and third terms are the rates at which energy is  
added to and removed from $V$ via the sound and diffusion fluxes.

In the steady state limit, it is easy to calculate the rate
at which energy is added to $V$ by the diffusion flux.
If $R \gg 1/T$, then one may compute the surface integrals
in Eq.~(\ref{dEdt2}) using hydrodynamics.
The long wavelength limit of the diffusion flux satisfies%
\footnote
   {%
   This equation was first obtained in
   Refs.~\cite{Gubser:2007ni,Gubser:2007ga} via a gravitational
   calculation.  However, it follows directly from  
   linearized hydrodynamics using the leading order effective source
   $J^\mu = F^\mu$.  Specifically, Eq.~(\ref{diffusecons})
   follows from the definition of the diffusion flux in
   Eq.~(\ref {Sdiffusion})
   plus the observation that
   $
   \del \cdot \bm S_{\rm nonlocal} = - F^0/v^2
   $,
   to leading order in gradients.
   }
\begin{equation}
\label{diffusecons}
\del \cdot \bm S_{\rm diffusion} = - \frac{1}{v^2} \, F^0 \,.
\end{equation}
{}From this expression we see that the diffusion flux {\em adds}
energy to the volume $V$ at the rate
\begin{equation}
\label{diffflux}
-\int_{\partial V} d \Sigma \cdot \bm S_{\rm diffusion} = \frac 
{f^0}{v^2} \,.
\end{equation}
The source of the energy supplied to $V$ via the diffusion flux
comes from the quark itself --- in the distant past when the quark was
outside of the volume $V$, it created the diffusion flux
which subsequently flows into $V$.
(The fact that the diffusion flux
adds energy to $V$ simply reflects the fact that the  
diffusion flux flows in the same direction as the quark's velocity.)
Note that the rate at which energy is added to $V$ via the diffusion flux
is independent of the size of $V$, and so is non-vanishing  
in the $R \rightarrow \infty$ limit.  Furthermore the rate (\ref{diffflux}) has a  
finite, non-zero limit as $v \rightarrow 0$
[recall that $f^0 \equiv \v \cdot \f = \O(v^2)$].
So the diffusion flux adds energy to $V$  
even when
the quark's velocity is taken to be arbitrarily small!  The origin of  
this peculiar
behavior comes from the assumption that the quark has been moving  
forever.
In this limit, diffusive energy flux deposited in the arbitrarily distant past  
can influence the rate of change
of the energy in the volume $V$.
Completely analogous conclusions also hold for the sound flux.%
\footnote
    {%
    To demonstrate this, one may solve the wave equation (\ref{soundflux})
    in the analytically tractable limit of small velocity.
    When $v \ll \cs$ no Mach cone is generated and the effects of viscosity on the
    long distance behavior of the sound energy flux are negligible.
    Taking the duration of time in which the
    quark has been moving $\Delta t \rightarrow \infty$ before taking $v \rightarrow 0$,
    one finds that the rate at which the sound flux removes energy from the volume $V$
    has a finite, non-zero limiting value.
    }

The assumption that the quark has been moving at constant velocity  
forever presents a
great technical simplification in both the gravitational and  
hydrodynamic computation of the quark wake.
However, the utility of this simplifying
assumption is limited to
questions which are insensitive to the details of the quark's  
trajectory in the distant past.
Asking how the  energy and momentum
lost by the quark are distributed into the sound and diffusion modes,
when the quark has been moving forever,
is not such a question.
A better and still analytically tractable question is to consider
is how much energy and momentum are deposited into each mode  when  
the quark has been
moving at constant velocity $v$ for a long but finite period of time
$ \Delta t = t_f - t_i$.
The total energy transfered to the plasma is then
\begin{equation}
\Delta E(t) = \int d^3 x \; T^{00}(t,\x) \,,
\end{equation}
where the integration is taken over all of space.
For late times after
the quark's motion has ceased,
all of the energy deposited by the quark will be transported away
via sound waves.  That is, because the quark has been
moving for a finite period of time, no fluxes at spatial infinity
contribute to $\Delta E(t)$.

The total momentum deposited by the quark is
\begin{equation}
\Delta \bm p = \Delta t \, \bm f \,.
\end{equation}
To see what fraction of the quark's momentum is deposited in each  
mode, we will examine
the diffusion mode and then infer the momentum deposited in the sound
mode via momentum conservation.
The total momentum transfered to the  diffusion mode is
\begin{equation}
\Delta \bm p_{\rm diffusion}(t) = \int d^3 x \,
\bm S_{\rm  diffusion}(t,\x) \,.
\end{equation}
If the interval during which the quark has been moving is sufficiently long,
then the total momentum transfered to the diffusion mode may be computed
with hydrodynamics.  Moreover, in the limit $\Delta t \gg 1/T$, the  
vast majority
of the volume of the diffusion wake will be well approximated by its  
steady state limit.  In the steady state limit it is easy to solve the  
diffusion equation (\ref{diffusionflux})
with the corresponding leading order effective source given in
Eq.~(\ref{Jdiff}).  Assuming that
the quark is at $\bm x = 0$ at time $t_f$, the result
reads
\begin{align}
\label{sdiffapprox}
\bm S_{\rm diffusion}(t{=}t_f,\x) = \frac{\bm f}{4 \pi D r} \,
\exp \left [ -\frac{v}{2 D} \left(x_{\|}+r \right ) \right ] + \cdots \,,
\end{align}
where $r = |\bm x|$
and the ellipsis denotes corrections suppressed by
inverse powers of $r$ (coming from higher order gradient corrections).
Eq.~(\ref{sdiffapprox}) provides a good approximation to the  
diffusion wake everywhere
except in a region of size $\sim \sqrt{D \Delta t}$ centered about  
the quark's starting
point.  In the $\Delta t \rightarrow \infty$ limit this region is  
negligible compared to
the \textit{length} of the diffusion wake, which is $v \, \Delta t$.  To  
obtain the total
momentum transfered to the diffusion wake, we integrate Eq.~(\ref 
{sdiffapprox})
over a length $v \, \Delta t$ in the $x_{\|}$ direction.  Specifically,  
we have
\begin{align}
\Delta \bm p_{\rm diffusion}(t_f) = \int_{-v \Delta t}^0 dx_{\|}  
\int d^2 x_\perp  \> \bm S_{\rm diffusion}(t_f,\x) \,.
\end{align}
The integral is elementary to carry out in the $\Delta t \rightarrow  
\infty$ limit.
The result reads
\begin{equation}
\Delta \bm p_{\rm diffusion}(t_f) = \Delta t \, \bm f.
\end{equation}
This shows that in the $\Delta t \rightarrow \infty$ limit all of the  
momentum lost
by the quark is transfered to the diffusion mode.  Correspondingly,  
the sound mode carries
no net momentum.  The qualitative origin of this statement can be  
understood by inspecting the
momentum density near the Mach cone in Figs.~\ref{v75}--\ref{vvs}.
In particular, the momentum density ({\it i.e.}, energy flux) flows  
predominately
parallel and anti-parallel to the Mach cone normal.
Evidently, when integrated over all space, the opposing flows cancel.

The direct applicability to RHIC physics of results obtained from
studying the wake produced by a quark moving for an unboundedly
long time through a plasma of infinite extent is, of course,
questionable at best.
The plasma produced at RHIC is very dynamic.
The expansion of the plasma and the resulting variable speed
of sound can have a dramatic effect on the structure of
the quark wake \cite{CasalderreySolana:2006sq}.
However, the agreement between hydrodynamics and gauge/string duality
in describing the structure of the quark wake
in strongly coupled SYM does have relevance for RHIC.
In particular,
at least for the  particular problem addressed in this paper,
it shows that one can use hydrodynamics to address physics
at length scales all the way down to distances less than two
times $1/T$.

Because the quark-gluon plasma produced in heavy ion  collisions is
believed to be strongly coupled \cite{Shuryak,Shuryak:2004cy},
the agreement between hydrodynamics and gauge/string duality
in a strongly coupled SYM plasma
bolsters the notion that one should be able to model accurately the wake
produced by a high energy quark (or gluon) moving through a QCD plasma
merely using hydrodynamics.
Of course,
the application of hydrodynamics requires 
among other things, knowledge of the viscosity of the 
plasma and knowledge of the drag force on the quark,
neither of which are under very good control for real quark-gluon plasma
at accessible temperatures.
(And, as noted earlier, neglecting viscosity will yield a poor approximation
in the vicinity of both the Mach cone and the diffusion wake.)
However, at least when $\Nc$ is large
(and the hydrodynamic equations are linear),
the \textit{structure} of quark's wake is rather 
insensitive to magnitude of the drag force.  For example,
in the steady state limit the magnitude of the drag force enters only
as an overall normalization of the quark wake.
As a next step toward a more realistic treatment,
it would be interesting, and should be feasible,
to study the wake produced by a quark traversing
an expanding and cooling $\Nfour$ SYM plasma
\cite{Janik:2006gp}.

\section{Conclusions} 
\label{Conclusions}

Using gauge/string duality,
we have evaluated the perturbation in the stress-energy tensor due
to the presence of a fundamental quark moving through a strongly coupled
large $\Nc$ maximally supersymmetric Yang-Mills plasma.
Our plots of the energy density and energy flux
in Figs.~\ref{v25}--\ref{vvs} clearly display the formation of Mach cone
for velocities $v \geq \cs$, together with the presence of a diffusive
wake in the energy flux at all velocities.

We have argued that the effective source for hydrodynamics is, to
leading order in gradients, simply determined by the microscopic
drag force acting on the quark.
By comparing the small momentum asymptotics of the stress-energy tensor
with the predictions of hydrodynamics, we were able to determine the first
subleading correction to the effective hydrodynamic source.

Using the leading order effective source for hydrodynamics, we compared
the hydrodynamic prediction to the complete result obtained via
gauge/string duality for the instantaneous angular distribution
of power radiated through spherical shells a distance $R$ from the quark.
The comparison showed remarkably good agreement between hydrodynamics and
the exact result at distances all the way down to $R\sim 1.6/T$ from the quark.
This reinforces the notion that wakes produced by high energy particles
traversing a real quark-gluon plasma can be well modeled
merely using hydrodynamics.

\begin{acknowledgments}
We thank A.~Andreev, C.P.~Herzog, A.~Karch, and D.T.~Son for many useful discussions.
This work was supported in part by the U.S. Department of Energy under
Grant No.~DE-FG02-96ER40956.

\end{acknowledgments}

\appendix

%%%%%%%%%%%%%%%%%%%%
\section{The Boundary Action}
\label{BoundaryAction}

The total gravitational action for the heavy quark effective theory is given by
\begin{equation}
S_{\rm G} =
    S_{\rm EH} + S_{\rm GH} + S_{\rm DBI} + S_{\rm NG} +S_{\rm CT} \,,
\end{equation}
where 
\begin{subequations}
\begin{align}
    S_{\rm EH} &\equiv \frac{1}{2 \kappa_5^2}
    \int d^5 x \> \sqrt{-G} \> (R {+} 2  \Lambda ) \,,
\\
    S_{\rm GH}
    &\equiv \frac{1}{2 \kappa_5^2}
    \int d^4 x \> \sqrt{-\gamma} \; 2 K \,,
\end{align}
\end{subequations}
and the other pieces were described in section~\ref{gravitational}.
We write the metric 
in the bulk as
\begin{equation}
G_{MN} = G_{MN}^{(0)} + h_{MN} \,,
\end{equation}
and choose to work in a gauge
where $h_{5 M } \equiv 0$.
We similarly write the metric on the boundary as
\begin{equation}
\gamma_{\mu \nu} = \gamma_{\mu \nu}^{(0)} + h_{\mu \nu} \,.
\end{equation}
To quadratic order in $h_{\mu \nu}$ we have
\begin{equation}
\sqrt{-G} \> (R +2  \Lambda) =
\sqrt{-G_{(0)} } \>
\Bigl(D_{P} W^{P} + \mathcal L_0 - \frac{2 d}{L^2} \Bigr) \,,
\end{equation}
where
\begin{align}
\label{L0}
\mathcal L_0  ={}&
\coeff{1}{4} D_{M}h D^{M} h +\coeff{1}{2}D_{P} h_{M N} D^{M} h^{N P} 
\nonumber\\ 
&{}-\coeff{1}{2} D_{P}h^{M P}D_{M} h -\coeff{1}{4} D_{P} h^{M N}D^{P} h_{M N} 
\nonumber\\ 
&{}+ \frac{d}{2 L^2} \left (\coeff{1}{2} h^2 - h^{M N}h_{M N} \right ) ,
\end{align}
and
\begin{align}
W^{P} =&
-\half h D^{P} h + \half D_{M} \left ( h h^{M P} \right) 
\nonumber \\ &{}
+ \coeff{1}{2} h^{P M} D_{M}h 
+ h^{M N} D^{P} h_{MN}
\nonumber \\ &{}
- D^{M} \left (h^{P N}h_{M N} \right )
  -D^{P} h + D^{M} h^{P}_{\ M} \,.
\end{align}

To vary the gravitational action, let
\begin{equation}
h_{\mu \nu} \rightarrow h_{\mu \nu} + \delta h_{\mu \nu} \,,
\end{equation}
where $h_{\mu \nu}$ satisfies the linearized equations of motion (\ref{lin1})
and $ \delta h_{\mu \nu}$ is infinitesimal.  The variation of the bulk
action density is
\begin{equation}
\label{deltaL0}
\sqrt{-G_{(0)}} \, \delta \mathcal L_0 =
\sqrt{-G_{(0)}} \left(
D_{P} \, \delta V^{P} - \kappa_5^2 \, \delta h_{M N} \, t^{M N}
\right) ,
\end{equation}
where
\begin{align}
\delta V^{P} ={}&
\coeff{1}{2} \delta h \, D^{P} h
- \coeff{1}{2} \delta h_{M N} \, D^{P} h^{M N}
+ \delta h_{M N} \, D^{M} h^{P N}
\nonumber\\ 
&{}
-\coeff{1}{2} \delta h \, D_{N} h^{P N}
-\coeff{1}{2} \delta h^{P N} \, D_{N} h  \,,
\end{align}
and $t^{MN}$ is the bulk stress-energy tensor, whose only source
(in the limit of large quark mass) is the string hanging down to the horizon.
The variation of the Nambu-Goto action is, by definition 
\begin{equation}
\delta S_{\rm NG} = \int d^5 x \sqrt{-G_{(0)}} \,
\coeff{1}{2} \delta h_{MN} t^{MN}
\,.
\end{equation}
Hence the variation in the Nambu-Goto action cancels
the last term in Eq.~(\ref{deltaL0}).
As discussed in section~\ref{gravitational}, in the large quark mass limit,
the DBI action may be replaced with the ordinary Maxwell action for the
electromagnetic field (residing on the boundary) which accelerates the quark.

The trace of the extrinsic curvature is given by 
\begin{equation}
K = \del_{M} n^{M}
\end{equation}
where $n^{M}= - \sqrt{G^{55}} \, \delta ^{5 M}$ 
is the outward pointing normal to the boundary and $\del_\mu$ denotes covariant
differentiation with respect to the full metric $G_{\mu \nu}$.  Using
\begin{equation}
\sqrt{\gamma} \, \del_M n^{M} = -\sqrt{G^{55}} \, \partial_u \sqrt{\gamma},
\end{equation}
we can write
\begin{equation}
S_{\rm GH} =-\frac{1}{\kappa_5^2}  \lim_{u \rightarrow 0}\sqrt{G^{55}} \>
\partial_u \int d^4 x \> \sqrt{- \gamma} \,.
\end{equation}

Assembling the pieces, and specializing to the AdS-Schwarzschild
bulk geometry, we have
\begin{equation}
\delta S_{\rm G} = \delta S_{\rm B} + \delta S_{\rm horizon} \,,
\end{equation}
where 
\begin{align}
\nonumber
\delta S_{\rm B} &= \frac{1}{2 \kappa_5^2} \lim_{u\rightarrow 0}
\int d^4x \bigg[
-\frac{L^3 f }{2 u^3} \, \delta h \, \partial_u h 
+\frac{L^3 f }{2 u^3} \, \delta h^{\mu}_{\ \nu} \, \partial_u h^{\nu}_{\ \mu} 
\\ \nonumber
&+
\bigg \{
\chi \, \delta h_{\mu \nu} \, h^{\mu \nu}
{-}\frac{ \chi }{2} \, \delta h \, h
{-}\frac{L f'}{4 f u} \, \delta h_{00} \, h
{-}\frac{L^3 f'}{2 u^3} \, \delta h_{0i} \, h^{0i}
\bigg \}
\\ 
\label{sb1}
&
-\frac{3 \left(\sqrt{f}-1\right) L}{\sqrt{f} \, u^2} \, \delta h_{00}
-\chi \, \delta h^{i}_{\ i}
\bigg ]
+\delta S_{\rm EM} \,,
\end{align}
with
\begin{equation}
\chi \equiv \frac{L^3 \left(-6 f+6 \sqrt{f}+u f'\right)}{2 u^4} \,,
\end{equation}
and $\delta S_{\rm horizon}$ is a surface term at the black-brane horizon
whose explicit form will not be needed.
{}From the linearized field equations (\ref{lin1}) it is easy to see that near the 
boundary $h_{\mu \nu} \sim u^2$.  Using 
$\delta h_{\mu \nu } \equiv \frac{L^2}{u^2} \, \delta H_{\mu \nu}$,
we see that all of the terms in the curly braces in Eq.~(\ref{sb1}) scale like
$u^4 \, \delta H_{\mu \nu}$ as $u \rightarrow 0$.  It follows that these terms
will not contribute to $\delta S_{\rm G}/ \delta H_{\mu \nu}(x,u)$ in the 
$u \rightarrow 0$ limit.
Neglecting terms which vanish in the $u \rightarrow 0$
limit we therefore have
\begin{align}
\nonumber
\delta S_{\rm B} ={}& \lim_{u\rightarrow 0}
\int d^4x \bigg[ \frac{L^3 }{4 \kappa_5^2 u^3}
\left ( \eta^{\alpha \mu} \eta^{\beta \nu} - \eta^{\alpha \beta} \eta^{\mu \nu} \right)
\\ 
&{}
\times \delta H_{\alpha \beta} \, \partial_u H_{\mu \nu} 
+\coeff{1}{2} \delta H_{\mu \nu} \, T^{\mu \nu}_{\rm eq} \bigg]
+\delta S_{\rm EM} \,,
\end{align}
where $\eta^{\mu \nu} = {\rm diag}(-1,1,1,1)$ 
is the Minkow\-ski space metric and
\begin{equation}
T^{\mu \nu}_{\rm eq} =
\epsilon \, {\ts {\rm diag}(1,\frac{1}{3},\frac{1}{3},\frac{1}{3} ) }
\end{equation}
is the stress-energy tensor of the equilibrium $\Nfour$ SYM plasma.
The explicit value of the equilibrium energy density is
\begin{equation}
\epsilon \equiv \frac{3 L^3}{2 \kappa_5^2 \uh^4} = \coeff{3}{8} \Nc^2 \pi^2 T^4 \,.
\end{equation}

We write the variation of the boundary action in Fourier space and
wish to express it in terms of the gauge invariants $Z_{s}$.
The variation in the boundary action is 
\begin{align}
\nonumber
\delta S_{\rm B} ={}& \lim_{u\rightarrow 0}
\int \frac{d^4 q}{(2 \pi)^4} \bigg[ \frac{L^3 }{4 \kappa_5^2 u^3}
\left (
    \eta^{\alpha \mu} \eta^{\beta \nu}
    - \eta^{\alpha \beta} \eta^{\mu \nu}
\right)
\\ \label{quadbdaction}
&{} \times \delta \H_{\alpha \beta} \, \partial_u \H_{\mu \nu} 
+\coeff{1}{2} \delta \H_{\mu \nu} \, \T^{\mu \nu}_{\rm eq} \bigg]
+\delta S_{\rm EM} \,.
\end{align}
Here products are understood to be 
evaluated at antipodal momentum as dictated by the reality condition 
$\H_{\mu \nu}^*(\omega,\q)=\H_{\mu \nu}(-\q,-\omega)$.

\subsection{Tensor Mode}
The terms in Eq.~(\ref{quadbdaction}) constructed out 
of radial derivatives of $Z_2^{ab} \equiv \H_{ab}-\half  \delta_{ab}\H$ are
\begin{equation}
S_{\rm B}^{2} = \lim_{u \rightarrow 0} 
\int \frac{d^4 q}{(2 \pi)^4} \,
\mathcal A_2   \, \delta Z^{ab}_{2} \, \partial_u Z^{ab}_{2}  \,,
\end{equation}
where 
\begin{equation}
\mathcal A_2 = \frac{L^3}{4 u^3 \kappa_5^2} \,.
\end{equation}

\subsection{Vector Mode}
The terms in Eq.~(\ref{quadbdaction}) constructed out 
of radial derivatives of $\H_{aq}$ and $\H_{0a}$ are
\begin{equation}
\delta S_{\rm B}^{1} = \lim_{u \rightarrow 0} \int \frac{d^4q}{(2 \pi)^4} \>
\frac{L^3}{2 \kappa_5^2 u^3} \,
\big [\delta \H_{0a} \, \partial_u \H_{0a} 
+ \delta \H_{qa} \, \partial_u \H_{qa} \big ] \,.
\end{equation}
We now use Eq.~(\ref{vec1}) to eliminate $\partial_u \H_{qa}$.  Doing so,
we find
\begin{equation}
\delta S_{\rm B}^{1} = \lim_{u \rightarrow 0} \int \!
\frac{d^4q}{(2 \pi)^4} \big [
\mathcal A_{1} \, (q \delta \H_{0a}
+ \omega \delta \H_{aq} ) \, Z_1^a 
+ \delta \H_{aq} \, \mathscr J^{aq} \big ] \,,
\end{equation}
where 
\begin{equation}
\mathcal A_1 = -\frac{L^3}{2 q \kappa_5^2 u^3} \,,
\end{equation}
and
\begin{equation}
\label{jaq}
\mathscr J^{aq} \equiv \frac{L^3}{i q u^3} \> \t_{a5} \,.
\end{equation}

\subsection{Scalar Mode}
The terms in Eq.~(\ref{quadbdaction}) constructed out 
of radial derivatives of $\H_{00}$, $\H_{aa}$, $\H_{qq}$ and $\H_{0q}$ are
\begin{align}
\delta S_{\rm B}^{0} =
\lim_{u \rightarrow 0} \int & \frac{d^4q}{(2 \pi)^4}
\frac{L^3}{8 \kappa_5^2 u^3} \,
\Bigl[ 2 \delta \H_{00} \, \partial_u \H_{00}
  -  4 \delta \H_{0q} \, \partial_u \H_{0q}
\nonumber\\ &{}
  +2 \delta \H_{qq} \, \partial_u \H_{qq}
  - \delta \H_{aa} \, \partial_u \H_{aa} 
\Bigr]\,.
\end{align}
Using the first order equations (\ref{sound3})--(\ref{sound5}),
this is seen to be
\begin{align}
\delta S_{\rm B}^{0} ={}& \lim_{u \rightarrow 0} \int \frac{d^4q}{(2 \pi)^4}
\Bigl [ 
\mathcal A_0 \, \delta Z_0 \, \partial_u Z_0
 +\coeff{1}{2} \delta \H_{00} \, \mathscr J^{00}
 \nonumber\\  &{}
 + \coeff{1}{2} \delta \H_{qq} \, \mathscr J^{qq}
 +\coeff{1}{2} \delta \H_{aa} \, \mathscr J^{aa}
 +\delta \H_{0q} \, \mathscr J^{0q}
\Bigr ] \,,
\end{align}
where
\begin{equation}
\mathcal A_0 = \frac{L^3}{6 \kappa_5^2 u^3 (q^2{-}\omega^2)^2} \,,
\end{equation}
and
\begin{subequations}
\label{jmunu}
\begin{align}
\label{j00}
\mathscr J^{00} &=
\frac{i L^3}{3 u^3 (q^2 {-} \omega^2)^2} 
\Big [
\omega \left (5 q^2  {-} 3 \omega ^2 \right )\t_{05}
\\ \nonumber &{}
- q\left (q^2 {-} 3 \omega^2 \right ) \t_{q5}
+ i u q^2 \left (q^2 {-} \omega^2 \right ) \t_{55},
\Big ],
\\ 
\mathscr J^{qq} &=
\frac{i  L^3}{3 u^3 (q^2 {-} \omega^2 )^2 }
\Big [
 -\omega \left (\omega^2 {-} 3 q^2 \right ) \t_{05}
\\ \nonumber &{}
 + q \left (5 \omega^2 {-} 3 q^2 \right) \t_{q5}
 +i u \omega^2 (q^2 {-} \omega^2) \, \t_{55} \Big ],
\\ 
\mathscr J^{aa} &=
\frac{i L^3}{3 u^3 (q^2 {-} \omega^2)}
\Big [ \omega \, \t_{05} + q \, \t_{q5} 
\\ \nonumber
&- i u (q^2 {-} \omega^2 ) \, \t_{55} \Big ],
\\ 
\label{j0q}
\mathscr J^{0q} &=
\frac{2 i L^3}{3 u^3 (q^2 {-} \omega^2)^2}
\Big [
  -q \left (\omega^2 {-} 3 q^2 \right ) \t_{05}
\\ \nonumber &{}
- \omega \left (q^2 {-} 3 \omega^2 \right ) \t_{q5}
+ i u q \left (q^2 {-} \omega^2 \right ) \t_{55} \Big].
\end{align}
\end{subequations}

%%%%%%%%%%%%%%

\section{Small Momentum Asymptotics} 
\label{smallq}
For very large or very small momentum (compared to $T$),
one may find explicit asymptotic expressions for the homogeneous solutions
$g_s^<$ and $g_s^>$,
and derive the resulting asymptotic
behavior of the SYM stress-energy tensor.
In what follows we only consider the small
momentum limit and, in particular, calculate $\T_s$ to $\O(1)$.

We write the homogeneous solutions $g^>_s$ and 
$g^<_s$ as a power series in $q$,
\begin{subequations}
\begin{align}
g_s^< &= \phi^{(0)}_s + q \, \phi^{(1)}_s + q^2 \, \phi^{(2)}_s + \O(q^3) \,,
\\
g_s^> &= \psi^{(0)}_s + q \, \psi^{(1)}_s + q^2 \, \psi^{(2)}_s + \O(q^3) \,.
\end{align}
\end{subequations}
The differential equations
(\ref{eqm2}), (\ref{eqm1}) and (\ref{eqm0}),
without sources,
are then solved order by order in $q$ subject to the boundary 
conditions discussed in Section \ref{Numerics}.  In what follows we 
define $r \equiv \omega/q$.

\subsection{Tensor Mode}
For the tensor mode it is sufficient to expand the homogeneous 
solutions to Eq.~(\ref{eqm2}) to $\O(q^0)$.  Doing so one obtains
\begin{equation}
\phi_2^{(0)} =
-\uh^4 \ln f ,
\end{equation}
and
\begin{equation}
\psi_2^{(0)} =1 \,.
\end{equation}
The corresponding small momentum limit of the Wronskian is
\begin{equation}
W_2 = -\frac{4 u^3}{f} + \O(q) \,.
\end{equation}

Substituting the above expansions into the definition of $\Delta \tensor Z_2^{(4)}$ 
in Eq.~(\ref{grnsfcnrep2}) and performing the radial integral, one finds
\begin{align}
\nonumber
\Delta \tensor Z_2^{(4)} ={}&
-\frac{\kappa_5^2 \sqrt{\lambda}}{8 \pi L^3 \uh \sqrt{1{-}v^2}} \,
\frac{(v q_{\perp})^2}{q^2} \, (2 \pi) \delta(\omega- \v \cdot \q)
\\ &{}
\times
(\eps_1 \otimes \eps_1 - \eps_2 \otimes \eps_2) + \O(q) \,.
\end{align}
Substituting $\Delta \tensor Z_2^{(4)}$ into Eq.~(\ref{t2}) yields
\begin{align}
\nonumber
\Delta \tensor \T_2 ={}& -\frac{ \sqrt{\lambda}}{4 \pi  \uh \sqrt{1{-}v^2}} \,
\frac{(v q_{\perp})^2}{q^2} \, (2 \pi) \delta(\omega- \v \cdot \q)
\\ &{}
\times
(\eps_1 \otimes \eps_1 - \eps_2 \otimes \eps_2) + \O(q) \,.
\end{align}

\subsection{Vector Mode}

For the vector mode it is necessary to expand the homogeneous 
solutions to Eq.~(\ref{eqm1}) to $\O(q^2)$.
Define for convenience  
\begin{equation}
f_{\pm} \equiv 1\pm \frac{u^2}{\uh^2} \,.
\end{equation}
The solutions $\phi_1^{(i)}$ and $\psi_1^{(i)}$ are given by 
\begin{subequations}
\begin{align}
\phi^{(0)}_1 &= u^3 \,,
\\
\phi^{(1)}_1 &= 0 \,,
\\
\phi^{(2)}_1 &=
\nonumber
\frac{u^3 \uh^2}{32}  
\Bigl ( 2 \tanh ^{-1}\frac{u^2}{\uh^2} +\ln \frac {f}{f_{-}^2} \Bigr )
\\ \nonumber &
- \frac{r^2 u^3 \uh^2}{32}
    \bigg (
    \frac{\pi^2}{3}
    + 4 \frac{\uh^2}{u^2} \, \ln f
     +8 \tanh^{-1} \frac{u^2}{\uh^2}
\\ \nonumber &\qquad{}
    -2 \ln^2 2
    - \ln \frac{f_{-}}{16} \ln f_{-}
    + \ln f_{+} \ln \frac{f_{+}}{f_{-}^2}
\\ &\qquad{}
    -2 \ln f \ln \frac{f_{+}}{f_{-}} 
    -4 \, \text{Li}_2 \frac{f_{-}}{2}
\bigg ) \,,
\end{align}
\end{subequations}
and
\begin{subequations}
\begin{align}
\psi^{(0)}_1 &= \frac{u^3}{\uh^3} \,,
\\
\psi^{(1)}_1 &= \frac{i r u^3}{2 \uh^2}
-\frac{i r u}{4} \, \Bigl(2+\frac{u^2}{\uh^2} \ln \frac{f_{-}}{f_{+}} \Bigr) \,,
\\ \nonumber
\psi^{(2)}_1 &= \frac{u \uh}{32 } 
\Bigl ( 4 f_{-} +\frac{2 u^2}{\uh^2} \tanh^{-1} \frac{u^2}{\uh^2} + \frac{u^2}{\uh^2} \ln \frac{f}{f_{+}^2}
\Bigr ) 
\\ \nonumber &
+ \frac{r^2 u^3}{32 \uh}
\bigg (
-4 \frac{\uh^2}{u^2} \, \ln \frac{f}{4}
-8 \tanh^{-1} \frac{u^2}{\uh^2}
\\ \nonumber &\qquad{}
+ \ln 2 \ln 4
-\ln \frac{16 f^2}{f_{-}} \ln f_{-}
+ \ln 16 \ln \frac{f_{-}}{f_{+}}
\\ &\qquad{}
+2 \ln (f f_{-} ) \ln f_{+}
- \ln^2 f_{+}
+ 4 \, \text{Li}_2 \frac{f_{-}}{2}
\bigg ) \,.
\end{align}
\end{subequations}
The corresponding Wronskian is 
\begin{equation}
W_1 = \frac{u^3}{f} \bigg [
i r q -\frac{\uh}{4} \left (1+r^2 \ln 4 \right ) q^2 + \O(q^3) \bigg].
\end{equation}
Substituting the above solutions into Eq~(\ref{grnsfcnrep1}), one finds
\begin{align}
\nonumber
\Delta \vec Z_1^{(3)} ={}&
\frac{\kappa_5^2 \sqrt{\lambda}}{\pi L^3 \uh \sqrt{1{-}v^2} } \,
\frac{v q_{\perp}}{q}
\bigg [\frac{1}{i r q \uh} - \frac{1{-}4 r^2}{4 r^2} 
+ \O(q) \bigg ] 
\\ &{} \times
(2 \pi) \delta(\omega-\v \cdot \q) \>\eps_1 \,.
\end{align}
{}From Eq.~(\ref{t1}) we therefore have
\begin{align}
\nonumber
\Delta \vec \T_1 ={}&  \frac{\sqrt{\lambda}}{2 \pi \uh \sqrt{1{-}v^2} } \,
\frac{v q_{\perp}}{q} \,
\bigg [\frac{i}{r q \uh} + \frac{1{-}4 r^2}{4 r^2} 
+ \O(q) \bigg ] 
\\ &{} \times
(2 \pi) \delta(\omega-\v \cdot \q) \>\eps_1 \,.
\end{align}

\subsection{Scalar Mode}

For the scalar mode it is sufficient to expand the homogeneous 
solutions to Eq.~(\ref{eqm0}) to $\O(q)$.  Doing so one obtains
\begin{subequations}
\begin{align}
\phi_0^{(0)} &= \frac{4  (2 {-} 3 r^2)}{9 (1{-}r^2)^2} \, u^4
- \frac{(1{-}3r^2) [u^4 {+} \uh^4 (1{-}3 r^2)]}{9 (1{-}r^2)^2} \, \ln f \,,
\\ 
\phi_0^{(1)} &= 0 \,,
\end{align}
\end{subequations}
and
\begin{subequations}
\begin{align}
\psi_0^{(0)} ={}& \frac{ u^4 - (1{-}3 r^2)\uh^4}{(2 {-} 3 r^2) \uh^4} \,,
\\ \nonumber
\psi_0^{(1)} ={}&\frac{i r}{4 \uh^3 (2 {-} 3 r^2)}
\bigg \{
    (4{+}\ln 4) u^4
    -\uh^4 \left(r^2 \ln64 {+}4{-}\ln 4\right) 
\\ &\kern 1.9cm {}
    -\left[u^4{+}\left(1{-}3 r^2\right) \uh^4\right] \ln f
\bigg \}\,.
\end{align}
\end{subequations}
The corresponding small momentum limit of the Wronskian is
\begin{align}
W_0 ={}& \frac{u^3 [u^4 {-} 3(1{-}r^2) \uh^4]^2}{(1{-}r^2)^2(2{-}3r^2) f}
\bigg[
-\frac{4(1{-}3 r^3)}{9 \uh^8}
\nonumber\\ &{}
+\frac{2 i r \left(r^2 \ln8 {-}\ln2{+}2\right)}{9 \uh^7} \, q
+\O(q^2)
\bigg] \,.
\end{align}
Substituting the above expansions into the definition of $\Delta Z_0^{(4)}$ in Eq.~(\ref{grnsfcnrep0})
and performing the radial integral, one finds
\begin{align}
\Delta Z_0^{(4)} ={}& \frac{\kappa_5^2 \sqrt{\lambda}}{2 \pi L^3 \sqrt{1{-}v^2}}
\frac{2 - 3 r^2 +v^2}{(1{-}3 r^2)\uh} \,
(2 \pi) \delta(\omega - \v \cdot \q)
\nonumber\\ &{} \times
\bigg [ -\frac{i r q}{\uh} 
+\frac{9 r^2 (1{-}r^2)^2}{4 (1{-}3 r^2)} 
+ \O(q^3)
\bigg ] \,.
\end{align}
Inserting $\Delta Z_0^{(4)}$ into Eq.~(\ref{t0}), we obtain
\begin{align}
\nonumber
\Delta \T_0 ={}& \frac{3 \sqrt{\lambda}}{2 \pi \uh \sqrt{1{-}v^2}} \>
(2 \pi) \delta(\omega - \v \cdot \q)
\\ &{}\times
\bigg [
\frac{-i r (1{+}v^2)}{(1{-}3 r^2) \uh q}
+\frac{r^2 (2 {-} 3r^2 {+}v^2)}{(1{-}3r^2)^2} 
+ \O(q)
\bigg] \,.
\end{align}

\vfill

\bibliographystyle{JHEP}
\bibliography{refs}

\providecommand{\href}[2]{#2}\begingroup\raggedright\begin{thebibliography}{10}

\bibitem{Shuryak}
E.~Shuryak, {\it Why does the quark gluon plasma at {RHIC} behave as a nearly
  ideal fluid?},  {\em Prog. Part. Nucl. Phys.} {\bf 53} (2004) 273--303,
  \href{http://xxx.lanl.gov/abs/hep-ph/0312227}{{\tt hep-ph/0312227}}.

\bibitem{Shuryak:2004cy}
E.~V. Shuryak, {\it What {RHIC} experiments and theory tell us about properties
  of quark-gluon plasma?},  {\em Nucl. Phys.} {\bf A750} (2005) 64--83,
  \href{http://xxx.lanl.gov/abs/hep-ph/0405066}{{\tt hep-ph/0405066}}.

\bibitem{Casalderrey-Solana:2006rq}
J.~Casalderrey-Solana and D.~Teaney, {\it Heavy quark diffusion in strongly
  coupled {$\mathcal N = 4$} {Y}ang {M}ills},  {\em Phys. Rev.} {\bf D74}
  (2006) 085012, \href{http://xxx.lanl.gov/abs/hep-ph/0605199}{{\tt
  hep-ph/0605199}}.

\bibitem{Herzog:2006gh}
C.~P. Herzog, A.~Karch, P.~Kovtun, C.~Kozcaz, and L.~G. Yaffe, {\it Energy loss
  of a heavy quark moving through {$\mathcal N = 4$} supersymmetric
  {Yang-Mills} plasma},  {\em JHEP} {\bf 07} (2006) 013,
  \href{http://xxx.lanl.gov/abs/hep-th/0605158}{{\tt hep-th/0605158}}.

\bibitem{Policastro:2002tn}
G.~Policastro, D.~T. Son, and A.~O. Starinets, {\it From {AdS/CFT}
  correspondence to hydrodynamics. {II:} {S}ound waves},  {\em JHEP} {\bf 12}
  (2002) 054, \href{http://xxx.lanl.gov/abs/hep-th/0210220}{{\tt
  hep-th/0210220}}.

\bibitem{Policastro:2001yc}
G.~Policastro, D.~T. Son, and A.~O. Starinets, {\it The shear viscosity of
  strongly coupled {$\mathcal N = 4$} supersymmetric {Yang-Mills} plasma},
  {\em Phys. Rev. Lett.} {\bf 87} (2001) 081601,
  \href{http://xxx.lanl.gov/abs/hep-th/0104066}{{\tt hep-th/0104066}}.

\bibitem{CaronHuot:2006te}
S.~Caron-Huot, P.~Kovtun, G.~D. Moore, A.~Starinets, and L.~G. Yaffe, {\it
  Photon and dilepton production in supersymmetric {Y}ang-{M}ills plasma},
  {\em JHEP} {\bf 12} (2006) 015,
  \href{http://xxx.lanl.gov/abs/hep-th/0607237}{{\tt hep-th/0607237}}.

\bibitem{Liu:2006ug}
H.~Liu, K.~Rajagopal, and U.~A. Wiedemann, {\it Calculating the jet quenching
  parameter from {AdS/CFT}},  {\em Phys. Rev. Lett.} {\bf 97} (2006) 182301,
  \href{http://xxx.lanl.gov/abs/hep-ph/0605178}{{\tt hep-ph/0605178}}.

\bibitem{Peeters:2006iu}
K.~Peeters, J.~Sonnenschein, and M.~Zamaklar, {\it Holographic melting and
  related properties of mesons in a quark gluon plasma},  {\em Phys. Rev.} {\bf
  D74} (2006) 106008, \href{http://xxx.lanl.gov/abs/hep-th/0606195}{{\tt
  hep-th/0606195}}.

\bibitem{Gubser:2006nz}
S.~S. Gubser, {\it Jet-quenching and momentum correlators from the gauge-
  string duality},  \href{http://xxx.lanl.gov/abs/hep-th/0612143}{{\tt
  hep-th/0612143}}.

\bibitem{Hatta:2007cs}
Y.~Hatta, E.~Iancu, and A.~H. Mueller, {\it Deep inelastic scattering off a
  {$\mathcal N=4$} {SYM} plasma at strong coupling},
  \href{http://xxx.lanl.gov/abs/arXiv:0710.5297}{{\tt arXiv:0710.5297
  [hep-th]}}.

\bibitem{Bak:2007fk}
D.~Bak, A.~Karch, and L.~G. Yaffe, {\it Debye screening in strongly coupled
  {$\mathcal N=4$} supersymmetric {Yang-Mills} plasma},  {\em JHEP} {\bf 08}
  (2007) 049, \href{http://xxx.lanl.gov/abs/arXiv:0705.0994}{{\tt
  arXiv:0705.0994 [hep-th]}}.

\bibitem{Peeters:2007ti}
K.~Peeters and M.~Zamaklar, {\it Dissociation by acceleration},
  \href{http://xxx.lanl.gov/abs/arXiv:0711.3446}{{\tt arXiv:0711.3446
  [hep-th]}}.

\bibitem{Caceres:2006dj}
E.~Caceres and A.~Guijosa, {\it Drag force in charged {$\mathcal N = 4$ SYM}
  plasma},  {\em JHEP} {\bf 11} (2006) 077,
  \href{http://xxx.lanl.gov/abs/hep-th/0605235}{{\tt hep-th/0605235}}.

\bibitem{CasalderreySolana:2007qw}
J.~Casalderrey-Solana and D.~Teaney, {\it Transverse momentum broadening of a
  fast quark in a {$\mathcal N = 4$} {Yang Mills} plasma},  {\em JHEP} {\bf 04}
  (2007) 039, \href{http://xxx.lanl.gov/abs/hep-th/0701123}{{\tt
  hep-th/0701123}}.

\bibitem{Janik:2006gp}
R.~A. Janik and R.~Peschanski, {\it Gauge / gravity duality and thermalization
  of a boost- invariant perfect fluid},  {\em Phys. Rev.} {\bf D74} (2006)
  046007, \href{http://xxx.lanl.gov/abs/hep-th/0606149}{{\tt hep-th/0606149}}.

\bibitem{Heller:2007qt}
M.~P. Heller and R.~A. Janik, {\it Viscous hydrodynamics relaxation time from
  ads/cft},  {\em Phys. Rev.} {\bf D76} (2007) 025027,
  \href{http://xxx.lanl.gov/abs/hep-th/0703243}{{\tt hep-th/0703243}}.

\bibitem{Aharony:1999ti}
O.~Aharony, S.~S. Gubser, J.~M. Maldacena, H.~Ooguri, and Y.~Oz, {\it Large
  {$N$} field theories, string theory and gravity},  {\em Phys. Rept.} {\bf
  323} (2000) 183--386, \href{http://xxx.lanl.gov/abs/hep-th/9905111}{{\tt
  hep-th/9905111}}.

\bibitem{Maldacena:1997re}
J.~M. Maldacena, {\it The large {$N$} limit of superconformal field theories
  and supergravity},  {\em Adv. Theor. Math. Phys.} {\bf 2} (1998) 231--252,
  \href{http://xxx.lanl.gov/abs/hep-th/9711200}{{\tt hep-th/9711200}}.

\bibitem{Meyer:2007ic}
H.~B. Meyer, {\it A calculation of the shear viscosity in {SU(3)}
  gluodynamics},  {\em Phys. Rev.} {\bf D76} (2007) 101701,
  \href{http://xxx.lanl.gov/abs/arXiv:0704.1801}{{\tt arXiv:0704.1801
  [hep-lat]}}.

\bibitem{Shuryak:2006ii}
E.~Shuryak, {\it The conical flow from quenched jets in {sQGP}},  {\em Nucl.
  Phys.} {\bf A783} (2007) 31--38,
  \href{http://xxx.lanl.gov/abs/nucl-th/0609013}{{\tt nucl-th/0609013}}.

\bibitem{Adler:2005ee}
{\bf PHENIX} Collaboration, S.~S. Adler {\em et~al.}, {\it Modifications to
  di-jet hadron pair correlations in au + au collisions at {$s_{NN}^{1/2} =
  200$} {GeV}},  {\em Phys. Rev. Lett.} {\bf 97} (2006) 052301,
  \href{http://xxx.lanl.gov/abs/nucl-ex/0507004}{{\tt nucl-ex/0507004}}.

\bibitem{Leitch:2006ex}
M.~J. Leitch, {\it Latest results on the hot-dense partonic matter at {RHIC}},
  {\em Eur. Phys. J.} {\bf A31} (2007) 868--874,
  \href{http://xxx.lanl.gov/abs/nucl-ex/0610015}{{\tt nucl-ex/0610015}}.

\bibitem{CasalderreySolana:2006sq}
J.~Casalderrey-Solana, E.~V. Shuryak, and D.~Teaney, {\it Hydrodynamic flow
  from fast particles},  \href{http://xxx.lanl.gov/abs/hep-ph/0602183}{{\tt
  hep-ph/0602183}}.

\bibitem{Herzog:2006se}
C.~P. Herzog, {\it Energy loss of heavy quarks from asymptotically {AdS}
  geometries},  {\em JHEP} {\bf 09} (2006) 032,
  \href{http://xxx.lanl.gov/abs/hep-th/0605191}{{\tt hep-th/0605191}}.

\bibitem{Friess:2006fk}
J.~J. Friess, S.~S. Gubser, G.~Michalogiorgakis, and S.~S. Pufu, {\it The
  stress tensor of a quark moving through {$\mathcal N = 4$} thermal plasma},
  \href{http://xxx.lanl.gov/abs/hep-th/0607022}{{\tt hep-th/0607022}}.

\bibitem{Gubser:2007nd}
S.~S. Gubser and S.~S. Pufu, {\it Master field treatment of metric
  perturbations sourced by the trailing string},
  \href{http://xxx.lanl.gov/abs/hep-th/0703090}{{\tt hep-th/0703090}}.

\bibitem{Yarom:2007ni}
A.~Yarom, {\it On the energy deposited by a quark moving in an {$\mathcal N =
  4$} {SYM} plasma},  \href{http://xxx.lanl.gov/abs/hep-th/0703095}{{\tt
  hep-th/0703095}}.

\bibitem{Chesler:2007an}
P.~M. Chesler and L.~G. Yaffe, {\it The wake of a quark moving through a
  strongly-coupled {$\mathcal N=4$} supersymmetric {Yang-Mills} plasma},
  \href{http://xxx.lanl.gov/abs/arXiv:0706.0368}{{\tt arXiv:0706.0368
  [hep-th]}}.

\bibitem{Gubser:2007xz}
S.~S. Gubser, S.~S. Pufu, and A.~Yarom, {\it Energy disturbances due to a
  moving quark from gauge-string duality},  {\em JHEP} {\bf 09} (2007) 108,
  \href{http://xxx.lanl.gov/abs/arXiv:0706.0213}{{\tt arXiv:0706.0213
  [hep-th]}}.

\bibitem{Gubser:2007ga}
S.~S. Gubser, S.~S. Pufu, and A.~Yarom, {\it Sonic booms and diffusion wakes
  generated by a heavy quark in thermal {AdS/CFT}},
  \href{http://xxx.lanl.gov/abs/arXiv:0706.4307}{{\tt arXiv:
  0706.4307 [hep-th]}}.

\bibitem{Gubser:2007ni}
S.~S. Gubser and A.~Yarom, {\it Universality of the diffusion wake in the
  gauge-string duality},  \href{http://xxx.lanl.gov/abs/arXiv:0709.1089}{{\tt arXiv:0709.1089 [hep-th]}}.

\bibitem{Arnold:2002zm}
P.~Arnold, G.~D. Moore, and L.~G. Yaffe, {\it Effective kinetic theory for high
  temperature gauge theories},  {\em JHEP} {\bf 01} (2003) 030,
  \href{http://xxx.lanl.gov/abs/hep-ph/0209353}{{\tt hep-ph/0209353}}.

\bibitem{Betz:2007ie}
B.~Betz, P.~Rau, and H.~Stocker, {\it Mach cones and hydrodynamic flow: Probing
  big bang matter in the laboratory},
  \href{http://xxx.lanl.gov/abs/arXiv:0707.3942}{{\tt arXiv:0707.3942
  [hep-th]}}.

\bibitem{Baeuchle:2007qw}
B.~Baeuchle, L.~P. Csernai, and H.~Stoecker, {\it {MACE} -- {Mach} cones in
  heavy ion collisions},  \href{http://xxx.lanl.gov/abs/arXiv:0710.1476}{{\tt arXiv:0710.1476 [nucl-th]}}.

\bibitem{Kovtun:2004de}
P.~Kovtun, D.~T. Son, and A.~O. Starinets, {\it Viscosity in strongly
  interacting quantum field theories from black hole physics},  {\em Phys. Rev.
  Lett.} {\bf 94} (2005) 111601,
  \href{http://xxx.lanl.gov/abs/hep-th/0405231}{{\tt hep-th/0405231}}.

\bibitem{Karch:2002sh}
A.~Karch and E.~Katz, {\it Adding flavor to {AdS/CFT}},  {\em JHEP} {\bf 06}
  (2002) 043, \href{http://xxx.lanl.gov/abs/hep-th/0205236}{{\tt
  hep-th/0205236}}.

\bibitem{Gubser:1998bc}
S.~S. Gubser, I.~R. Klebanov, and A.~M. Polyakov, {\it Gauge theory correlators
  from non-critical string theory},  {\em Phys. Lett.} {\bf B428} (1998)
  105--114, \href{http://xxx.lanl.gov/abs/hep-th/9802109}{{\tt
  hep-th/9802109}}.

\bibitem{Witten:1998qj}
E.~Witten, {\it Anti-de {S}itter space and holography},  {\em Adv. Theor. Math.
  Phys.} {\bf 2} (1998) 253--291,
  \href{http://xxx.lanl.gov/abs/hep-th/9802150}{{\tt hep-th/9802150}}.

\bibitem{Skenderis:2000in}
K.~Skenderis, {\it Asymptotically anti-de {S}itter spacetimes and their stress
  energy tensor},  {\em Int. J. Mod. Phys.} {\bf A16} (2001) 740--749,
  \href{http://xxx.lanl.gov/abs/hep-th/0010138}{{\tt hep-th/0010138}}.

\bibitem{Gibbons:1976ue}
G.~W. Gibbons and S.~W. Hawking, {\it Action integrals and partition functions
  in quantum gravity},  {\em Phys. Rev.} {\bf D15} (1977) 2752--2756.

\bibitem{Chesler:2006gr}
P.~M. Chesler and A.~Vuorinen, {\it Heavy flavor diffusion in weakly coupled
  {$\mathcal N = 4$} super {Yang-Mills} theory},  {\em JHEP} {\bf 11} (2006)
  037, \href{http://xxx.lanl.gov/abs/hep-ph/0607148}{{\tt hep-ph/0607148}}.

\bibitem{Callan:1997kz}
C.~G. Callan and J.~M. Maldacena, {\it Brane dynamics from the {B}orn-{I}nfeld
  action},  {\em Nucl. Phys.} {\bf B513} (1998) 198--212,
  \href{http://xxx.lanl.gov/abs/hep-th/9708147}{{\tt hep-th/9708147}}.

\bibitem{Gibbons:1997xz}
G.~W. Gibbons, {\it Born-{I}nfeld particles and {D}irichlet p-branes},  {\em
  Nucl. Phys.} {\bf B514} (1998) 603--639,
  \href{http://xxx.lanl.gov/abs/hep-th/9709027}{{\tt hep-th/9709027}}.

\bibitem{Liu:1998bu}
H.~Liu and A.~A. Tseytlin, {\it {$D = 4$} super {Yang-Mills}, {$D = 5$} gauged
  supergravity, and {$D = 4$} conformal supergravity},  {\em Nucl. Phys.} {\bf
  B533} (1998) 88--108, \href{http://xxx.lanl.gov/abs/hep-th/9804083}{{\tt
  hep-th/9804083}}.

\bibitem{Kovtun:2005ev}
P.~K. Kovtun and A.~O. Starinets, {\it Quasinormal modes and holography},  {\em
  Phys. Rev.} {\bf D72} (2005) 086009,
  \href{http://xxx.lanl.gov/abs/hep-th/0506184}{{\tt hep-th/0506184}}.

\bibitem{Son:2002sd}
D.~T. Son and A.~O. Starinets, {\it Minkowski-space correlators in {AdS/CFT}
  correspondence: Recipe and applications},  {\em JHEP} {\bf 09} (2002) 042,
  \href{http://xxx.lanl.gov/abs/hep-th/0205051}{{\tt hep-th/0205051}}.

\bibitem{Lin:2007pv}
S.~Lin and E.~Shuryak, {\it Stress tensor of static dipoles in strongly coupled
  $\cal{N}$=4 gauge theory},  \href{http://xxx.lanl.gov/abs/arXiv:0707.3135}{{\tt arXiv:0707.3135 [hep-th]}}.

\bibitem{Yarom:2007ap}
A.~Yarom, {\it The high momentum behavior of a quark wake},
  \href{http://xxx.lanl.gov/abs/hep-th/0702164}{{\tt hep-th/0702164}}.

\end{thebibliography}\endgroup

\end{document}